\documentclass[12pt,preprint,colordvi]{aastex}

\usepackage{epsf}

\usepackage{color}

\usepackage{amsmath}
\usepackage{amssymb}

\newcommand{\Msun}{\mbox{M$_\odot$}}

\newcommand{\sfrac}[2]{\mathchoice
                      {\kern0em\raise.5ex\hbox{\the\scriptfont0 #1}\kern-.15em/
                        \kern-.15em\lower.25ex\hbox{\the\scriptfont0 #2}}
                      {\kern0em\raise.5ex\hbox{\the\scriptfont0 #1}\kern-.15em/
                        \kern-.15em\lower.25ex\hbox{\the\scriptfont0 #2}}
                      {\kern0em\raise.5ex\hbox{\the\scriptscriptfont0 #1}\kern-.2em/
                        \kern-.15em\lower.25ex\hbox{\the\scriptscriptfont0 #2}}
                      {#1\!/#2}}

\newcommand{\xb}{\ensuremath{\mathbf{x}}}

\begin{document}

\title{The Deflagration Stage of Chandrasekhar Mass Models For Type Ia
  Supernovae: I. Early Evolution}

\shorttitle{High-Resolution Studies of Deflagration in SN Ia}
\shortauthors{Malone et al.}

\author{C.~M.~Malone\altaffilmark{1},
        A.~Nonaka\altaffilmark{2},
        S.~E.~Woosley\altaffilmark{1},
        A.~S.~Almgren\altaffilmark{2},
        J.~B.~Bell\altaffilmark{2},
        S.~Dong\altaffilmark{1},
        M.~Zingale\altaffilmark{3}
}

\altaffiltext{1}{Department of Astronomy \& Astrophysics,
                 The University of California, Santa Cruz,
                 Santa Cruz, CA 95064}

\altaffiltext{2}{Center for Computational Sciences and Engineering,
                 Lawrence Berkeley National Laboratory,
                 Berkeley, CA 94720}

\altaffiltext{3}{Department of Physics \& Astronomy,
                 Stony Brook University,
                 Stony Brook, NY 11794}

\begin{abstract}
We present high-resolution, full-star simulations of the post-ignition
phase of Type Ia supernovae using the compressible hydrodynamics code
Castro.  Initial conditions, including the turbulent velocity field
and ignition site, are imported directly from a simulation of the last
few hours of presupernova convection using a low Mach number code,
Maestro.  Adaptive mesh refinement allows the initial burning front to
be modeled with an effective resolution of 36,864$^3$ zones
(136~m zone$^{-1}$).  The initial rise and expansion of the
deflagration front are tracked until burning reaches the star's edge
and the role of the background turbulence on the flame is
investigated.  The effect of artificially moving the ignition location
closer to the star's center is explored.  The degree to which
turbulence affects the burning front decreases with increasing
ignition radius since the buoyancy force is stronger at larger radii.
Even central ignition --- in the presence of a background convective
flow field --- is rapidly carried off-center as the flame is carried
by the flow field. We compare our results to analytic models for
burning thermals, and find that they reproduce the general trends of
the bubble's size and mass, but underpredict the amount of buoyant
acceleration due to simplifying assumptions of the bubble's
properties. Overall, we find that the amount of mass that burns prior
to flame break out is small, consistent with a ``gravitationally
confined detonation'' occurring at a later epoch, but additional
burning will occur following breakout that may modify this conclusion.
\end{abstract}
\keywords{supernovae: general --- white dwarfs --- hydrodynamics ---
          nuclear reactions, nucleosynthesis, abundances --- convection ---
          methods: numerical}

\section{Introduction}

It is generally agreed that a Type Ia supernova (SN Ia) is the
thermonuclear explosion of a carbon-oxygen white dwarf driven to
instability by the accretion of mass from a binary companion, but the
nature of the progenitor star and just how the burning ignites and
spreads is debated
\citep[e.g.][]{hillebrandtniemeyer2000,hillebrandt:2013}.  Here we
focus on the ``single degenerate, Chandrasekhar mass'' (MCh) model
where the white dwarf accretes material from its binary companion
slowly enough to reach 1.39 \Msun \ and ignite a subsonic carbon
runaway near its center.  As the temperature rises in response to the
burning, the white dwarf expands very little because the matter is
degenerate. A growing convection zone develops, however, and carries
the excess energy from the center outwards where it gradually heats an
increasing fraction of the white dwarf to high, but not explosively
high temperature \citep{Bar04}.  After about a century of this
simmering \citep{Woosley:2004}, the central energy generation rate
exceeds what can be carried away by convection and the burning becomes
increasingly localized. The temperature gradient steepens rapidly,
reflecting the high power upon which the carbon fusion rate
depends. Eventually a discontinuity develops between fuel and ash and a
``flame'', or ``deflagration'' is born.

Extensive literature exists regarding the subsequent propagation of
these flames, once ignited, but the geometry assumed by various
authors for that ignition varies greatly. Some assume ignition at a
single point at, or very near the center
\citep{khokhlov:1995,niemeyer:1996,gamezo:2005,townsley_etal:2007,
  jackson2010,bkk2012,Ma:2013}; others employ many ignition kernels
distributed throughout an extended region of the stellar interior
\citep{garciasenz:2005,roepke2005,roepke-multispot,roepke2007,Kromer_etal2013}.
It is generally agreed, however, that in the absence of detonation,
the ensuing deflagration by itself produces too little $^{56}$Ni and
ejects matter that is too well-mixed when compared to the stratified
ejecta observed in typical SN Ia spectra (although see
\cite{Kromer_etal2013} for a recent pure deflagration model applied to
a class of sub-luminous SNe Ia).  Simulations that successfully match
supernova observations therefore usually insert, by hand, a transition
to detonation once the deflagration has reached some threshold
condition.  This ``deflagration-to-detonation transition'' (DDT)
\citep{khokhlov:1991} produces an ejecta velocity structure with
sufficient intermediate mass elements (IME) and iron-group elements
(IGE) to reasonably match observations of typical SN Ia.

Since the detonation always follows a deflagration that sets the
initial conditions, the final SN Ia model is sensitive to how the
subsonic burning is modeled and, as we shall see, to exactly how it is
ignited. The simplest assumption would be to ignite the runaway at the
dead center, but this overlooks the fact that matter there is not at
rest. As the matter moves around, it carries within it the ``sparks''
that can potentially develop into an explosion. The idea that the
convective flow would naturally lead to off-center ignition was first
explored by \citet{garciasenzwoosley1995} who estimated a displacement
from the center of several hundred km.  \cite{hoflichstein:2002}
modeled a two-dimensional wedge of the white dwarf with the center cut
out, and found inwardly moving convective plumes caused compressional
heating that triggered a runaway in a single downward moving hot spot
very near the center.  Three-dimensional simulations using an
anelastic hydrodynamics code \citep{kuhlen-ignition:2005} painted a
somewhat different picture however, finding an overall dipolar nature
for the convective flow and ignition in an {\sl outwards} bound flow,
again about 100 km off center. This was consistent with analytic
estimates \citep{Woosley:2004,wunschwoosy:2004}, but was somewhat
sensitive to the need to prescribe an artificial ``hole'' in the center
of the anelastic hydro grid (which used spectral coordinates), and thus
questionable.

More recently, full star simulations have been performed
\citep[hereafter, N12]{paper4,wdconvect,wdturb} of the last few hours
of convection using the low Mach number stellar code Maestro
\citep{multilevel} and a Cartesian grid with adaptive mesh refinement
(AMR).  These calculations, done so far only for non-rotating or
slowly rotating white dwarfs, clearly show ignition only happening a
single time and in a very small region. The most likely off-center
displacement for the ignition lies in a range 40 to 75 km, with a
small probability extending inwards to the origin and outwards to 110
km. That is, the final runaway, while always starting at a single
point, is chaotic in its location. There is not a single prescribed
radius, but a range of possibilities depending on random flows just
prior to runaway. The distinction between central ignition and
ignition even slightly off center is dramatic. Once the star begins to
burn on one side, it never burns, as a deflagration, in the inner
regions of the other side. The resulting supernova is inherently
asymmetric.

In this paper, we use the compressible hydrodynamics code, Castro
\citep{castro}, to follow the evolution from the time the first point
runs away up to break out of the deflagration through the surface of
the star. While only part of a full explosion model, this is a
sufficiently complex problem to justify its own careful analysis.  It
is also an important step in our ultimate goal, a full, end-to-end
treatment of the explosion of Chandrasekhar mass white dwarfs, from
convection through homologous coasting as supernovae.  In this paper,
however, we do not include any deflagration-to-detonation transition
and do not calculate ejecta composition or light curves.

Since the Maestro and Castro codes both use the same BoxLib
software framework \citep{BoxLib,rendleman-hyper,scidac-scaling}, it
is straightforward to import the turbulent ignition conditions
directly from the convection simulation into Castro,
while retaining the same initial grid structure and data.  Previous
work of this sort was done in two dimensions with an artificial
velocity field \citep{Livne:2005}. Here we investigate the role of a
realistic, three-dimensional background convective flow field on the
flame as it rises and expands towards the stellar surface.  We also
compare our results to analytic models developed for burning thermals
on small scales in the presence of turbulence and gravitational
stratification from \cite{BurningThermals}.

An additional advantage of the new study is the very fine spatial
resolution employed. This is accomplished by using several levels of
AMR and made feasible by early access to the NSF Blue Waters machine.
Although still many orders of magnitude too coarse to resolve the flame
thickness, the resolution used here is fine enough that the initial
laminar propagation of the flame is well resolved before it begins to
float and become highly turbulent. During this laminar phase, existing
perturbations burn out and the ash exists for a time as an almost
perfectly spherical bubble. This is important because, as we shall
see, the outcome is quite sensitive not only to the location of the
ignition point, but how perturbations of the flame surface evolve.

\section{Numerical Setup}
\label{sec:numerics}

For our study, we used the compressible multi-dimensional
hydrodynamics code Castro \citep{castro}.  This code has been
optimized to scale well on upwards of 100,000 processors
\citep{scidac-scaling} and is capable of employing multiple levels of
AMR with subcycling in time.  As initial input, the 4.34 km
zone$^{-1}$ model (hereafter the N12 model) of \cite{wdturb} was
employed. This model of a Chandrasekhar mass white dwarf, calculated
using the Maestro code, ignited a runaway at a representative radius
of 41.3 km.  The Maestro simulation used two levels of AMR with a
finest resolution of 4.34~km (effectively 1152$^3$ zones) and that is
the same zoning used in the initial remap into Castro.  The Maestro
data is employed for the entire star, including the turbulent velocity
field, temperature, density, and pressure. The equation of state is
the same but the treatment of nuclear burning differs. Maestro used a
simple reaction network adequate only for carbon burning. The present
study uses tables evaluated using a much larger network (see Section
\ref{sec:microphysics}). For simplicity, the composition of the N12
model was reset everywhere to be a 50-50 mixture of $^{12}$C and
$^{16}$O.  This is a reasonable approximation since the burning during
the simmering phase of the N12 simulation only decreased the
carbon/oxygen mass fractions from 0.5 to 0.49.

The link time, when the transition was made from Maestro to Castro,
was when the temperature in any zone in the N12 simulation first
exceeded $8\times10^8$ K. Other nearby zones were too cool to run away
in the same time frame. As discussed in \cite{wdturb}, this transition
time was 0.57 s earlier than the time when the actual runaway
occurred, as defined by the first production of iron-group nuclei and the
formation of a flame at $T\sim 1.1\times10^9$ K. Following the
evolution with Maestro beyond this point would have violated the time
step limits in the code (burning time shorter than the time for a
fluid element to cross the zone) and that was when the data dump was
made.  Continued evolution with this criterion relaxed showed that the
hot spot moved less than two zones before flashing to very high
temperature. We do not think this 0.57 s (one to two zone) offset
affects our results in any important way.

\subsection{Castro Initialization from Maestro Data}
\label{sec:init-m2c}

Although they share the same underlying data structures, Maestro and
Castro solve fundamentally different hydrodynamic equations.  Unique
to Maestro is the concept of a one-dimensional, radial \textit{base
  state} of density, $\rho_0$, and pressure, $p_0$, that are in
hydrostatic equilibrium and represent the ``average'' state as a
function of radius (see \citealt{multilevel} for details).  Using low
Mach number asymptotics, the pressure is decomposed into a base state
pressure and a perturbational pressure, $p(\xb,t) = p_0(r,t) +
\pi(\xb,t)$, where the base state pressure is used in place of the
full state pressure everywhere except in the momentum equation.  This
replacement effectively allows the base state pressure to govern the
thermodynamics of the fluid while the dynamics are driven by the
perturbational pressure.  Next, the equation of state is recast as a
divergence constraint on the velocity field.  Using this model, sound
waves are filtered from the system, thus allowing for larger time
steps under the {\it advective}, rather than acoustic, Courant
condition.

When transforming the low Mach number results to a fully compressible
code, one has several options for determining the thermodynamic state
of the fluid \citep{scidac-endtoend}.  For example, the pressure field
in the compressible data set could be set equal to $p_0$, $p_0+\pi$,
or determined from the equation of state (EOS) given the composition,
temperature, and density from the Maestro data set.  Therefore,
although the various methods of mapping from the Maestro code to the
Castro code may be analytically equivalent (to
$\mathcal{O}\left(M^2\right)$), there are inherent numerical
differences that may lead to thermodynamic inconsistencies.  Also, one
must take care when interpolating data from the one-dimensional,
radial base state to the three-dimensional Cartesian grid used to hold
the full state quantities. The approach we took, which minimizes the
development of spurious pressure waves from hot spots at the beginning
of the Castro simulation, is as follows:
\begin{enumerate}
  \item Extract the velocity, density, base state and dynamical
    pressure from the Maestro simulation at the time of ignition.
  \item Reset the composition to a 50 - 50 mixture of $^{12}$C and $^{16}$O.
  \item Use quadratic interpolation of the radial base state pressure
    and add the Cartesian grid-based dynamical pressure to formulate the full
    pressure field on the Cartesian grid.
  \item Use the pressure, density, and composition as input to the EOS
    to obtain the temperature and internal energy fields.
\end{enumerate}
We note that resetting the initial composition from the N12 model's
values effectively decreases the mean molecular weight in regions
where significant carbon burning was present in the N12 simulation.
This decrease in molecular weight corresponds to a decrease in
temperature of less than 5\% in those regions.  The thermal gradient
is much more sensitive to the pressure and density than the
composition, and therefore the thermal gradient is left essentially
unchanged with our resetting of the composition.  Regions that were
convectively unstable in the N12 simulation remain unstable in the
simulations presented here; however, we do not capture the central
carbon burning that is driving the convection.  In addition, the
resetting of the composition alters the $^{22}$Ne content of the N12
ash state, but this is likely to have little affect on the
deflagration evolution \citep{townsley2009}.

\begin{figure}
  \begin{center}
    \includegraphics[width=6.0in]{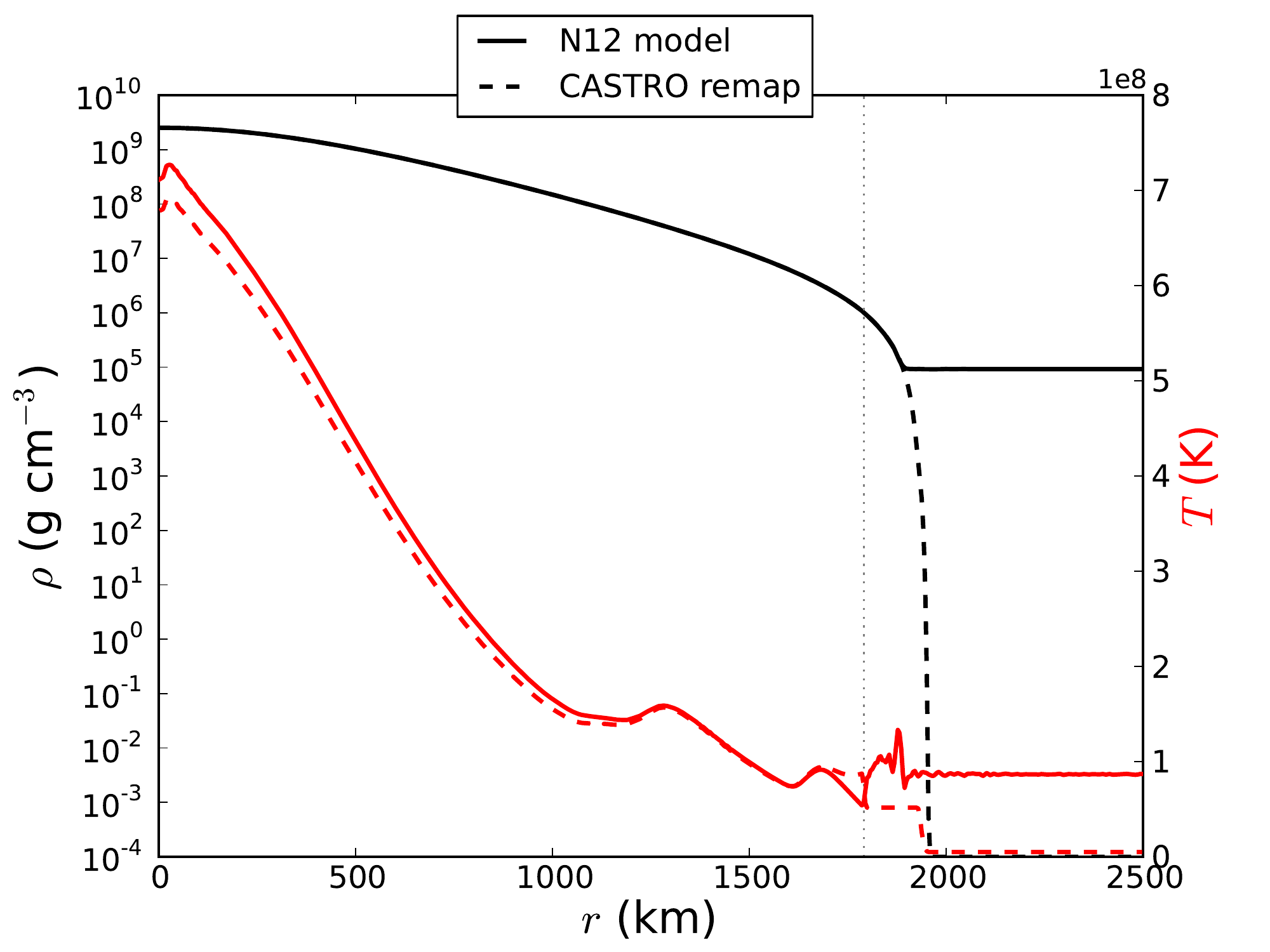}
    \caption{\label{fig:initial_models} Average radial profiles of
      density and temperature for the N12 model (solid lines) and our
      mapping of the N12 model into the Castro code (dashed lines).
      The slight decrease in temperature, especially towards the core,
      is a result of resetting the composition to a 50/50 mix of
      carbon and oxygen as described in the text.  The vertical dotted
      line shows the location where we reconstructed a model in HSE.}
  \end{center}
\end{figure}

In addition, care must be taken when importing data near the edge of
the star.  To prevent large velocities from developing in Maestro
below a specified low density cutoff, $\rho_c=10^6$ g cm$^{-3}$, the
algorithm changes its velocity divergence constraint to behave more
like that of an anelastic approximation.  Also, once the density has
fallen below $\rho=10^5$ g cm$^{-3}$, both $\rho_0$ and $p_0$ are held
constant (see \citealt{multilevel} for details).  Castro can handle
densities much lower than the typical cutoff density used in Maestro.
We therefore overwrote the Maestro data at the edge of the star to
drop the density down to a much smaller value than the Maestro cutoff
density.  The first step is to examine the radial base state density
from Maestro, $\rho_0$.  We find the radial coordinate, $r_c$, where
the density drops below $\rho_c$, and we use the base state density
and temperature in a 50 - 50 mix of $^{12}$C and $^{16}$O to create a
one-dimensional model in hydrostatic equilibrium (HSE) outward from
$r_c$, following the procedure of \cite{ppm-hse}.  The one-dimensional
model is then interpolated onto the full Cartesian grid, overwriting
the Maestro data for zones with $r \ge r_c$.  Figure
\ref{fig:initial_models} shows the average radial profiles of both
density (black) and temperature (red; right axis) for the N12 model
(solid) and our mapping of this model into Castro (dashed).  The
vertical dotted line marks the location where the density fell below
$\rho_c=10^6$ g cm$^{-3}$ in the N12 model, and we reconstruct the HSE
model.

\subsection{Microphysics}
\label{sec:microphysics}

The N12 model used as input and the simulations presented in this
paper both use the Helmholtz EOS of \cite{timmes_swesty:2000}, which
includes contributions from relativistic and non-relativistic
electrons of arbitrary degeneracy, ideal ions, radiation, and Coulomb
corrections.  Thermal diffusion was not present in the simulations of
N12, but is included here in order to move the flame (see Section
\ref{sec-flame}).

The treatment of nuclear reactions differs substantially between those
used to calculate the N12 model and this paper and warrants
explanation.  The N12 model used a very sparse reaction network
described in \cite{wdconvect}, which was based upon the work of
\cite{chamulak2008}.  This network included only the $^{12}$C+$^{12}$C
heavy ion reaction and a multiplier to account for the energy release
of subsequent reactions.  The ``ash'' composition of this artificial
reaction was a mixture of $^{13}$C, $^{16}$O, $^{20}$Ne, and $^{23}$Na
with an average atomic weight and charge of $A_\textrm{ash} = 18$ and
$Z_\textrm{ash} =8.8$.

Here, depending upon the density, burning may proceed through carbon,
neon, oxygen, and silicon burning, all the way to nuclear statistical
equilibrium (NSE). The ``ash'' can continue to evolve via electron
capture, alpha capture, and photodisintegration reactions even after
the burning front has passed. The necessary nuclear physics is
therefore more complex.  To make this problem tractable in three
dimensions, the tables described in \cite{Ma:2013} were employed.  To
briefly review, the nuclear burning is divided into two phases:
reactions inside the flame and reactions after the flame has passed.
Inside the flame, burning is assumed to occur at constant pressure,
with results that are most sensitive to the density and electron
fraction, $Y_e$, of the degenerate gas and very little to the initial
temperature.  The initial density and composition of the fuel thus
determine the energy yield and final composition which are computed
off-line using a large reaction network containing 188 isotopes up to
$A=72$.  To save space these abundances are packed into just seven
abundance variables: ``helium'' (for $A<12$), ``carbon'' ($12\le
A<16$), ``oxygen'' ($16\le A<20$), ``neon'' ($20\le A<24$),
``magnesium'' ($24\le A< 28$), ``silicon'' ($28\le A <48$, IME), and
``iron'' ($A>47$, IGE).  This treatment of the burning is applied when
the temperature rises by conduction to above $2\times10^9$ K in zones
with a density above $3\times10^6$ g cm$^{-3}$ and a carbon mass
fraction over 1\%.

After the flame has passed, energy can still be generated or absorbed
by further evolution of the NSE ashes of the flame.  This proceeds by
a combination of photodisintegration, recombination, $\beta$-decay,
positron decay and $e$-capture reactions, which depend on the
temperature, density, and $Y_e$ of the NSE ashes.  For this phase a
separate table was constructed offline using a NSE network with 127
isotopes up to $A=60$.  The changes from the NSE evolution are
interpolated from the table and then packed into the helium, silicon,
and iron groups of our abundance variables.  The table was called for
``ash'' zones defined by high temperature, the presence of iron, and
the absence of carbon. In particular the NSE tables were used when the
temperature was above $3\times10^9$~K, density above $\sim10^8$~g
cm$^{-3}$, and the combined helium and iron group abundance above
88\%.  The cutoff in helium plus iron abundance was determined
empirically to assure that the NSE table was only called on zones that
were actually in NSE, even at very high temperature where silicon and
calcium can exist in abundance..

\subsubsection{Thickened Flame Model}
\label{sec-flame}

For the conditions characterizing ignition inside a nearly
Chandrasekhar mass white dwarf, the thickness of the laminar
thermonuclear flame is less than a millimeter
\citep{timmeswoosley1992}.  Turbulence and other instabilities twist
and churn this thin flame sheet into a complex topology that increases
the area and the effective burning rate. Resolving these structures
while simulating large fractions of the star in three dimensions is
not computationally feasible.  In the astrophysical and chemical
combustion communities, such ``turbulent flames'' are usually handled
in one of two ways: 1) by spreading the flame over several grid zones,
creating a {\em thickened flame} whose propagation is determined by
advection, the energy release from burning, and thermal diffusion
across the flame front \citep[e.g.][]{khokhlov:1995}, or 2) by
treating the flame as a discontinuity, using {\em level sets} to
identify the flame surface, and solving advection equations to
propagate the flame \citep[e.g.][]{reinecke1999}.  Both methods have
their strengths. For simplicity, we follow here the work of
\citet{Ma:2013} and use the thickened flame model described therein.
Here we briefly review the basic principles.

In a thickened flame model, the combustion region is usually
identified by a {\em progress variable}, which characterizes the
degree of completion of the burning
\citep[e.g.][]{khokhlov:1995,townsley_etal:2007}.  In our model, the
progress variable is the carbon mass fraction which ranges from fuel
($X(^{12}\rm{C})=0.5$) to ash ($X(^{12}\rm{C})\lesssim0.01$).  The
thickened flame is artificially spread over a number of computational
zones, $n$, such that the flame thickness, $\delta = n\Delta x$, can
be resolved on a grid with spacing $\Delta x$.  By specifying both $n$
and the flame speed, $v_f$, a time scale is determined,
$\tau=\delta/v_f$, that allows the calculation of an effective
``opacity''
\begin{equation}  \label{eq:opacity}
  \kappa = \frac{c\tau}{3\rho\left(n\Delta x\right)^2},
\end{equation}
and coefficient of thermal conduction,
\begin{equation}\label{eq:thermalcond}
  K = \frac{4acT^3}{3\rho\kappa} = 4aT^3\frac{\left(n\Delta x\right)^2}{\tau},
\end{equation}
where $c$ is the speed of light and $a$, the radiation constant.  The
energy generation rate is then given simply by the need to burn up the 
carbon in the defined region on the specified time scale,
\begin{equation}\label{eq:enuc}
  \epsilon_{\rm nuc} = 9.64\times10^{23} \frac{d\left[BE/A\right]}{d\left[X\left(^{12}{\rm C}\right)\right]}
  \frac{d\left[X\left(^{12}{\rm C}\right)\right]}{dt} 
       {\rm \ erg\ g}^{-1} {\rm \ s}^{-1},
\end{equation}
where $d\left[BE/A\right]/d\left[X\left(^{12}{\rm
    C}\right)\right]\simeq\Delta \left(BE/A\right)/0.5$, and
$\Delta\left(BE/A\right)$ is the total change in nuclear binding
energy (in MeV) per nucleon when carbon fuel burns to ash at the given
density and $^{12}$C mass fraction. This quantity is determined from
interpolation in a table.  The rate of change of carbon abundance in
the flame is then $d\left[X\left(^{12}{\rm
    C}\right)\right]/dt\simeq-0.5/\tau$. For dense regions with very
little residual carbon ($X\left(^{12}{\rm C}\right)\lesssim0.02$),
this prescription gives too slow a decrease, so the remaining trace of
carbon is burned more quickly by using $d\left[X\left(^{12}{\rm
    C}\right)\right]/dt\simeq-X\left(^{12}{\rm C}\right)/\Delta t$,
with $\Delta t$ the time step.  Given some desired flame speed and
thickness, the local density and the nuclear yield table thus
determine a self-consistent thermal diffusion coefficient and energy
generation rate that move the flame front accordingly. In practice,
$n$ varied between three and eight.

It remains to choose the effective speed of the flame. While the
laminar speed sets a lower bound, there are other important
factors. The buoyant growth of Rayleigh-Taylor modes or local shear
and turbulence can wrinkle or stretch the flame front, increasing the
overall surface area for heat transport resulting in an increased
burning rate.  There are several ways this enhancement has been
treated in past supernova models \citep[see][for a
  discussion]{Ma:2013}. The speed could be chosen as to burn out any
sub-grid structure assuming the dominance of the Rayleigh-Taylor
instability in creating that structure
\citep{khokhlov:1995,zhang:2007}, i.e. $v_{\rm eff} \approx F (g_{\rm
  eff} \Delta x)^{1/2}$ with $g_{\rm eff}$ the acceleration due to
gravity times the Atwood number ($\sim10^9$ cm s$^{-2}$), $\Delta x$,
the grid scale (typically 10$^5$ to 10$^6$ cm), and $F$, a constant
less than 1. This gives flame speeds $\sim 100$ km s$^{-1}$. Or,
perhaps more physically, one could analyze the turbulent flows on the
grid in the vicinity of the flame and obtain a representative turbulent
speed at the grid scale
\citep[e.g.][]{schmidt2006a,schmidt2006b}. \cite{roepke:flame} finds
values for $v_{\rm eff}$ that vary with location, but whose
distribution peaks, again, typically at 100 km s$^{-1}$, but with an
extended tail well past 200 km s$^{-1}$. This approach, however,
requires additional coding that has not yet been implemented and
tested in Castro. In \citet{Ma:2013}, a constant value was assumed for
$v_{\rm eff}$, and it was found that the answer for turbulent
deflagration in the interior of the white dwarf was insensitive to the
value chosen for $v_{\rm eff}$ in the range 50 to 200 km s$^{-1}$.
Furthermore, Model BT of \citet{Ma:2013} used a simplified model
turbulent flame speed constructed from the local grid-scale turbulent
intensity and found that the resulting nucleosynthesis and bulk flame
properties depended more upon the initial conditions than on the choice
of a fixed or turbulence-based variable flame speed.

Here we continue to follow the constant flame speed procedure of
\citet{Ma:2013}.  The flame speed is set to a constant value, $v_f =
\max\left(50{\rm~km~ s}^{-1},v_l\right)$, where $v_l$ is the laminar
flame speed as determined in \citet{timmeswoosley1992}.  This speed is
consistent with the A50 simulations of central ignition from
\citet{Ma:2013}, but is a bit slower than typical turbulent speeds
($\sim 100$ km s$^{-1}$). The smaller value was used in order to
retain additional structure on our very fine grid and avoid undue
``fire polishing''.  In our next paper, we will explore a more
realistic flame speed distribution based upon a sub-grid turbulence
model.  In the Appendix to this paper we have included a few test
problems that ensure the flame is moving as expected under various
velocity-field conditions.  As discussed later in the paper, both the
Rayleigh-Taylor based model and the constant flame speed model may
become highly suspect once the burning breaks out and starts to spread
over the surface of the white dwarf.

\section{Simulations}\label{sec:simulations}

Our simulations begin with output from the N12 model calculated using
the Maestro code at $t=10562.5$ seconds, corresponding to the time
where the temperature first exceeds $8\times10^8$ K in a single
ignition zone.  The coordinate system has the center of the star as
its origin so the hottest point in the domain is located $r=41.34$ km
off-center.  Two main runs are carried out (hereafter the A series)
that use this point as the ignition location, one with the background
convective flow field (Model AV) and one with the initial velocity
field set to zero (Model A0).  Four similar calculations were also
carried out with the ignition point arbitrarily moved to 10 km
off-center (Models BV and B0) and to the center of the star (Models CV,
C0).  The A series of models was evolved the longest ($\sim0.8$ s)
while the series B and C simulations were only used to investigate
differences in the early evolution.  For computational efficiency, all
simulations calculated gravity using a monopole approximation as
described in \citep{castro}.  Table \ref{tab:models} summarizes the
properties of the various models.

The N12 output file from Maestro had one level of refinement (i.e., a
total of two levels), with its finest level having a resolution of
$4.34$~km zone$^{-1}$ in all regions with $\rho>5\times 10^7$~g
cm$^{-3}$. In the present study, the single ignition zone of N12 is
replaced by a highly resolved spherical bubble with an average radius
of $\sim2$ km, but with a perturbed surface. In order to have the
necessary $\sim10$'s of zones to resolve the bubble surface,
additional refinement well beyond that in the Maestro initial model
was necessary.  In addition, our thickened flame model (Section
\ref{sec-flame}) spreads the flame over several zones such that a
higher resolution corresponds to a more realistic (albeit still
under-resolved) flame.  This particular flame model has been used at
much coarser resolution \citep{Ma:2013,Dong:2013}, and the general
evolution and features of the burning model do not depend sensitvely
on the resolution used with the caveat of the initial laminar burning
phase discussed below discused in Section \ref{sec:seriesA}.  We note
that this laminar stage is very quickly burned through within a few
time steps (likely even unnoticed) in simulations with coarser
resolution, such as in \cite{Dong:2013}, for example, where the
ignition spot was much larger ($r=20$ km, compared to our $\sim2$ km),
but with perturbations of the same relative size.

\begin{figure}
  \begin{center}
    \includegraphics[width=0.32\textwidth]{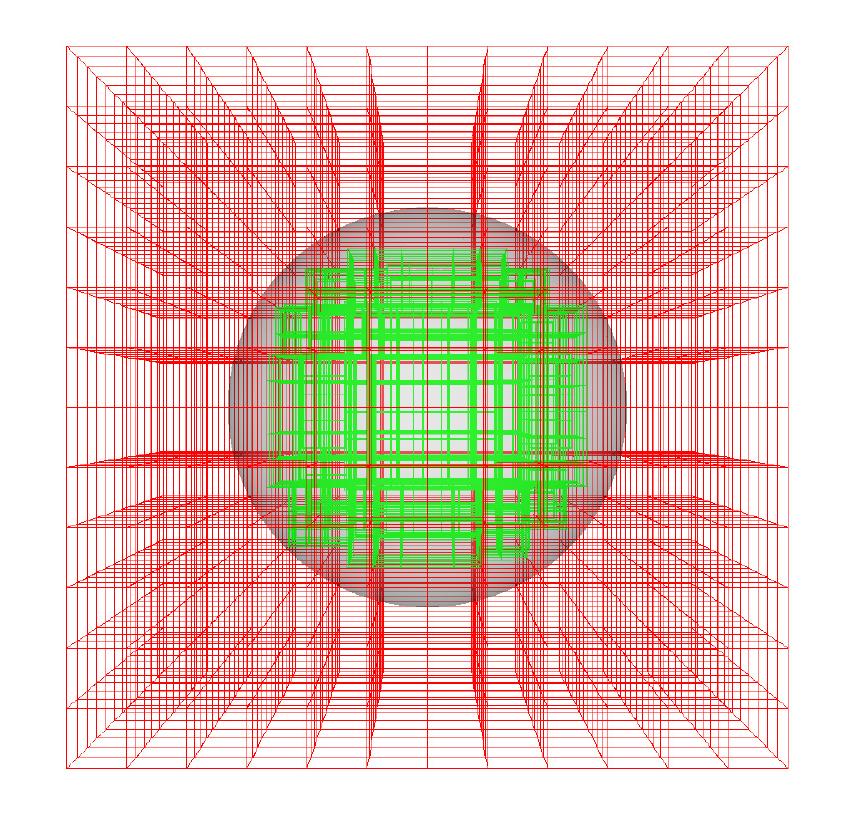}
    \includegraphics[width=0.32\textwidth]{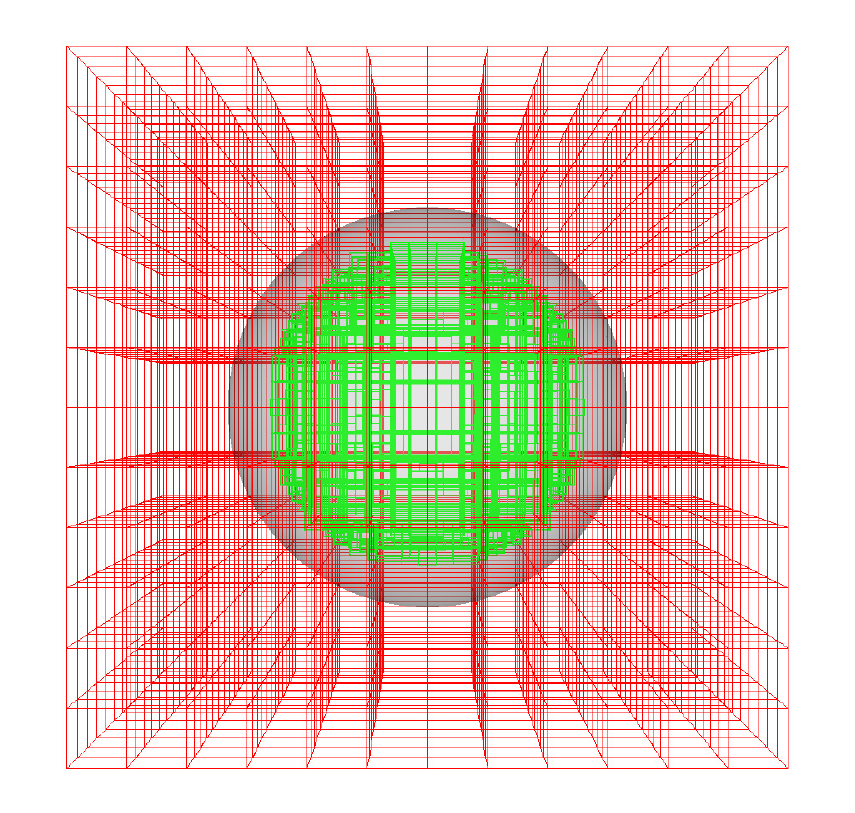}
    \includegraphics[width=0.32\textwidth]{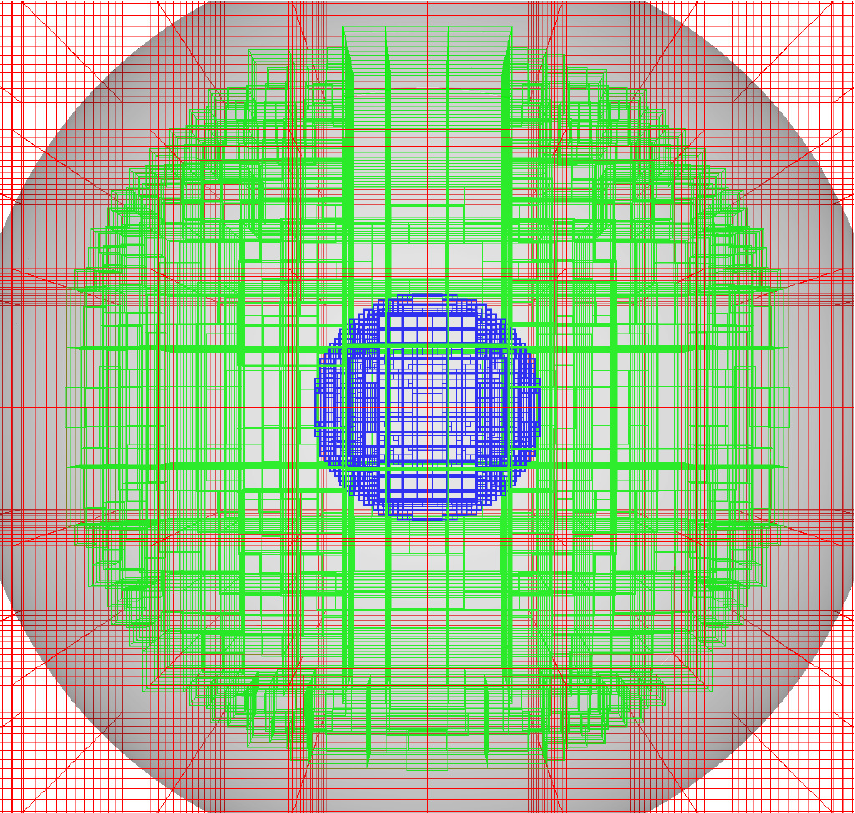}
    \includegraphics[width=0.32\textwidth]{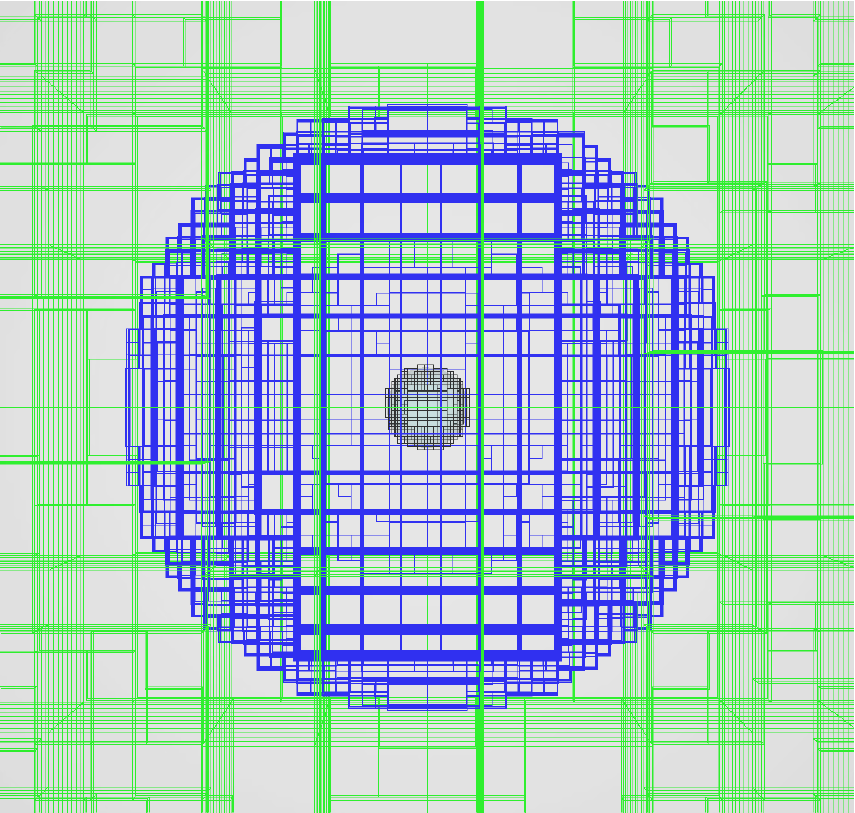}
    \includegraphics[width=0.32\textwidth]{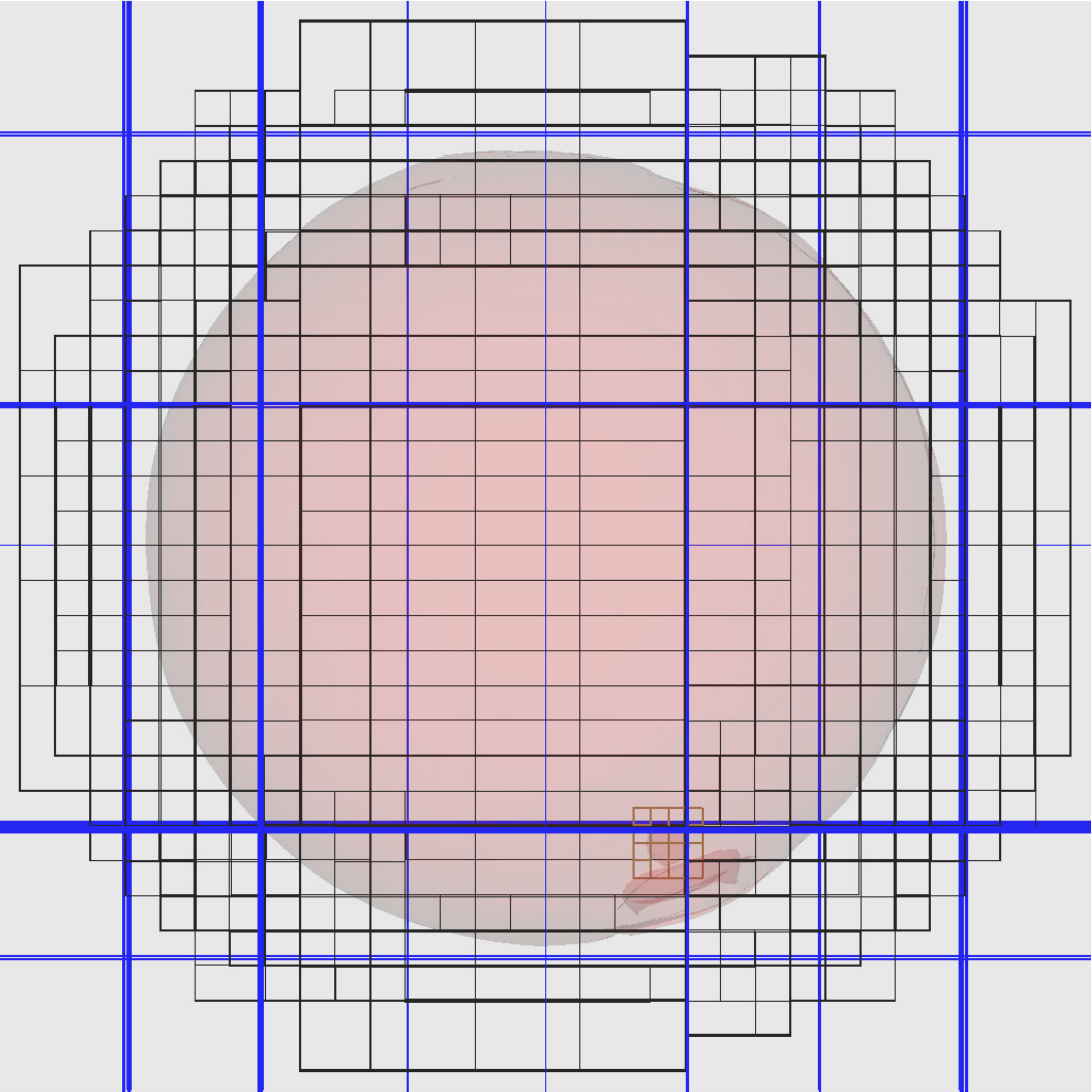}
    \includegraphics[width=0.32\textwidth]{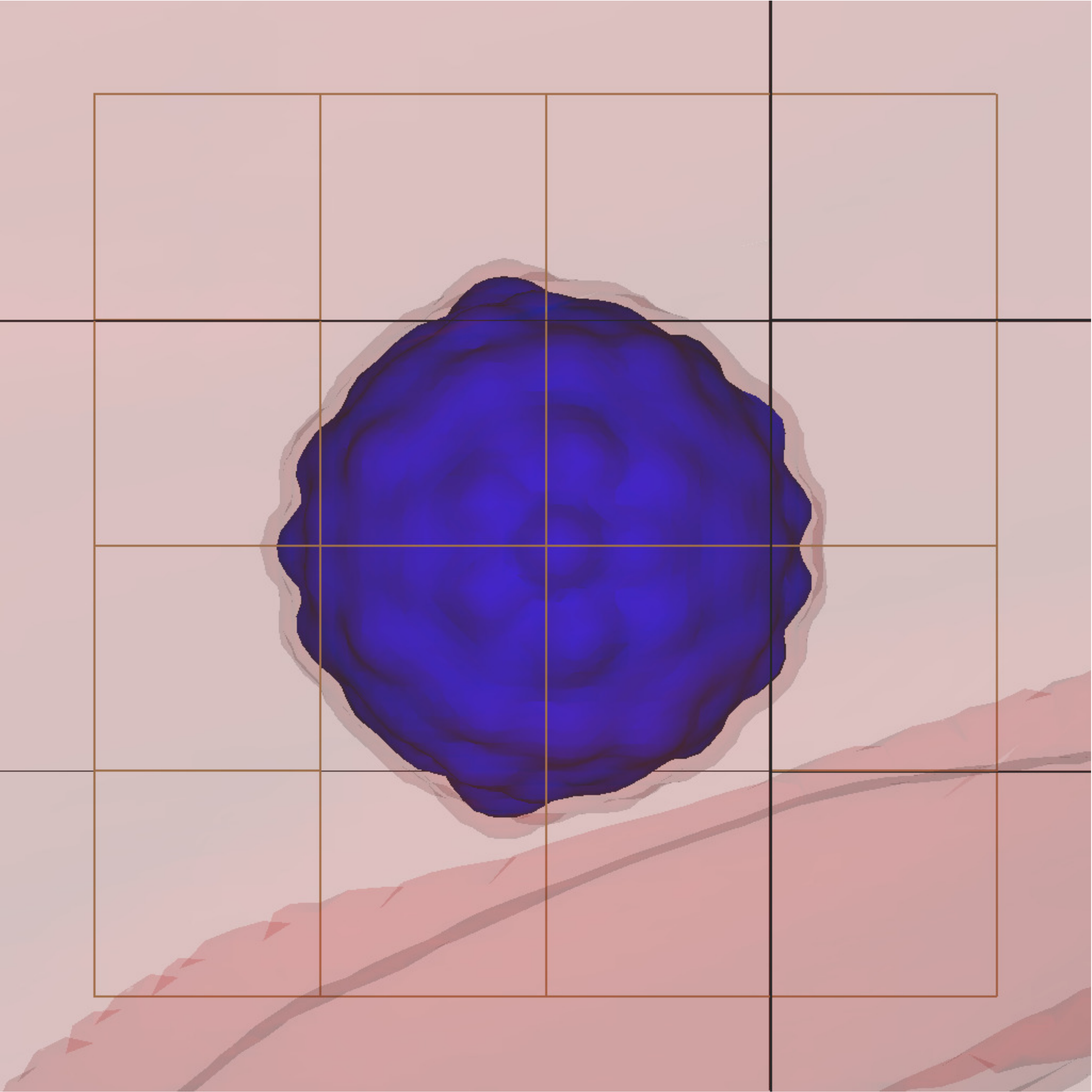}
    \caption{\label{fig:mesh_plots} Grid layout with each additional
      level of refinement; each box represents a block of $16^3$ to
      $48^3$ zones, depending on the size.  The top left figure shows
      the grid structure for the N12 calculation in Maestro, and the
      top center figure shows this grid after 20 coarse time steps in
      Castro.  The
      red and green lines indicate the base ($8.68$ km zone$^{-1}$)
      and first ($4.34$ km zone$^{-1}$) levels; the grey contour is an
      isodensity surface at $\rho=10^4$ g cm$^{-3}$.  The next two
      plots zoom in on the next two levels of refinement, with blue
      (black) at $\sim1085$ ($\sim271$) m zone$^{-1}$.  The next plot
      shows the final level of refinement in brown at $\sim135$ m
      zone$^{-1}$ and a density contour in red at $\rho=10^9$ g
      cm$^{-3}$.  The region of ignition is evident in the southeast
      quadrant.  The final figure is a zoom in on the added ignition
      point where the carbon mass fraction is shown as a blue
      contour.}
  \end{center}
\end{figure}

The addition of levels of refinement began by first mapping the N12
data into the Castro framework, as described in Section
\ref{sec:init-m2c}, onto a {\em fixed} grid structure that was exactly
the same as in the N12 simulation (refer to Figure
\ref{fig:mesh_plots}).  The system was then evolved for ten
coarse-level time steps (0.5 ms) to adjust to the new framework.
Next, the system was evolved for an additional ten coarse-level time
steps (1.3 ms), while allowing the grids to change based on the same
AMR refinement criteria as before (we note that Castro uses tighter
grid generation tolerances, so that the Castro grid structure is
slightly different).

Additional levels of refinement were then added one at a time, with
ten coarse-level time steps between each addition.  To reach our
target resolution of $\sim135$~m zone$^{-1}$, we used levels with
varying refinement ratios --- the jump in resolution when going from a
coarse grid to the next finer grid.  A total of five levels was
finally employed with refinement ratios of either two or four, giving
the resolution as $8.68$, $4.34$, $1.09$~km zone$^{-1}$, $271.3$, and
$135.6$~m zone$^{-1}$ for each level from coarsest (level 0) to finest
(level 4).  The initial refinement criteria for selecting the zones to
be refined to a specific level was based on the density.  Zones were
placed in level $l$ if their density is greater than $\rho_l$, with
$\rho_1 = 5\times10^7,\ \rho_2 = 1.5\times10^9,\ \rho_3 =
2.49\times10^9,$ and $\rho_4 = 2.5\times10^9$~g cm$^{-3}$.  These
particular values of $\rho_l$ were carefully chosen along with the
number and size of grids that get blocked together on a single MPI
task to ensure efficient computational load balancing.

After all the levels of refinement had been added, the system was
evolved for an additional 50 coarse-level time steps in Castro to
guarantee relation to the new grid.  The CFL condition for the size of
the time step is different between Maestro and Castro.  In particular,
the low Mach number approximation in Maestro allows a stable time step
using an advective CFL condition rather than an acoustic CFL
condition, so that the Maestro time step is {\em at least} a factor of
$1/M$ greater than the allowable time step in Castro, where $M$ is the
maximum Mach number in the simulation.  At the time of ignition, the
Maestro time step is a factor of 18 greater than the maximum allowable
Castro time step.  Furthermore, we limit the initial time step size in
Castro to be a factor of $0.05$ times the allowable time step and
allow it to grow by no more than $10\%$ per time step.  This implies
that even though the system has been evolved for 50 Castro time steps
before we have reached our final grid resolution, the total elapsed
time (0.017 s) is less than a single Maestro time step, and the
conditions on the grid have not changed appreciably.

It is important to note that adding levels of refinement alters the
turbulent structure of the high Reynolds number velocity field.  In
particular, increasing the resolution increases the simulation's
inertial subrange of the energy cascade in wavenumber space (see
Figure 20(c) of N12, for example).  Once a level is added, the
small-scale eddies present before the addition can break down to
smaller scales.  There is a time scale associated with the traversal
of an eddy through the inertial range.  In high Reynolds number flow,
large-scale eddies (size $L$) take approximately an eddy turnover time
($t_L$) to break down to the Kolmogorov dissipation scale.  Smaller
eddies have smaller turnover times ($t_l/t_L \sim
\left(l/L\right)^{2/3}$), and they breakdown to the dissipation scale
in less than an eddy turnover time.  The cascade of turbulence to
``fill in'' the added level in our simulations should happen quicker
than the typical small-scale eddy turnover time from N12 ($\lesssim
0.1$ s).  That said, AMR levels were added on a shorter timescale ---
the turbulent energy in our simulations may not have had sufficient
time to completely cascade to the smallest scales.  A further
resolution study is required to determine the degree of incomplete
energy cascade.  We note, however, that the turbulent velocities on
the 4 km-scale in N12 were of order $14$ km s$^{-1}$, much less than
the laminar flame speed ($\sim 100$ km s$^{-1}$).  The turbulent
fluctuations on our finest grid (a factor of 32 more refined than N12)
should then be even smaller due to Kolmogorov scaling arguments
($u_l/u_L\sim\left(l/L\right)^{1/3}$).

For the A series simulations, the location of the ignition point was
determined by the hottest grid zone in the initial Maestro data; for
the B(C) series, the ignition point was artificially placed at $x=y=0$
and $z=10$(0) km.  The ignition point for series A was naturally in a
region of outflow with radial velocity $\sim 10$ km s$^{-1}$ (see
N12).  The artificially placed ignition point for series B was also in
a region of outward radial velocity, however there was a
non-negligible ($\sim30\%$) azimuthal velocity contribution to the
local velocity, the total speed of which was approximately $\sim40$ km
s$^{-1}$.  The ignition spot was defined as a sphere of ash having
radius $2$~km perturbed by random spherical harmonics. The
thermodynamic state was set by the properties of ashes from nuclear
burning as given by the table described in Section
\ref{sec:microphysics} with temperature $T = 8.5\times 10^9$ K,
average binding energy per nucleon $BE/A = 8.166$ MeV, and average
atomic mass $\bar{A} = 11.6$.  The density of the ash is determined
cell-by-cell from the isobaric burning condition and the EOS; typical
values are around $\rho\sim2.15\times10^9$~g cm$^{-3}$.  The $^{12}$C
mass fraction jumps from $0.5$ in the fuel to $\sim 10^{-5}$ in the
ash.  We use the $^{12}$C mass fraction and its gradient as refinement
criteria for AMR to better restrict our finest level coverage to the
ash and its surrounding region.  Figure \ref{fig:mesh_plots} shows the
progression of adding levels from the initial N12 grid structure (top
left) through the inclusion of our hot spot, shown as a blue contour
of $X\left(^{12}\textrm{C}\right) = 0.49$ (lower right).

For comparison to other simulations in the literature, we
consider the fully refined grid with added ignition point as our
initial conditions for the simulations presented here.  That is, any
evolution of the system during the addition of levels of refinement
is ignored, and $t=0$ s once the ignition spot is in place.

The simulations presented here used the resources of several
supercomputing facilities.  Most of the calculations were done using
the Early Science System of the Blue Waters supercomputer at NCSA.
While this machine was closed for upgrades to the full Blue Waters
machine, which was also used in later simulations, the calculations
were continued on the Titan machine at OLCF and on the Hopper
supercomputer at NERSC.  It should be noted that although the
performance of the pre-production Blue Waters systems did not
represent the full scalability of the complete machine, the validity
and integrity of the data and calculations performed therein have been
verified to be sufficiently consistent with those on other
full-production machines.  The calculations were run on a maximum of
65,536 cores (4096 MPI tasks, each with 16 OMP threads).  For
computational efficiency, the finest level of refinement was derefined
once it contained $\sim1.1\times10^9$ zones.  For example, in the A
series of runs this happened at 0.4 s and again at 0.59 s.  The
maximum number of grid zones at any given time was just over
2$\times10^9$.  The two A series runs used approximately 20 million
CPU hours each; the B and C series of simulations together used
approximately 30 million CPU hours.

\section{Results}
\subsection{Series A: Natural Ignition Location}
\label{sec:seriesA}

The most direct mapping from the N12 simulations is Model AV.  The
left panel of Figure \ref{fig:early-burn} shows the initial flame
surface of Model AV as denoted by an isocontour of $X(^{12}\rm{C}) =
0.45$.  The initial hot spot has a diameter of $\sim4$ km plus random
perturbations on the scale of a few hundred meters.  The middle
($t=0.0059$ s) and right ($t=0.0119$ s) panels of Figure
\ref{fig:early-burn} show the evolution of the flame surface as it
quickly burns through the initial perturbations.  At the ambient
density of the flame's initial location, the laminar flame speed is
about 95 km s$^{-1}$ \citep{timmeswoosley1992}, which is greater than
the constant (lower bound) speed we assumed, and therefore the flame
initially burns laminarly.  Since the density is nearly constant in
the small region involved, the bubble remains nearly spherical and
increases in volume by $\sim80\%$ during the first ten milliseconds of
evolution, although there is less expansion on the bottom of the
bubble where the burning has to directly compete with buoyancy.
Indeed, the perturbations on the underside of the bubble are not
completely burned out this early in the evolution.  The bubble does
not experience significant buoyant rise for another twenty
milliseconds, however, after which its volume has increased another
order of magnitude. An initial ignition point of $\sim5$ km in radius
with smaller relative perturbations would probably yield a similar
starting point (see Section \ref{sec:early}).

Figure \ref{fig:later-burn} shows the same isocontour as in Figure
\ref{fig:early-burn}, but at later times.  The persistence of
perturbations on the underside is seen in the left panel ($t=0.15$ s)
of Figure \ref{fig:later-burn}.  Around $t=0.26$ s (middle panel),
instabilities start to wrinkle the flame's surface.  By this time, the
cap of the bubble has also rolled up into itself creating a more
traditional ``mushroom'' shape.  By $t=0.47$ s (right panel), the
outer edge of the bubble is a little more than half way through the
star's convective region, and the volume enclosed by its highly
wrinkled surface is $\sim1.5\times10^5$ times the original hot spot's
volume.

Figure \ref{fig:volrend-vort} shows volume renderings of the magnitude
of vorticity at the same time and spatial scales as used in Figure
\ref{fig:later-burn}.  Large values of vorticity are shown as the
yellow-white tubes and indicate regions of high turbulence.  Early in
the evolution, the vorticity is confined to a single torus similar to
the behavior found in previous simulations \citep{BurningThermals}.
In this stage, the cap of the buoyant bubble is forming as fresh fuel
is being entrained from the edges and rolled under the brim of flame's
surface.  Later, the vortex tube bundle comprising the single dominant
torus begins to break apart and spawns smaller vortex tubes that flow
along the bubble's surface.  Eventually there are several vortex tubes
lying on the flame's surface and they dominate over the now nearly
destroyed torus-shaped bundle of vortex tubes making the brim of the
bubble, as seen in the middle panel of Figure \ref{fig:volrend-vort}.
At even later time shown in the right panel, the strong turbulence has
warped high vorticity tubes into the interior of the bubble as well as
the trailing tail of the plume.

Model A0 was similar to Model AV.  Figures \ref{fig:slice5},
\ref{fig:slice6}, and \ref{fig:slice75} show a comparison as slices of
various quantities at $t=0.5$ s, $t=0.6$ s, and $t=0.75$ s,
respectively, for Model AV (top two rows) and Model A0 (bottom two
rows).  The plots in each column are, from left to right: temperature,
magnitude of vorticity, and energy generation rate.  For all three
figures (\ref{fig:slice5},\ref{fig:slice6}, and \ref{fig:slice75}),
the color range for temperature varies smoothly and linearly between
blue ($10^9$ K) and the maximum temperature in the slice, which was
several billion Kelvin.  Also in these figures, the color range for
vorticity is logarithmically spaced between 100 s$^{-1}$ (blue) and
the maximum (white) value in each slice, which, as in the volume
renderings of Figure \ref{fig:volrend-vort}, was several thousand
s$^{-1}$.  The energy generation rate color maps are divided between
endoergic reactions (blue) and exoergic reactions (red); only values
within 0.1\% of each extrema are colored.  The thin, horizontal white
lines across the top row for each model show the location of the
orthogonal slice depicted in the bottom row; the slice location was
chosen to be through the region of largest lateral extent.  The top
panels have an aspect ratio shown with distance indicators, whereas
the bottom panels have a one-to-one aspect ratio using the horizontal
scale indicated.

At $t=0.5$ s in Figure \ref{fig:slice5}, the bulk of the high
vorticity lies just interior of the outer edges of the flame, although
some turbulence is deeper within the ash, for both Models AV and A0.
There are pockets of high temperature and exoergic reactions, but also
pockets where cold, fresh fuel has been entrained under the roiling
lip of the flame.  Some of these pockets are on the order of 50 km in
size, but do not last very long as turbulence quickly pulls them apart
into ever smaller regions that can burn quickly as heat diffuses
through their surface.  Reactions within the flame itself are always very
exoergic and most of the energy from nuclear burning is released
there, as indicated by the solid red tracing the flame's outline in
all energy generation rate (right) panels.  There is a small amount of
photodisintegration (blue) just interior to the outer flame surface.
Interior to this endoergic reaction region, the exoergic reactions
become much more prominent as expansion and cooling lead to reassembly
of alpha particles to iron.  Both Models AV and A0 have a similar
shape and structure, but the vertical extent of Model A0 is
slightly less owing to the lack of an initial outward velocity field
present in Model AV.  Furthermore, the stem and cap of the bubble of
Model A0 tend to be more slender than the equivalent features of Model
AV.  In both cases, the aspect ratio of the bubble's shape is nearly
$1:2$.

In Figure \ref{fig:slice6} ($t=0.6$ s), regions of high vorticity are
again largely confined interior to the flame's surface.  Endoergic
reactions are now nearly exclusively found just behind the periphery
of the flame with very little in the interior, even in the vicinity of
pockets of fresh fuel.  This shift in burning behavior occurs as the
bubble reaches lower density where the cross sections for endoergic
reactions drop appreciably and NSE processing is stunted.  Also at
lower density and larger extent, the bubble more easily expands
laterally due to lateral pressure gradients (see Section
\ref{sec:geometry}), leading to an accelerated increase in the
bubble's aspect ratio to nearly $2:3$.  At this time the cap of the
bubble is nearly isothermal at $\sim 3\times10^9$ K for both models.
Again, we notice that the features of Model A0 tend to be narrower
than those of Model AV, however the disparity appears to be
decreasing.  The bubbles are just about to cross the original location
of the convective boundary.

Figure \ref{fig:slice75} ($t=0.75$ s) also shows the $\rho=10^8$ g
cm$^{-3}$ (interior) and $\rho=10^7$ g cm$^{-3}$ (exterior)
isocontours as dashed grey lines; the $\rho=10^6$ g cm$^{-3}$ contour
lies just outside the edge of the image and is slightly puffed out due
to the expansion caused by the bubble.  One feature that is
immediately noticeable is that below $\rho=10^8$ g cm$^{-3}$, there is
very little energy generation; indeed, only exoergic reactions occur
below this density (see Section \ref{sec:microphysics}), and only in
the vicinity of fresh fuel.  The caps of the bubbles are no longer
nearly isothermal, but have a temperature gradient that goes from
$\sim 2.4\times10^9$ K in the center to just over $10^9$ K near the
perimeter.  Both Models A0 and AV have the bulk of their high
vorticity regions again on the periphery of the ash, and there is a
decrease in vorticity in regions absent of strong reactions.  The
disparity in the features of the two models has diminished.  Both
models have reached about the same radial extent, $\sim1700$ km, and
have assumed a nearly $1:1$ aspect ratio due to significant lateral
expansion.

One source of the vorticity generated in Figures \ref{fig:slice5},
\ref{fig:slice6}, and \ref{fig:slice75} is due to the baroclinicity,
$\psi$:
\begin{equation}\label{eq:baro}
  \vec\psi = \frac{1}{\rho^2}\nabla p \times \nabla \rho.
\end{equation}
The burning within the flame front itself is isobaric, but a density
gradient forms as cold fuel is burned to hot ash.  This causes a
misalignment of the local pressure and density gradients and thus a
nonzero baroclinicity, which generates vorticity.  Figure
\ref{fig:baro} shows an example of the magnitude of baroclinicity for
Model AV at $t=0.8$ s.  The colorbar goes from $\psi$ less than 1
(white) to $3.3\times10^6$ s$^{-2}$ (black).  It is interesting that
the strongest baroclinicity occurs within the tail of the flame and
along the underside of the cap of the bubble.  However, in the tail,
$\vec\psi$ is orthogonal to the flame surface, whereas the cap has
$\vec\psi$ nearly parallel to its surface.  In the tail, this
alignment creates flow mainly along the direction of the plume.  The
vorticity generated near the cap tends to entrain fresh fuel across
the flame surface.  Fuel is still entrained into the tail of the
flame, but not nearly as vigorously as in the cap.

In addition to the minor morphological differences between Models AV
and A0, there are also small differences in the nucleosynthetic
yields.  Figure \ref{fig:mass-burned} shows the evolution of the
production of iron-group (top panel) and IME (bottom panel) elements
for both Model AV (blue) and Model A0 (red).  The thin, dashed, grey
line is the percent difference between the two models.  In general,
Model A0 burns more material than Model AV, likely owing to the lack
of an initial push from the outward plume present in the N12 data.
However, given the stated assumptions of our thickened flame model,
the background turbulence affects the nucleosynthetic yields on the
10\% level for an ignition point placed at the runaway location
indicated by our low Mach number calculations.  It is important to
note that this is only the affect during the evolutionary period when
the bubble is buoyantly rising towards the surface.  Any subsequent
burning after the flame has broken through the white dwarf's surface
(e.g DDT) can either increase or decrease the final yield difference
depending on the details of the local density and composition.  The
convective flow field in the interior likely has a negligible effect
on the post-breakout burning.  Table \ref{tab:models} lists the latest
time evolved for these models, $t_{\textrm{max}}$, including the
farthest distance reached by the burning, $r_{\textrm{max}}$, and the
total amount of IME, IGE, and nuclear energy release produced at the
end of the simulations.  The total amount of material burned during
the deflagration depends quite sensitively on the initial conditions
(i.e. location and size --- or solid angle) of the ignition point;
this is being explored with our flame model in an upcoming paper
\citep{Dong:2013}.  We note that the nuclear energy released in our A
series of simulations is comparable to that of the deflagration phase
of Models 16b100o8r, 25b100o6r, and 25b100o8r of \cite{jordan:2008}.
Also, for the A series of simulations, $t_{\rm max}$ is approximately
equal to the time the flame first hits the $\rho = 10^7$ g cm$^{-3}$
surface; at this point, the central density was $\rho_c =
2.27\times10^9$ g cm$^{-3}$.

While it is of paramount importance to capture the progress of the
flame itself, it is also necessary to consider in detail the nuclear
reactions that go on in the ash after the flame has passed. These are
important not only to the nucleosynthesis, but to the dynamics. While
the net burning from fuel to ash is always exoergic, there are several
phases to consider. First, carbon and oxygen burn isobarically in the
flame to a given composition that, at high temperature in nuclear
statistical equilibrium (NSE), may contain substantial helium and
protons from photodisintegration (the protons come especially from the
photodisintegration of $^{56}$Ni into $^{54}$Fe + 2p). This NSE
composition continues to evolve in response to the changing
temperature and density after the flame itself has passed. If the
temperature goes down or the density up, reassembly of
photodisintegration products, especially helium, can give a large
energy yield. Conversely, if the temperature increases or density
declines too rapidly, the energy generation can be locally endoergic
as iron-group elements are photodisintegrated to light
species. Generally, adiabatic expansion gives a large positive energy
generation because the decreasing temperature has a larger effect than
the declining density. In addition, electron capture changes the
makeup of the iron group turning $^{56}$Ni into $^{54}$Fe and
$^{58}$Ni and even $^{56}$Fe. These more neutron-rich nuclei are more
tightly bound, so electron capture tends to increase the energy yield,
but it also decreases the electron density and pressure causing the
matter to be compressed and less buoyant.

Figures \ref{fig:bea_ye} and \ref{fig:t9_he} illustrate the evolution
of various quantities from $t=0.65$ s (top row) to $t=0.8$ s (bottom
row) in a typical run (specifically, model AV here).  Figure
\ref{fig:bea_ye} shows BE/A in the range [7.82,8.68] MeV/nucleon in
the left column and $Y_e$ in the range [0.47,0.50] in the right
column, with the color maps increasing from blue to red/pink.  Figure
\ref{fig:t9_he} shows $T_9 = T/10^9$ K in the range [2,9] in the left
column and $X(^4{\rm He})$ in the range [0.0,0.25] in the right
column.

Initially burning produces a very high temperature and yields a
composition with comparable amounts of helium and iron-group
elements. This gives a relatively low prompt energy yield as seen in
the plot of BE/A, but creates ashes that can yield more energy upon
later expansion. This is apparent in the low value for BE/A in the
tail of the rising plumes in Figure \ref{fig:bea_ye}.  Later however,
the burning goes to lower temperature and produces less helium and
greater BE/A as in the bottom left panel of Figure
\ref{fig:bea_ye}. This is most apparent along the rising head of the
ashes where the temperature is lowest and recombination has gone the
farthest.  This reassembly creates a fountain effect in that matter
rising along the axis of the plume gets an extra boost of energy that
causes greater expansion and more rapid floatation. The expansion
gives more cooling and therefore still more energy generation.  As
expected, the greatest amount of electron capture occurs deep in the
star where the density is high. The electron mole number, $Y_e$,
reaches a minimum of 0.467 there. Burning at lower density gives $Y_e$
very close to 0.50, its initial value for a mixture of $^{12}$C and
$^{16}$O. However there is still substantial neutron-rich material in
the outer part of the star along the axis of the rising plume due to
the fountain effect mentioned above. Upon ejection this material will
be $^{54}$Fe and $^{58}$Ni, not $^{56}$Ni.

\subsection{Series B and C: Artificial Ignition Locations}
\label{sec:unnatural}

To observe the effect of buoyant bubbles under slightly different
conditions, we artificially moved the ignition spot closer to the
center of the star where the flow pattern is different from the nearly
radially outward flow field from Model AV.  Furthermore, the decrease
in buoyancy force as the ignition point is placed closer to the star's
center implies that the bubble should interact with the background
turbulence for a longer period of time before the bubble rise speed
becomes comparable to the background velocity.

Figure \ref{fig:10-slice} shows slices of the fluid speed for models
BV (top row) and B0 (bottom row) at three different times (from left
to right): $t=0.1$, $0.2$, and $0.3$ s.  The blue color scale
logarithmically spans $\leq$ 10 km s$^{-1}$ in black to 10$^4$ km
s$^{-1}$ in white; the red contour marks the flame surface.  For each
plot, the center of the star is located at the center of the bottom
edge, and each edge spans 280 km.  As stated in Section
\ref{sec:simulations}, the ignition point was placed 10 km off-center
--- i.e., from the bottom of these slices.  The initial burning is
laminar at around 95 km s$^{-1}$, but the bottom of the bubbles only
expand about a kilometer radially inward before stalling against
buoyancy.  These bubbles do not burn back through the center, even in
the case where there is no initial velocity field (B0); this is in
contrast to the resolved study of buoyant flames in simplified,
constant gravity backgrounds of \cite{zingaledursi}.

The presence of the strong non-radial initial velocity component near
the ignition location of Model BV is clearly visible.  In these
slices, the background velocity has a general flow that is towards the
upper left corner of the frame, which is along the direction of the
stretch and deformation of Model BV at later times.  It should be
noted that the same ignition shape was used in the B series runs as in
the A series; however, the overall orientation with respect to the
local gravity vector is different between the two series.  In
particular, two of the more dominant spherical perturbations to the
initial spherical flame surface are seen in this particular viewing
angle as the two nodules on the lateral sides of the flame at $t=0.1$
s in both models.  These tend to burn somewhat laterally away from the
bulk of the bubble before buoyantly rising and becoming distorted by
the turbulent wake of the leading edge of the flame.

Figure \ref{fig:cent-slice} shows similar slices as in Figure
\ref{fig:10-slice} but now for Models CV (top) and C0 (bottom) at
$t=0.1$, $0.2$, and $0.4$ s from left to right.  The color scale here
spans from under 1 km s$^{-1}$ (black) to $10^3$ km s$^{-1}$ (white).
In these panels, the center of the star is at the center of the image,
and each edge of the panel spans 150 km.  As in series A and B, the
bubble quickly burns through the initial perturbations.  In the case
of no background flow field, Model C0, the flame remains nearly
spherical in its early evolution, until it begins to interact with the
velocity field generated from its own expansion.  In the presence of
the background velocity field, Model CV shows a highly distorted flame
surface that takes on a shape consistent with the general
directionality of the initial velocity field.  The very early
evolution of centrally located bubbles --- during the time period
where the local buoyant rise/expansion speed is less than the
turbulent intensity --- will likely play a role in determining the
degree of asymmetry of the explosion.  Given that ignition
calculations show a non-isotropic convective flow field
\citep{kuhlen-ignition:2005,wdconvect,wdturb}, it is likely that even
if ignition of a perfect sphere occurred at the exact center of a white
dwarf, the subsequent deflagration should be highly
asymmetric. 

Figure \ref{fig:10-cent-massburned} compares the nucleosynthetic
yields of iron-group elements (top row) and IME (bottom row) as a
function of time for both the B series (left column) and C series
(right column) of simulations in a similar fashion as Figure
\ref{fig:mass-burned} for the A series.  Again, the dashed grey line
shows the percent difference where positive values indicate when the
model without the background flow field has produced more material
than the model with an initial velocity field.  Note that the total
mass produced in these simulations is significantly less than that of
the A series because they were not evolved nearly as far.  With the
exception of iron-group elements in the C series of models, both
series show a clear trend towards the models with a convective flow
pattern producing increasingly more material than the simulations with
a static initial velocity field.  Furthermore, the percent differences
seem much larger here than in the further off-center simulations of A
series seeming to suggest that the turbulence affects the burning more
strongly for more centrally located ignition.  For Models CV and C0,
the iron-group element yield curves have similar percent differences
and relative slopes as those found for series A (top of Figure
\ref{fig:mass-burned}) where the subsequent evolution reversed the
roles, and the model without the initial convective flow pattern ended
up producing more material.

\subsection{Comparison To Analytic Thermals}
\label{sec:thermals}

Based upon resolved simulations of small-scale burning thermals,
\cite{BurningThermals} developed a one-dimensional model of reacting,
buoyant bubble evolution.  The model was then extended to a full star
environment where background stratification becomes important for
governing bubble expansion.  \cite{BurningThermals} then used the
extended model to compared to three-dimensional calculations similar
to simulations presented here, though the simulations in
\cite{BurningThermals} were less resolved and started with a larger
ignition spot.  The reader is referred to the \cite{BurningThermals}
paper for details, but here we give a brief review of their
one-dimensional model for comparison with our three-dimensional
simulations.

The general idea of the Aspden model is that there exists a {\em
  spherical} bubble of radius (to use the \citealt{BurningThermals}
notation) $b$ and {\em uniform} density $\rho$ whose center of mass is
located a distance $z$ from the star's origin.  This bubble is
buoyantly rising, with speed $u$, through a stratified, hydrostatic
background of fuel with density $\rho_f$ and gravitational
acceleration $g$.  The fuel is also assumed to be governed by a
relativistic degenerate polytrope of order $n=3$ ($\gamma = 4/3$),
which, when combined with the equation of hydrostatic equilibrium,
yields a Lane-Emden equation that requires numerical integration.  The
fuel within the flame front is assumed to be burned instantaneously to
ash with density $\rho_a$.  One can then write a simple set of coupled
evolution equations for $z,g,\rho_f,b,\rho,$ and $u$ that take into
account entrainment, stratification, and the effective gravity as fuel
is burned to ash.  The model has two parameters, which are calibrated
in their paper: 1) $\alpha = 0.17$, which is the ratio of the
entrainment rate to the bubble rise speed, and 2) $\beta = 0.5$, which
is an empirical constant that relates the bubble's effective buoyancy
to acceleration of the center of mass of the bubble.

To apply this model to our three-dimensional simulations, we must
first determine the initial conditions.  We find the bubble's
properties by finding those zones inside the bubble, defined by
$X(\rm{C}^{12}) \leq 0.1$.  We then integrate the quantities 1,
$\rho$, and $\rho r$ over this region to obtain the bubble's volume,
mass, and center of mass location, respectively.  At a time, $t_0$,
when the bubble is well-formed, we use the bubble's mass and volume to
obtain its initial radius, $b_0$, and density $\rho_0$, assuming it to
be a sphere.  The initial location, $z_0$ is given by the bubble's
center of mass location, and the initial velocity, $v_0$, is taken to
be the center of mass velocity at $t=t_0$.  From the star's central
density and pressure, we can determine the normalization constant for
the polytropic equation of state, and then the Lane-Emden equation can
be integrated to obtain $g_0$ and $\rho_{f,0}$ at $z_0$, thus
completing the initial conditions.  Table \ref{tab:init-cond} shows
the initial conditions used for the semi-analytic models for both
simulations in the A series.

One also needs to specify how the density jumps across the flame ---
how $\rho_a$ changes as a function of $\rho_f$.
\citet{BurningThermals} fit a power law to the properties of laminar
flames from \cite{timmeswoosley1992}.  Here, we use a power law fit to
the valid density region of the results from flames of \cite{Ma:2013},
on which our burning network and thickened flame are based.  Our power
law takes the form $\rho_a = 0.168\rho_f^{1.075}$.

Figure \ref{fig:anal-model} shows the application of Aspden's model to
our Model A0 (left column) and Model AV (right column) data.  The
three-dimensional data are given as blue crosses, while the lines mark
Aspden's model with different parameters integrated until $t=1$ s.
The red lines are Aspden's model using the fit parameters as described
above.  The one-dimensional model is most sensitive to the choice of
the empirical constant, $\beta$, which governs the fraction of
buoyancy that goes towards moving the bubble's center of mass.  The
green and black curves show the effects of changing $\beta$ from its
initial value of $0.5$ to $0.75$ and $1.2$, respectively.  Each of the
parametrizations follows the general trend of the bubble height as a
function of radius (top row), including the turn-over as the bubble
reaches lower density where steep density gradients exist.  It is
immediately clear, however, that the original parametrization from
\cite{BurningThermals} simply does not rise quick enough compared to
our simulations.  This fact is more evident in the plots showing the
evolution of the bubble's height as a function of time (middle row).
None of the parametrizations tend to rise nearly as quickly as our
three-dimensional simulations.  As expected, a larger value of $\beta$
increases the rise speed, and indeed the $\beta=1.2$ curve comes
closest to the three dimensional data, although its gradient is much
steeper towards the end of the integration.  In terms of the total
mass burned (bottom row), all parametrizations do a decent job
tracking the three-dimensional results. None of them, however,
reproduce the sudden (more so in Model A0) increase in height after
$\sim10^{-3}\ M_\odot$ of material has burned and the bubble entered a
lower density region where the burning takes on a different behavior.

The semi-analytic model has several simplifying assumptions
that make the inherently three-dimensional problem tractable, namely: 
\begin{enumerate}
  \item the bubble is always a sphere of radius $b$,
  \item the bubble has a uniform density, $\rho$, at any given time,
  \item the bubble entrains fuel at a rate proportional to it's rise speed with proportionality constant $\alpha$,
  \item the only burning is that which occurs in the vicinity of the
    flame's surface, instantaneously changing fuel to ash,
  \item and the bubble rises isentropically such that the pressure
    equilibrium is maintained with the ambient fuel.
\end{enumerate}
Here we discuss the validity of these assumptions as applied to the
interpretation of the differences between the model and simulation
data.

The first assumption --- sphericity --- is certainly only valid at
early times as evidenced by Figures \ref{fig:slice5},
\ref{fig:slice6}, and \ref{fig:slice75}.  Indeed, given the same
volume, a plume-shaped object will experience less fluid drag than a
perfect sphere, and should thus rise more quickly as is the case of
our simulation data compared to the one-dimensional model.  The second
assumption --- uniform density --- is also only valid early on, with
the error worsening during the evolution, but at a much slower rate
than the error from the first assumption.

The third assumption is reasonable to first order --- one expects the
amount of fuel crossing the flame front to be proportional to the
expansion rate of the front through the fuel.  In the presence of
turbulence the flame front becomes wrinkled, and in the same way that
the increased surface area effectively increases the burning rate, so
too will the entrainment be increased.  The degree of turbulent
entrainment will also increase with the decrease in density as the
bubble transitions from the laminar to the flamelet to the distributed
burning regimes.  Therefore, even though the buoyant bubble will reach
a terminal velocity, the turbulent entrainment rate is likely (on
average) a monotonically increasing function with time until the
density is low enough that reactions cease.  This effect may, in some
sense, be captured by using a turbulent flame speed instead of the
constant assumed in both our three-dimensional simulations and the
Aspden models.  A possible alternative assumption would be to say that
the entrainment rate was proportional to some turbulent velocity in
the vicinity of the flame.  One then has the difficulty of forming a
meaningful average over the flame's surface of this turbulent velocity
to use in the one-dimensional model.  We found no such simple model,
but save this for further exploration. 

The fourth assumption --- instantaneous, local burning--- is certainly
not the situation within our three-dimensional simulations.  As
outlined in Section \ref{sec:microphysics}, there is an extensive
amount of burning that occurs in the ashes of the flame (see also
Figures \ref{fig:slice5}, \ref{fig:slice6}, and \ref{fig:slice75}).
These reactions will slightly alter the local density, feeding back
into the violation of the second assumption. It is not entirely clear
how this affects the bubble's buoyancy, on average, over the course of
its evolution.

The last assumption --- isentropic rise --- is likely quite accurate
until the bubble has expanded enough such that the sound crossing time
becomes comparable to the timescale for heat diffusion across the
bubble, and a lateral pressure gradient forms.  This gradient then
causes significant lateral expansion, which is not captured by the
model.  This disparity is evident in plots of height versus radius
(top row) of Figure \ref{fig:anal-model} where the simulation data has
a stronger turnover compared to the models once the spherical bubble
is about 100 km in radius.  

\subsection{Characteristic Evolutionary Phases}
\label{sec:geometry}

During its evolution, the burning bubble goes through several phases
that can be characterized by their geometry, especially their solid
angles, $\Omega$, as viewed from the center of the star.  The solid
angle, since it generally increases here, is related to the total
amount of material being burned at a function of time. That is, for a
given star and flame physics, a larger {\em initial} solid angle will
give more burning.  This simplification neglects secondary effects,
such as the degree to which the burning rate is decreased due to the
lower density if significant expansion occurs, i.e. central ignition.
For off-center ignition though, such a parametrization is robust.

The solid angle of a spherical cap is given by $\Omega =
2\pi\left(1-\cos\theta\right)$, where $\theta$ is the angular radius
of the cap in radians.  Central ignition, if it continued to burn
isotropically, would have $\Omega = 4\pi$, but the solid angle
resulting from off-center ignition is much smaller.  Figure
\ref{fig:geo} shows the evolution of the bubble's solid angle as a
function of time for Model AV (red), Model A0 (blue), and the
one-dimensional model based on simulation AV with varying values of
the empirical constant $\beta$ (black).  The curves for the
one-dimensional model based on simulation A0 look similar.  The
thin, grey vertical lines separate the five distinctive phases of the
bubble's evolution to be discussed.

During the first phase ($t\lesssim0.05$ s), the evolution is dominated
by laminar burning.  Since the density and pressure are very nearly
constant, the bubble burns isotropically at a constant laminar speed
producing a sphere. Little buoyant rise or distortion occurs
(e.g. Figure \ref{fig:early-burn}).  For Model A, after the initial
laminar phase, which burned through the imposed surface perturbations,
but just before the flame began to float rapidly ($t\sim30$ ms), its
solid angle was $\Omega/4\pi=4.4\times10^{-3}$.  For comparison, the
3B25d100 and 3B25d200 models of \cite{roepke-gcd} had {\em initial}
solid angles at ignition of $\Omega/4\pi = 1.5\times10^{-2}$ and
$3.86\times10^{-3}$, respectively, and Model N1 of \cite{ivo:2013} had
$\Omega/4\pi = 1.8\times10^{-2}$. The series of single-point,
off-center models of \cite{jordan:2008} had values that ranged from
$6.28\times10^{-3}$ to $1.10\times10^{-1}$. As we shall see,a smaller
solid angle implies that less burning occurs in the deflagration,
especially early on.

During the next phase ($0.05\lesssim t \lesssim 0.1$ s), buoyancy
becomes important, and by $0.06$ s, the bubble's surface has become
elongated along the radial direction.  Its top is rising faster than
the local laminar speed. As it becomes increasingly distorted, the
bubble takes on the characteristic mushroom shape of Rayleigh-Taylor
instability while the effects of turbulence remain small.  Although
the flame is still expanding laterally via burning, the increase in
solid angle slows during this phase due to the accelerating
radial floatation.

The third phase ($0.1\lesssim t \lesssim 0.18$ s) is characterized by
the formation and evolution of a single dominant torus quite visible
in a plot of vorticity (see the first panel of Figure
\ref{fig:volrend-vort}).  Buoyancy still pushes the widest part of the
bubble to larger radii, but there is also a brim forming near the
bubble's top as the flame begins to roll into itself.  Around $t=0.14$
s, buoyancy dominates sufficiently that the solid angle reaches a
local maximum.  At about the same time, the high-vorticity brim of the
bubble has become well-formed.  Up until this point, the widest part
of the flame --- the part that sets the solid angle as seen from the
center of the star --- occurred at a distance somewhere near the
middle of the bubble's radial extent.  Between $t=0.14$ and $0.17$ s,
however, the brim of the bubble undergoes a single overturn, rolling
the flame surface such that the widest part occurs closer to the top
of the bubble.  This causes an abrupt drop in the apparent solid angle
of the flame.  Over the next 0.01 s, the brim undergoes a second
overturn causing an even sharper drop in solid angle. The torus looks
like it is pulsating.

This high-vorticity torus does not remain the only vortex tube for
another turnover of the brim, however.  Indeed, the next phase
($0.18\lesssim t\lesssim 0.5$ s) marks the transition to turbulence as
the single torus breaks down into several vortex tubes. These tubes
initially lie near the brim and upper regions of the flame surface,
but later progress into the ash.  During this period, the flame
expands in a nearly conical, self-similar fashion with a nearly
constant solid angle.  There are small fluctuations about the constant
solid angle as roiling turbulence in the cap occasionally causes
large-scale wrinkles to form on the surface, such as those seen in
Figures \ref{fig:slice5}, \ref{fig:slice6}, and \ref{fig:slice75}.
These turbulent fluctuations are experienced throughout the remainder
of the bubble's evolution as it makes its way to lower density
regions, where the final phase occurs.

During the last, fifth phase ($t\gtrsim 0.5$ s), the bubble expands
laterally much more rapidly than in its earlier buoyant rise (e.g. top
panel of Figure \ref{fig:anal-model}). This causes the solid angle to
increase dramatically. Of the five phases modeled, most of the burning
happens here. Figure \ref{fig:contours} shows contours of pressure
(left) and density (right) for Model AV at $t=0.8$ s; each slice is
2400 km on a side, and the red contours are highlighted to aid the
eye.  Early in the bubble's evolution --- or, equivalently, near the
center of the star --- the pressure scale height and the sound speed
are large.  Here the bubble is small and floating slowly so that sound
waves have sufficient time to maintain pressure equilibrium across the
plume as it ascends.  That is, the sound crossing time for the ash is
short compared with the time for the plume to rise a fraction of a
pressure scale height.  Consequently, near the center of the star, the
isobars in Figure \ref{fig:contours} are aligned with gravitational
equipotential lines, which lie on spheres.  Later in the evolution,
farther from the center, the situation is reversed. The bubble has
become large and is floating at a fraction of the sound speed. The
pressure scale height has also become shorter.  During the time it
takes a sound wave to cross the ash, the elongated plume moves an
appreciable distance (compared with the pressure scale height). The
isobars at larger radii in Figure \ref{fig:contours} thus no longer
lie on lines of constant radius across the flame.  Sound waves can no
longer maintain pressure balance along equipotentials, and pressure
gradients form across the flame.

The lateral acceleration from the pressure gradients can be estimated
by considering an idealized plume of ash with a maximum cylindrical
extent, $\Delta L$, rising through a region with a background pressure
gradient $dP/dr|_{\rm fuel}$. Because the unburned part of the star is
not yet expanding rapidly, local hydrostatic equilibrium is still a
good approximation, so $dP/dr|_{\rm fuel} \approx -\rho g$, where $g$
is the local acceleration due to gravity.  From the isobars at large
radii in Figure \ref{fig:contours}, one sees that the pressure
gradients within the ash are shallower than in the fuel, but still
very roughly in hydrostatic equilibrium.  It is the difference in
these pressure gradients, $dP/dr|_{\rm ash} - dP/dr|_{\rm fuel}\approx
f\rho g$ that drives the lateral acceleration.  The factor $f \ll 1$
is included to account for the difference in radial pressure gradients
across the growing flame. Initially $f$ is zero, but it increases as the
flame grows.  If the local floatation speed is $v_{\rm float}$, the
plume rises a radial distance $\Delta r = v_{\rm float} \Delta L/c_s$
during a sound crossing time and sees an effective pressure change
$\Delta P \approx f\rho g v_{\rm float} \Delta L/c_s$. The pressure
gradient $\Delta P/\Delta L$ thus causes a lateral expansion driven by
an acceleration
\begin{equation}\label{eq:gamma}
  \gamma \approx f\left(\frac{v_{\rm float}}{c_s}\right) g.
\end{equation}  
The baroclinicity discussed in Section \ref{sec:seriesA} (and evident
in the misalignment of lines of constant pressure and density in
Figure \ref{fig:contours}) drives turbulence that allows the expansion
to digest the entrained matter, but the baroclinicity and the
vorticity it drives are both very small in the fuel.  The burning
expands sideways because it is pushed; the turbulent entrainment
digests the fuel that is swept up. The sideways pressure gradient
force is no different, in principle, from the floatation caused by
buoyancy.

\subsubsection{Time Scales and Scaling Relations for Early Burning}
\label{sec:early}

The above physical description allows some simple analytic
approxmations and yields some interesting scaling relations.
At sufficiently early times, burning in a bubble close to the star's
center occurs in a region of nearly constant density and pressure.
The bubble experiences a buoyancy force given by the local effective
gravity
\begin{equation} \label{eq:geff}
  g_{\rm eff}(r) = \frac{4}{3}\pi G\rho r A_t,
\end{equation}
where $r$ is the radial distance, $\rho$ the fuel density, and $A_t$
the Atwood number of the flame.  Figure \ref{fig:geff} shows $g_{\rm
  eff}$ for our white dwarf using the Atwood number as a function of
density given by \citet{timmeswoosley1992} (note that the quantity
$\Delta \rho/\rho$ in that paper is the difference between density in
the fuel and the ash density divided by the fuel density,
$\Delta\rho/\rho = 2A_t/(1+A_t)$; Timmes, private communication).  The
grey vertical line marks the radius at which our Model A simulations
were ignited.  For our density profile, the Atwood number is close to
$0.1$ for radii less than $\sim300$ km (about a pressure
scale-height), so one may write
\begin{equation}\label{eq:geff_approx}
  g_{\rm eff}(r_7) \approx 7.0\times10^8\ r_7\ {\rm cm\ s}^{-2},
\end{equation}
where $r_7 = r / 10^7$ cm.  The curve in Figure \ref{fig:geff}
maintains a linear relationship ($g_{\rm eff}\propto r_7^{0.92}$) with
radius until the density noticeably drops.  At the ignition location
for Model A, $r_7 = 0.413$, Equation \ref{eq:geff_approx} gives
$g_{\rm eff} \sim 3\times10^8$ cm s$^{-1}$, consistent with Figure
\ref{fig:geff}.

The bubble initially expands isotropiclly due to laminar burning.  The
time scale for floatation to stretch the bubble in the radial
direction is roughly equal to the time it takes the bubble to float
the same distance it burns, $t_{\rm fl} = 2v_l / g_{\rm eff}$.  After
$t_{\rm fl}$, the bubble responds to forces that, until lateral
pressure gradients develop in stage five (Section \ref{sec:geometry}),
are strictly radial. Within the first pressure scale height, the near
constant density profile yields a constant $v_l$ profile, such that
$t_{\rm fl}$ takes on a simple relationship with ignition radius
\begin{equation}\label{eq:t_fl}
  t_{\rm fl} \approx 27\ (r_7)^{-1}\ {\rm ms}.
\end{equation}
During $t_{\rm fl}$, the flame grows in radius by $l_{\rm fl} =
v_lt_{\rm fl}$ or
\begin{equation}\label{eq:l_fl}
  l_{\rm fl} \approx 2.58\ (r_7)^{-1}\ {\rm km}.
\end{equation}
For our ignition radius, Equation \ref{eq:t_fl} gives 65 ms, which is
just after the onset ($t\sim 50$ ms) of our qualitative phase two
shown in Figure \ref{fig:geo}.  Equation \ref{eq:l_fl} gives an
increase in bubble radius during this time of  6.25 km for our
ignition radius; the three-dimensional simulations gave an average
increase in radius of about 7.75 km at $t_{\rm fl}$, which agrees with
Equation \ref{eq:l_fl} to an accuracy of about 25\%.  By $t_{\rm fl}$ in
the simulation, the bubble has begun to interact with the velocity
field generated from its own rise, something that is neglected in the
simple scaling analysis analysis.

The quantity $l_{\rm fl}$ sets the grid resolution required to treat
the early burning correctly. Whether the burning ignites at a
geometrical point or in a spherical region with radius 1 km does not
matter. Laminar burning will quickly turn a point into a sphere of at
least this size, but the scale where the bubble starts to float and
deform must be resolved.  It is promising to note that similar scaling
relations for the time scale for float speed to equal flame speed,
$t_{\rm fl}/2$, show that a bubble ignited less than $\sim10$ km
off-center will burn through the center before it floats away,
consistent with more detailed estimates \citep[e.g.][]{zingaledursi}.

It is also curious that the solid angle at $t_{\rm fl}$,
$\Omega/4\pi\approx 0.013$ from the simulation, is approximately
equal to the near constant solid angle attained later during the
turbulent phase four in Figure \ref{fig:geo}.  The local maximum in
solid angle is a transient event related to the formation and
breakdown of the high-vorticity torus.  More simulations are required
to determine if this is a general property of burning bubbles in their
early evolution or just a coincidence.  If this is a property of
buoyant flames, then our scaling relations can be used to approximate
the solid angle of the bubble before it begins lateral expansion in
phase five.  Indeed, Equation \ref{eq:l_fl} gives a solid angle at
$t_{\rm fl}$ of
\begin{equation}\label{eq:omega}
\frac{\Omega}{4 \pi} \ \approx \ \frac{1}{4} \left(\frac{l_{\rm fl}}{
    r_7}\right)^2 \approx 5.7\times10^{-3}\left(\frac{0.413}{r_7}\right)^4,
\end{equation}
where, as in Equations \ref{eq:t_fl} and \ref{eq:l_fl}, $r_7$ is the
{\em ignition} radius; this is reasonably consistent with the
numerical simulation.  Equation \ref{eq:omega} is also an estimate of
the fraction of the star that burns during phase four.  The amount of
burning during this phase is thus very sensitive to the radius where
the initial burning is ignited. In the next subsection, however, we
shall see that this sensitivity is diminished by the rapid lateral
expansion that happens in stage five.

Different white dwarf models will give similar scaling relations as
Equations \ref{eq:geff_approx}, \ref{eq:t_fl}, and \ref{eq:l_fl}, just
with different normalization constants.  The Atwood number and laminar
flame speed both have known scalings with density and carbon abundance
from \cite{timmeswoosley1992}.  The general trend is that a bubble in
a more dense (or carbon-rich) white dwarf will grow larger in size
before buoyancy dominates and the bubble moves radially than it will
in a less dense (or carbon-poor) white dwarf.

\subsubsection{Expansion During Late Burning}

The bulk of the energy release and nucleosynthesis occurs during the
fifth stage outlined above; indeed, nearly 80\% of the burning occurs
during this phase (see, Figure \ref{fig:mass-burned}).  As previously
discussed, this stage is marked by significant lateral expansion,
which allows for the processing of a large volume of fuel, as is
clearly seen around $t=0.5$ s where the solid angle behavior changes
abruptly from nearly constant to a profile proportional to
$t^{\textrm{3--4}}$.

The solid angle scales roughly as $\Omega\propto(L/r)^2$, where $L$ is
the lateral size of the plume and $r$ its radial distance from the
center of the star.  In Section \ref{sec:geometry} it was shown that
lateral pressure gradients form within the bubble causing an
acceleration $\gamma$ given by Equation \ref{eq:gamma}.  As the bubble
grows ever larger, the time for sound waves to communicate the change
in pressure to the edge of the bubble grows.  The lateral
acceleration, $\gamma$, itself is {\em increasing} on a sound-crossing
time scale --- i.e. $f$ and $v_{\rm float}$ in Equation \ref{eq:gamma}
increase with time.  The ratio $c_s/g\approx0.04$ s is roughly
constant throughout the bubble's evolution.  This gives a growth in
plume size somewhat faster than that of constant acceleration motion:
\begin{equation}\label{eq:L}
  L \sim \frac{1}{2}\gamma_0 t^{2+n},
\end{equation}
where $\gamma_0$ is the lateral acceleration of Equation
\ref{eq:gamma} at the start of phase five, and $n \sim 0.7$ from
inspection of average values of $fv_{\rm float}$.  The radial distance
of the bubble is also increasing with time due to the effective
gravitational acceleration, $g_{\rm eff}$.  At the start of stage
five, the bubble's radial distance is $r\sim475$ km; Figure
\ref{fig:geff} shows that $g_{\rm eff}$ itself is {\em decreasing}
steeply for larger radii.  In a similar manner as in Equation
\ref{eq:L}, the radial distance of the bubble grows as
\begin{equation}\label{eq:r}
  r \sim \frac{1}{2}g_{{\rm eff},0}t^{2-m},
\end{equation}
where $g_{{\rm eff},0}$ is the effective gravity at the start of phase
five, and fits to the data in Figure \ref{fig:anal-model} give $m \sim
0.95$.  This yields a solid angle profile of
\begin{equation}\label{eq:solid_angle}
  \Omega\propto t^{2(n+m)} \sim t^{3.3}.
\end{equation}

During phase four, the solid angle is roughly constant with time,
$\Omega/4\pi \approx 0.013 \approx 0.25\left(L/r\right)^2$.  This
implies $v_{\rm lateral} \approx v_{\rm float}\left(L/r\right)$, where
$v_{\rm lateral}$ is the lateral spread speed.  At the point when
phase four transitions to phase five, $v_{\rm lateral}$ is determined
approximately from the derivative of Equation \ref{eq:L}.  This yields
\begin{equation}\label{eq:teq}
  t \approx 0.053 \left(\frac{1}{f_0}\right)^{0.59} {\rm \ s},
\end{equation}
where the float speed cancels out from Equation \ref{eq:gamma}, and we
have used the above values for the solid angle and $c_s/g$.  Here,
$f_0$ is the value of $f$ in Equation \ref{eq:gamma} at the transition
from phase four to phase five.  At the bubble's radial distance at the
phase four to phase five transition (equivalently, about a third of
the way up the flame in Figure \ref{fig:contours}), isobars slightly
deviate from equipotential lines and $f_0 \lesssim 2\%$, small but
non-zero.  Plugging this into Equation \ref{eq:teq} yields $t\gtrsim
0.53$ s, which is close to the transition time of about 0.5 s from
Figure \ref{fig:geo}.  This very simple analysis, which ignores some
of the finer details of viscosity and shear discussed in
\cite{zingaledursi}, agrees well with our simulation data for our two
Model A runs.

While the rapid lateral expansion at the end reduces the strong
sensitivity of the mass burned to ignition radius seen in phase four
($r_{\rm ign}^{-4}$), it probably does not remove it entirely. A plume
with a larger solid angle to begin with will begin the rapid expansion
associated with phase five earlier since the sound waves will
find it more difficult to maintain lateral pressure balance deeper in
the star. Additional study is needed to determine the actual
scaling \citep[see][]{Dong:2013}.

\section{Conclusions}

We have extended our low Mach number studies of white dwarf ignition
with Maestro \citep{wdturb} into the regime of flame propagation using
the fully compressible code, Castro.  The new simulations directly map
the turbulent convective flow field from the earlier study onto the
initial grid to determine its effect on buoyant flame propagation.
This allows, for the first time, a study of the deflagration phase of
the traditional single degenerate, $M_{\rm{Ch}}$-mass model of SNe Ia
using a realistic background convective flow field in three
dimensions.  One goal of this study was to assess the degree to which
this realistic flow field would affect the evolution of the buoyant
flame as it made its way to the stellar surface.  To this end, we used
a simple thickened flame model coupled with nuclear reaction tables
generated from detailed calculations of isobaric carbon burning plus
post-flame burning in the NSE regime, similar to that used in
\cite{Ma:2013}.

We find that the background turbulence from the convective flow field
is generally a minor modification to the hydrodynamics, once the
burning enters a strongly buoyant phase. Characteristic turbulent
speeds from prior convection are comparable to the laminar speed early
on, but not to floatation speeds. Thus, except for determining the
site where the explosion originates, the turbulence from prior
convection does not play an important role in our standard run ignited
41 km off center. However, closer to the center, the buoyancy force is
smaller and the bubble interacts with the turbulence for a longer
period of time. This leads to more distortion and wrinkling that is
compounded by Rayleigh-Taylor instabilities, which grow with time.  In
the extreme case of central ignition, the background turbulence has a
profound effect on the outcome. Since the typical convective speeds
and laminar speeds are nearly the same, an outbound eddy can carry an
ignited region off-center before its floatation speed becomes
dominant. Since hot ignition sites almost universally exist in
outflows, not inflows, even ``central'' ignition becomes non-central
and produces an asymmetrical explosion. This makes it even more
difficult to burn a large solid angle of the star and cause
significant expansion prior to detonation.  Less white dwarf expansion
implies the detonation burns at higher density, producing more
$^{56}$Ni and a brighter SN Ia.

At all times, even in the presence of the background flow field, our
calculations show that the high vorticity flow indicative of strong
turbulence exists only within the ashes of the flame.  {\sl There is
  very little turbulence outside or ahead of the flame}.  This could
have consequences for more sophisticated turbulent flame models that
average the local turbulent fluctuations over some region in the
vicinity of the flame to obtain a turbulent flame speed.  If the
region over which the average is formed includes a large fraction of
material outside of the flame, this will result in a turbulent flame
speed that is too low, which might alter the total mass burned and the
nucleosynthesis.

Using a simplified flame model allows us to compare the evolution of
our buoyant flames to one-dimensional semi-analytic work of buoyant
thermals presented in \cite{BurningThermals}.  We find these
semi-analytic models approximate the overall behavior of our
three-dimensional simulations quite well, although the analytic
approximations produce bubbles with somewhat slower rise speeds and
not quite enough lateral expansion at later times.  A possible remedy
to the analytic model's slower rise speed compared to our simulations,
although complicating the simple model, would be to use a non-constant
entrainment coefficient ($\alpha$) or non-constant flame speed, or
both.  The lack of late time expansion is likely due to the neglect of
pressure gradients in the analytic approximation.

Lastly, we find that the buoyant flame bubble undergoes at least five
different phases of evolution as it progresses towards the surface.
We have distinguished these phases based upon the behaviour of the
bubble's geometric solid angle as seen from the center of the star.
In the first phase, the bubble simply burns isotropically at the
laminar speed as there is little to no turbulence.  After a short
time, the bubble has expanded enough that buoyancy starts to dominate
the upper cap, causing the bubble to extend radially.  This
buoyancy-driven phase is ended with the development of significant
vorticity in the cap of the bubble.  The high-vorticity torus that
forms in the bubble creates a brim where the flame rolls over into
itself.  This tight torus lasts for about two turnover times before
splitting up into several smaller vortex tubes, marking the transition
to a turbulence-dominated phase.  Turbulence is important to the
flame's propagation during the remainder of its evolution.
Eventually, the flame reaches a low enough density region where
lateral pressure gradients form, leading to the final phase of
significant lateral expansion lasting through breakout.  These five
phases appear to be rather robust as they are experienced by both of
our A series simulations.  That is, the background turbulence has
almost no effect on the presence or duration of these different
buoyant flame phases.

The very early laminar burning phase is important as it sets the scale
for resolution of the initial floatation and deformation of the flame.
Simple arguments relating the floatation speed to the burning speed,
show that the early growth of a bubble is inversely proportional to
its off-center ignition radius, out to about 300 km.  A point source
ignited at a typical off-center ignition location will grow in size to
a sphere of radius of a few kilometers before beginning to float.
This should be resolved.  The solid angle of the ignition point as it
starts to float should correlate to the total amount of burning
during the deflagration; our simple scaling relations are therefore
promising for model parameterization.  This behavior appears robust,
even in the presence of background turbulence for ignition
sufficiently off-center.  Closer to the center, turbulence will affect
the smooth laminar burning that is assumed in our simple description.
Further studies are required to confirm and calibrate the simple
arguments made in the text, but it is promising that one may be able
to approximately relate the total burning and expansion during the
deflagration simply to the ignition radius.  We leave this for a
future paper.

There is also a very important sixth phase of burning that occurs
after the ash erupts from the white dwarf and spreads, at sound speed,
over its surface. Hot, light ash is now on top of cold fuel and any
Rayleigh-Taylor-based description of the flame is invalid. There is,
however, still a great deal of shear between the spreading ash and the
fuel beneath. This will drive turbulent mixing that, in the extreme
case, might cause detonation, but at a minimum causes continued
turbulent burning. Our modeling of this important phase is deferred to
subsequent papers. It has been treated previously by \citet{roepke-gcd}
and \citet{jordan:2008} among others, but with different results.

If this additional burning is ignored, then our results are consistent
with the amount of pre-break out burning seen in a recent similar
study of single point ignition by \cite{ivo:2013} (their Model
N1). They are also consistent with the relatively small amount of
burning seen in this phase by the Chicago FLASH team
\citep[e.g.]{jordan:2008} and thus tend to support their conclusion
that gravitationally confined detonation will occur in this model. The
main underlying cause of the small amount of burning is our previous
result \citep{wdturb} that ignition only occurs once at a single
point, well off center. 

However, it is possible that the complete neglect of post-break out
burning is unrealistic.  Different groups treat the uncertain flame
physics after breakout in different ways.  Some simulations show very
little burning at low density ($\lesssim$ 1\%) \citep{Dong:2013} while
others show substantially more (N1 model of \cite{ivo:2013}; private
communication).  Furthermore, the burning in the surface shear layer
mentioned above poses an opportunity for a DDT from shear-heated fuel.
Further calculations of the surface burning in a shear layer are
needed to settle this important question.

\acknowledgements We thank the referee for their thorough reading of
our manuscript and their constructive comments.  We would also like to
thank Andy Aspden for insights into the nature of burning thermals and
his one-dimensional models.  We also thank Kalyana Chadalavada for
help in getting us (and keeping us) running on the Blue Waters Early
Science System.  CMM would also like to thank Rob Sisneros and Dave
Semararo for discussions of large-scale visualization and the inherent
problems therein.  Our discussions with Rainer Moll regarding
fluid-flame interactions have been extremely insightful, and we thank
him.  Correspondene with Ivo Seitenzahl was helpful in understanding
unpublished details of the MPA models. This research is part of the
Blue Waters sustained-petascale computing project, which is supported
by the National Science Foundation (award number OCI 07-25070) and the
state of Illinois.  Blue Waters is a joint effort of the University of
Illinois at Urbana-Champagne and its National Center for
Supercomputing Applications.  This work is also part of the ``PRAC
Type Ia Supernovae'' PRAC allocation support by the National Science
Foundation (award number OCI-1036199).  Additional computer time for
the calculations in this paper was provided through a DOE INCITE award
at the Oak Ridge Leadership Computational Facility (OLCF) at Oak Ridge
National Laboratory, which is supported by the Office of Science of
the U.S. Department of Energy under Contract No. DE-AC05-00OR22725.
This research used resources of the National Energy Research
Scientific Computing Center, which is supported by the Office of
Science of the U.S. Department of Energy under Contract
No. DE-AC02-05CH11231. This work was also supported, at UCSC, by the
National Science Foundation (AST 0909129), the NASA Theory Program
(NNX09AK36G), and especially by the DOE HEP Program through grant
DE-FC02-06ER41438.  MZ is supported by DOE/Office of Nuclear Physics
grant No.~DE-FG02-06ER41448 to Stony Brook.

\clearpage

\clearpage

\begin{table}
  \caption{Properties of the various models in this paper.\label{tab:models}}
  \begin{center}
    \begin{tabular}{cccccccc}
      & Initial &  & & & & &\\
      & Flow & $r_{\textrm{ign}}$\tablenotemark{a} 
      & $t_{\textrm{max}}$\tablenotemark{b} 
      & $r_{\textrm{max}}(t_{\textrm{max}})$\tablenotemark{c} 
      & $M_{\textrm{IME}}(t_{\textrm{max}})$
      & $M_{\textrm{IGE}}(t_{\textrm{max}})$ 
      & $E_{\textrm{nuc}}(t_{\textrm{max}})$ \\
      Model & Field? & (km) & (s) & (km) & ($10^{-3}M_\odot$) & ($10^{-3}M\odot$) & (erg)\\
      \hline\hline
      A0 & N & 41.3 & 0.784 & 1947 & 1.46 & 26.1 & 3.8$\times10^{49}$\\
      AV & Y & 41.3 & 0.798 & 1970 & 1.43 & 25.8 & 3.8$\times10^{49}$\\
      B0 & N & 10.0 & 0.335 & 300.8 & 0.20 & 2.63 & 3.5$\times10^{48}$\\
      BV & Y & 10.0 & 0.365 & 519.3 & 0.53 & 6.15 & 2.3$\times10^{49}$\\
      C0 & N & 0.00 & 0.424 & 53.85 & 0.02 & 0.41 & 3.8$\times10^{47}$\\
      CV & Y & 0.00 & 0.497 & 101.1 & 0.05 & 0.29 & 7.2$\times10^{47}$\\
      \hline
      \tablenotetext{a}{Off-center ignition distance}
      \tablenotetext{b}{Maximum evolution time}
      \tablenotetext{c}{Maximum radial extent of burning at $t=t_{\textrm{max}}$}
    \end{tabular}
  \end{center}
\end{table}

\clearpage

\clearpage

\begin{figure}
  \begin{center}
    \includegraphics[width=0.32\textwidth]{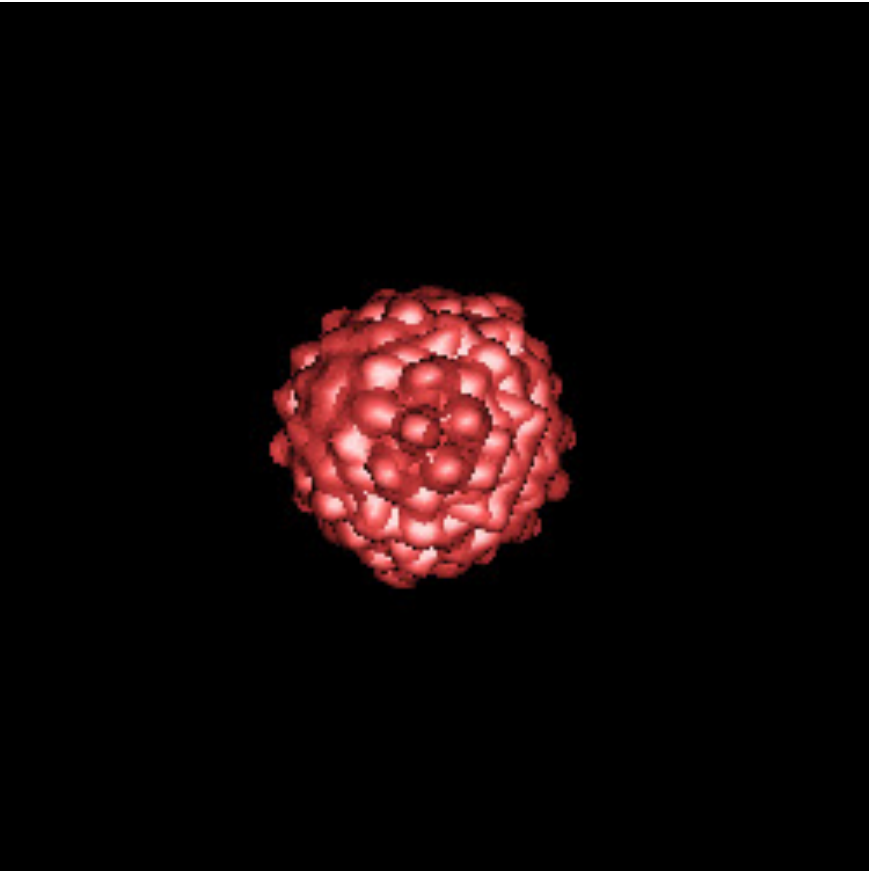}
    \includegraphics[width=0.32\textwidth]{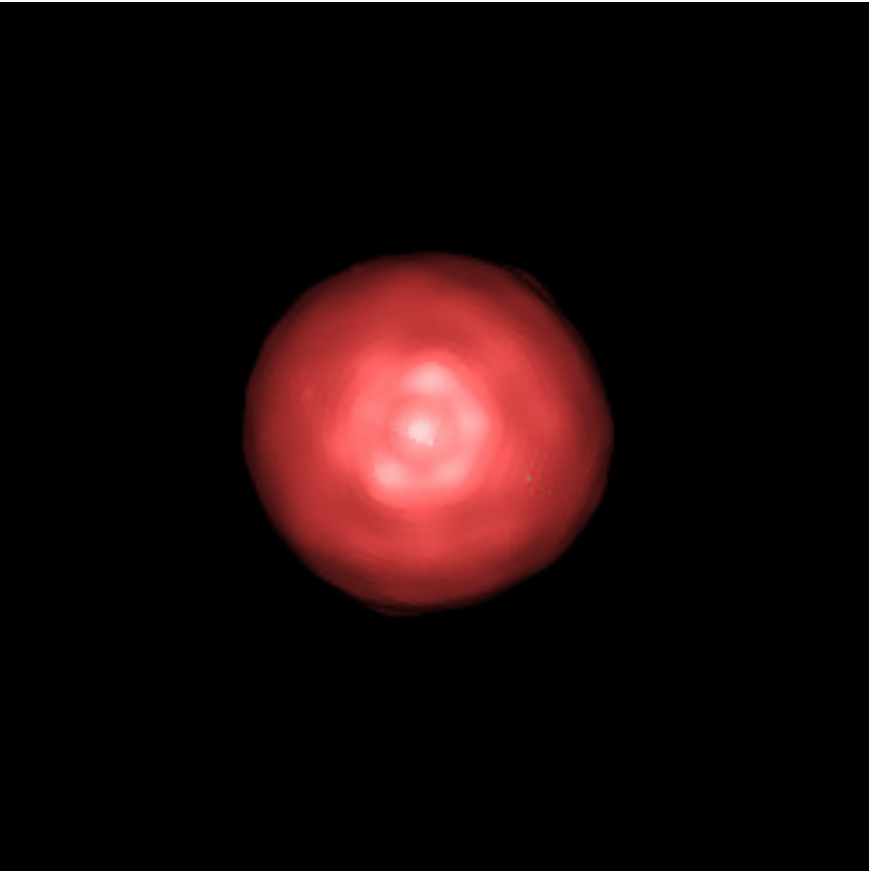}
    \includegraphics[width=0.32\textwidth]{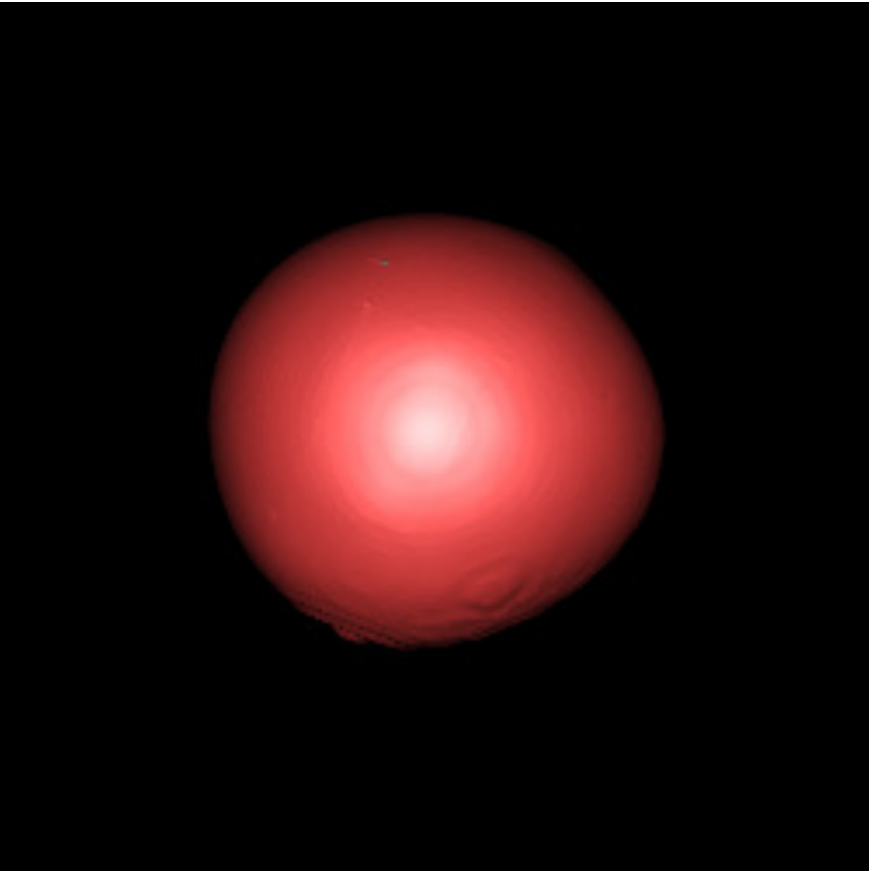}
    \caption{\label{fig:early-burn} Flame surface
      ($X(^{12}\rm{C})=0.45$ isocontours) during early evolution of
      Model AV at $t=0.00$, $t=0.0059,$ and $t=0.0119$ s of evolution,
      from left to right.  Each frame has the same spatial scale.  The
      flame quickly burns through the initial perturbations on the top
      side of the bubble.  The direction towards the center of the
      star is downward in these frames.}
  \end{center}
\end{figure}

\clearpage

\begin{figure}
  \begin{center}
    \includegraphics[width=0.32\textwidth]{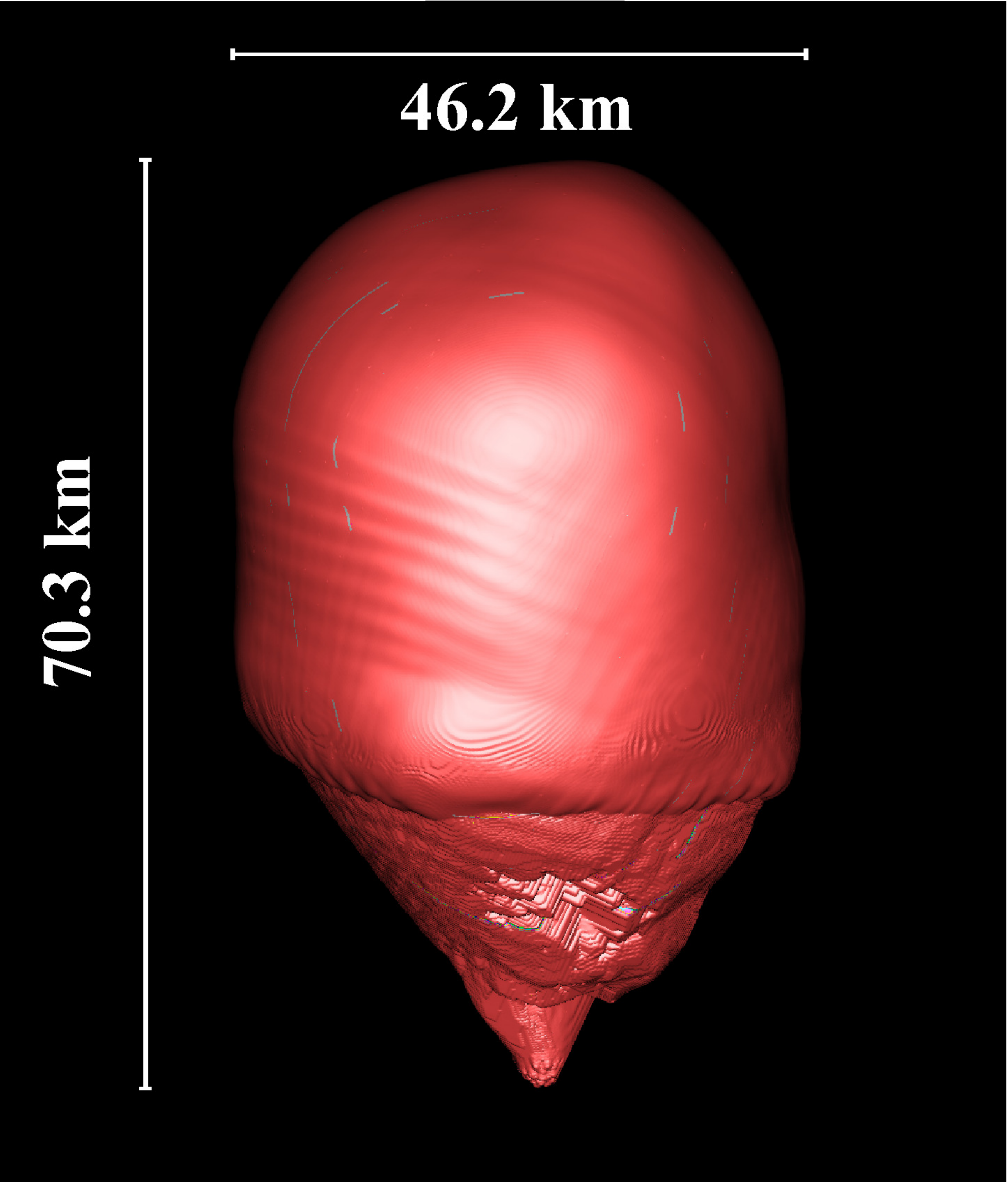}
    \includegraphics[width=0.32\textwidth]{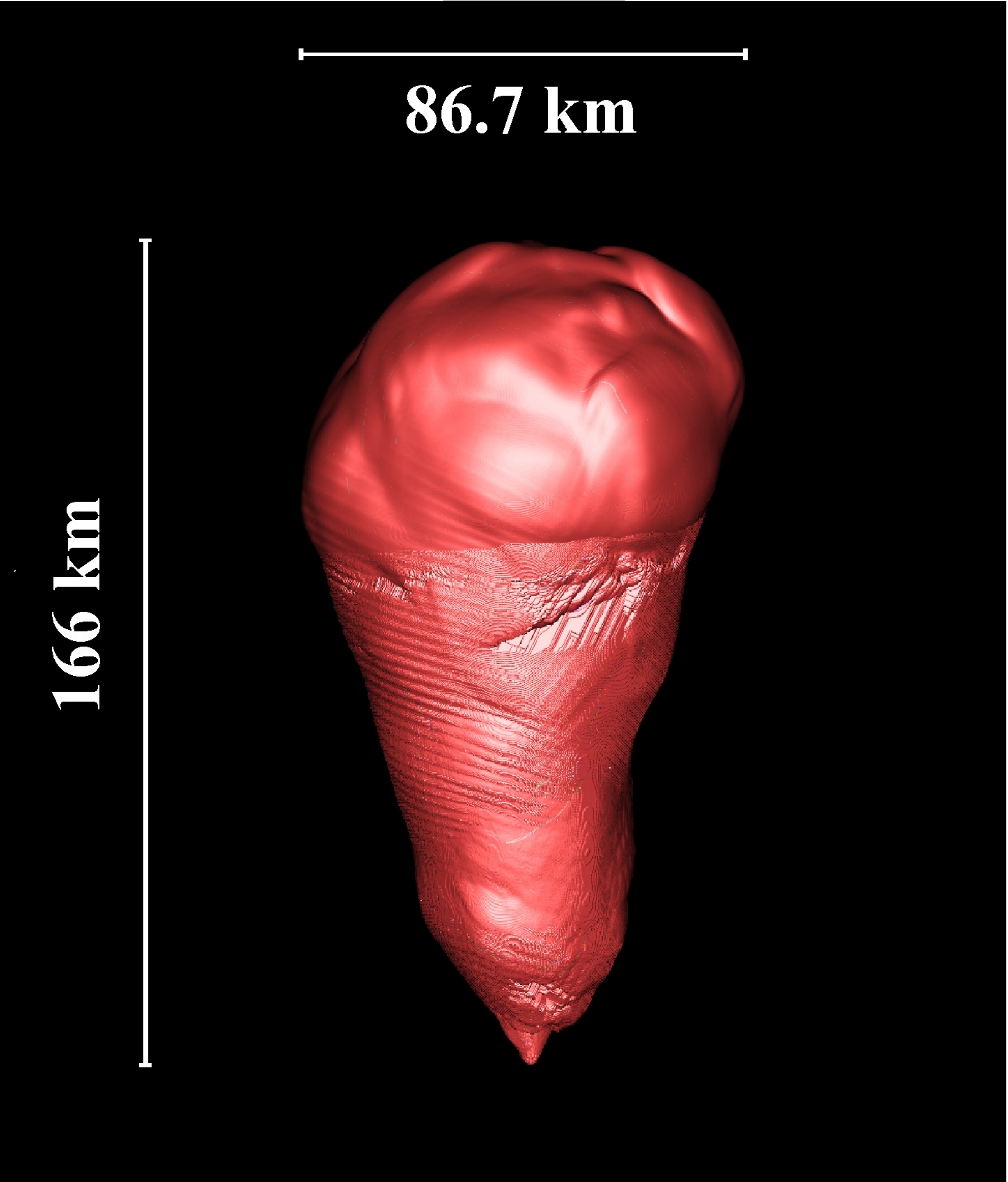}
    \includegraphics[width=0.32\textwidth]{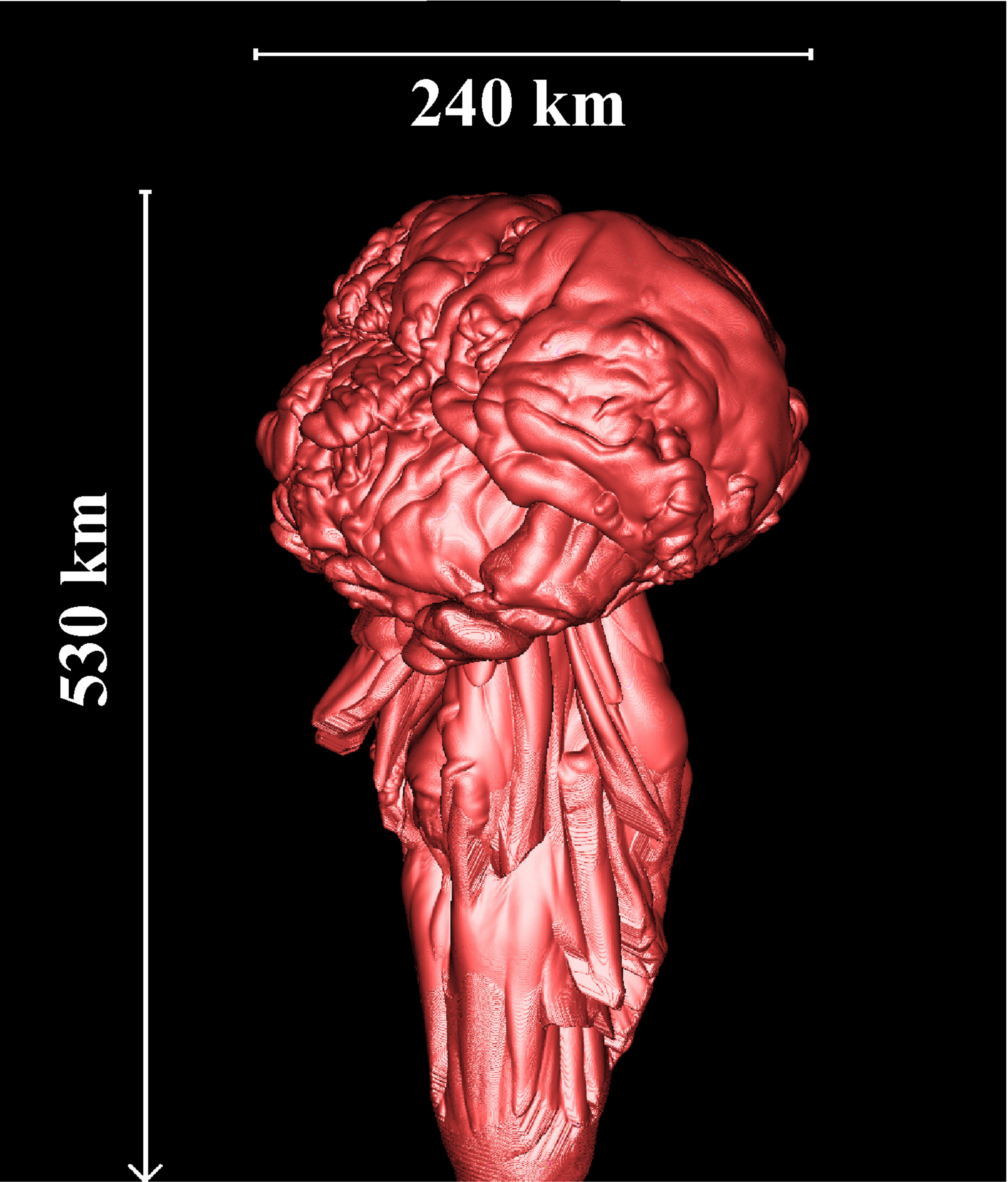}
    \caption{\label{fig:later-burn} Same as Figure
      \ref{fig:early-burn} --- $X(^{12}\rm{C})=0.45$ isocontours
      (flame surface) --- but at $t=0.150,\ 0.265,\ 0.469$ s from left
      to right.  Each frame has a different zoom level, with the
      spatial extents of the flame surface indicated.  The direction
      towards the center of the star is downward in these frames.}
  \end{center}
\end{figure}

\clearpage

\begin{figure}
  \begin{center}
    \includegraphics[width=0.32\textwidth]{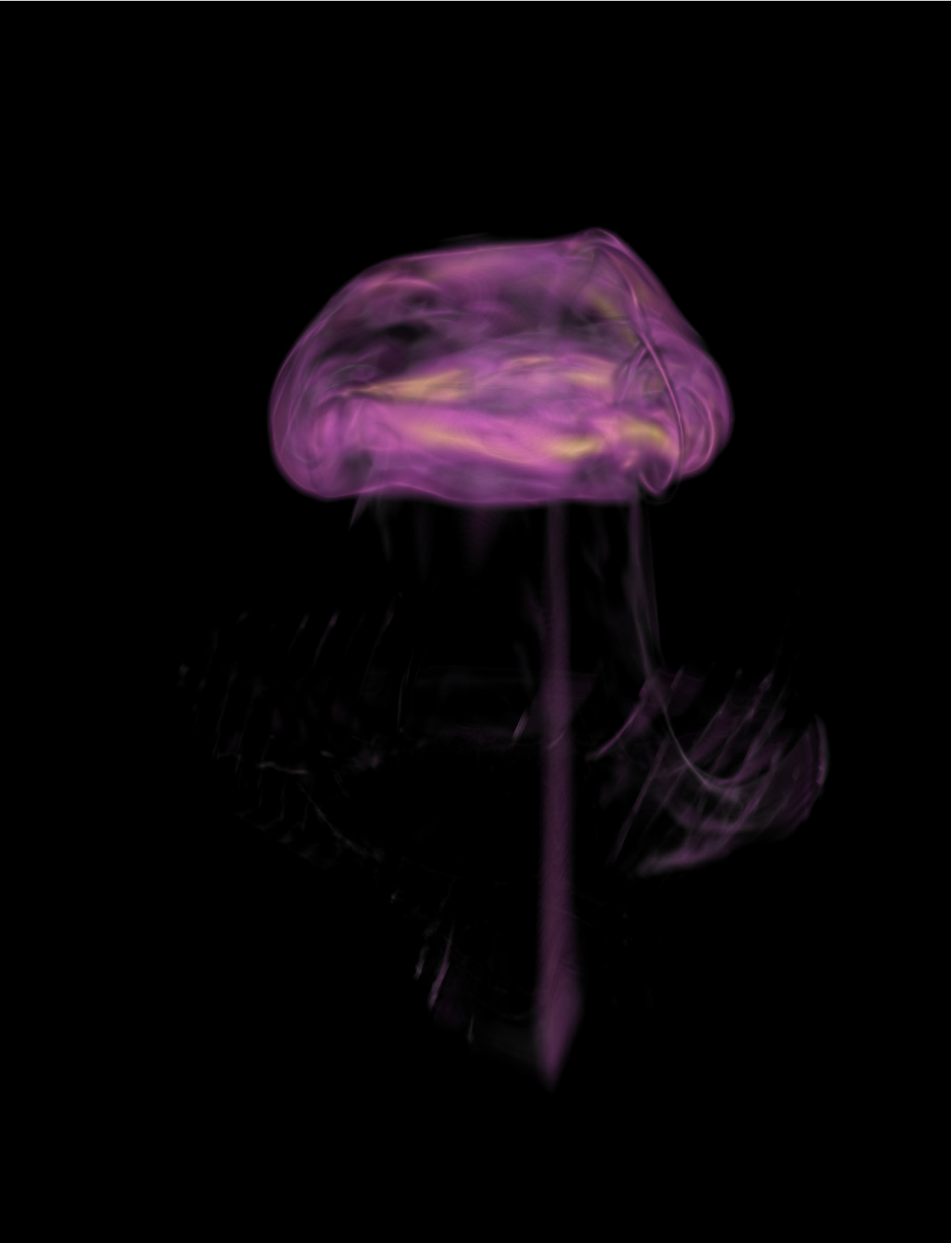}
    \includegraphics[width=0.32\textwidth]{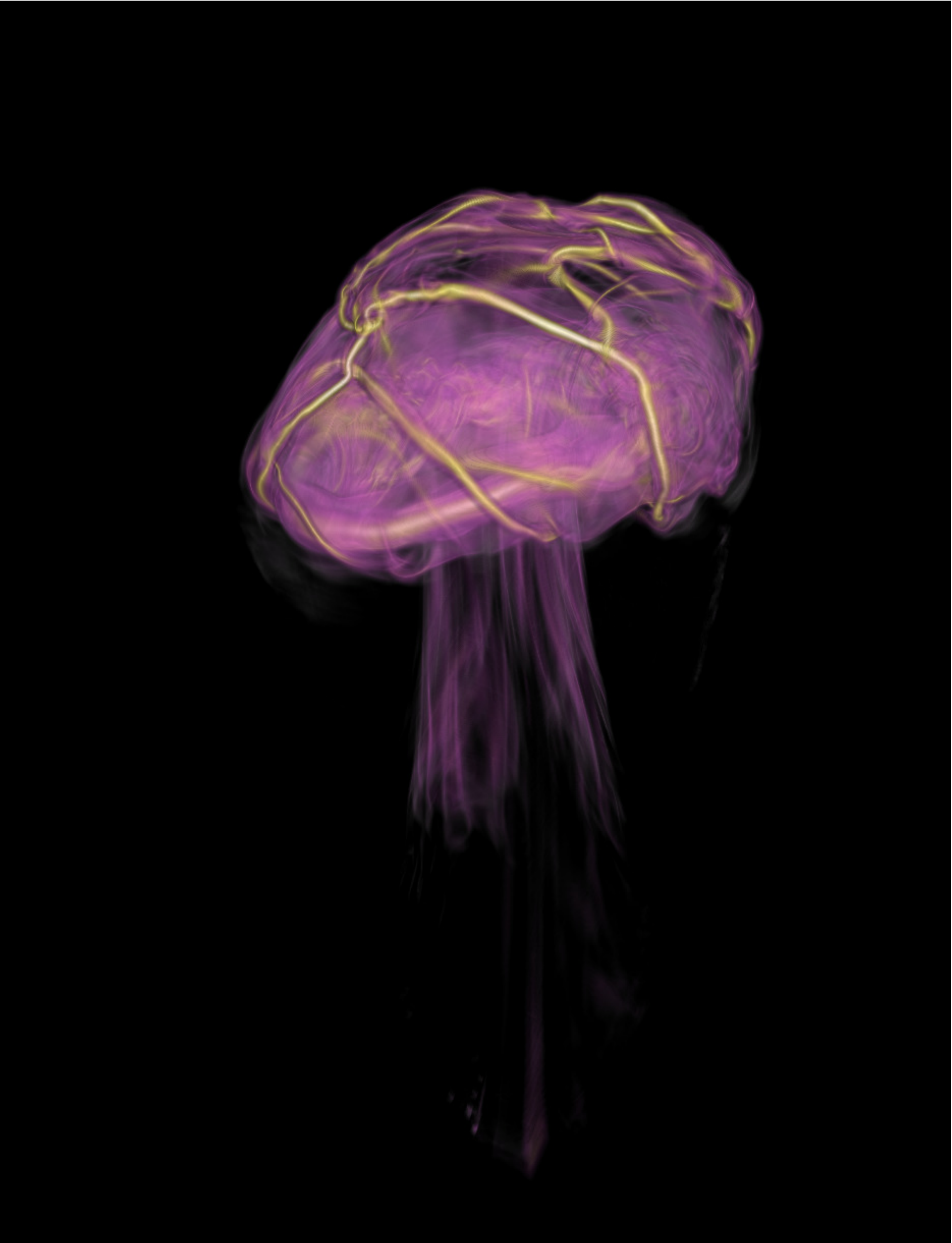}
    \includegraphics[width=0.32\textwidth]{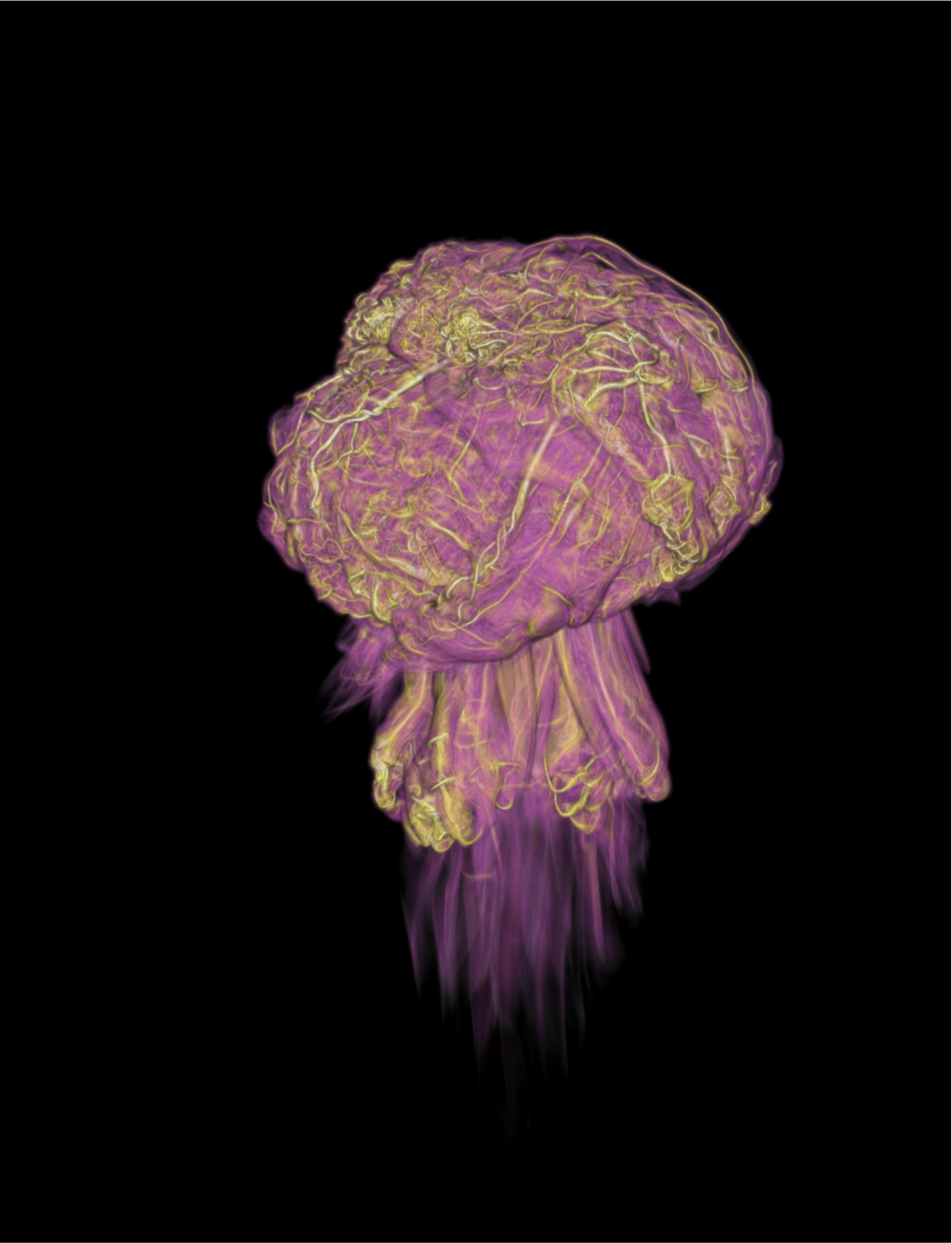}
    \caption{\label{fig:volrend-vort} Volume renderings of the
      magnitude of vorticity at the same times ($t=0.150,\ 0.265$, and
      $0.469$ s, from left to right)and spatial scales as in Figure
      \ref{fig:later-burn}.  The color scheme goes smoothly from grey,
      to blue, to purple, to yellow-white as the vorticity increases
      logarithmically from 100 s$^{-1}$ to several thousand s$^{-1}$.}
  \end{center}
\end{figure}

\clearpage

\begin{figure}
  \begin{center}
    \includegraphics[width=0.8\textwidth]{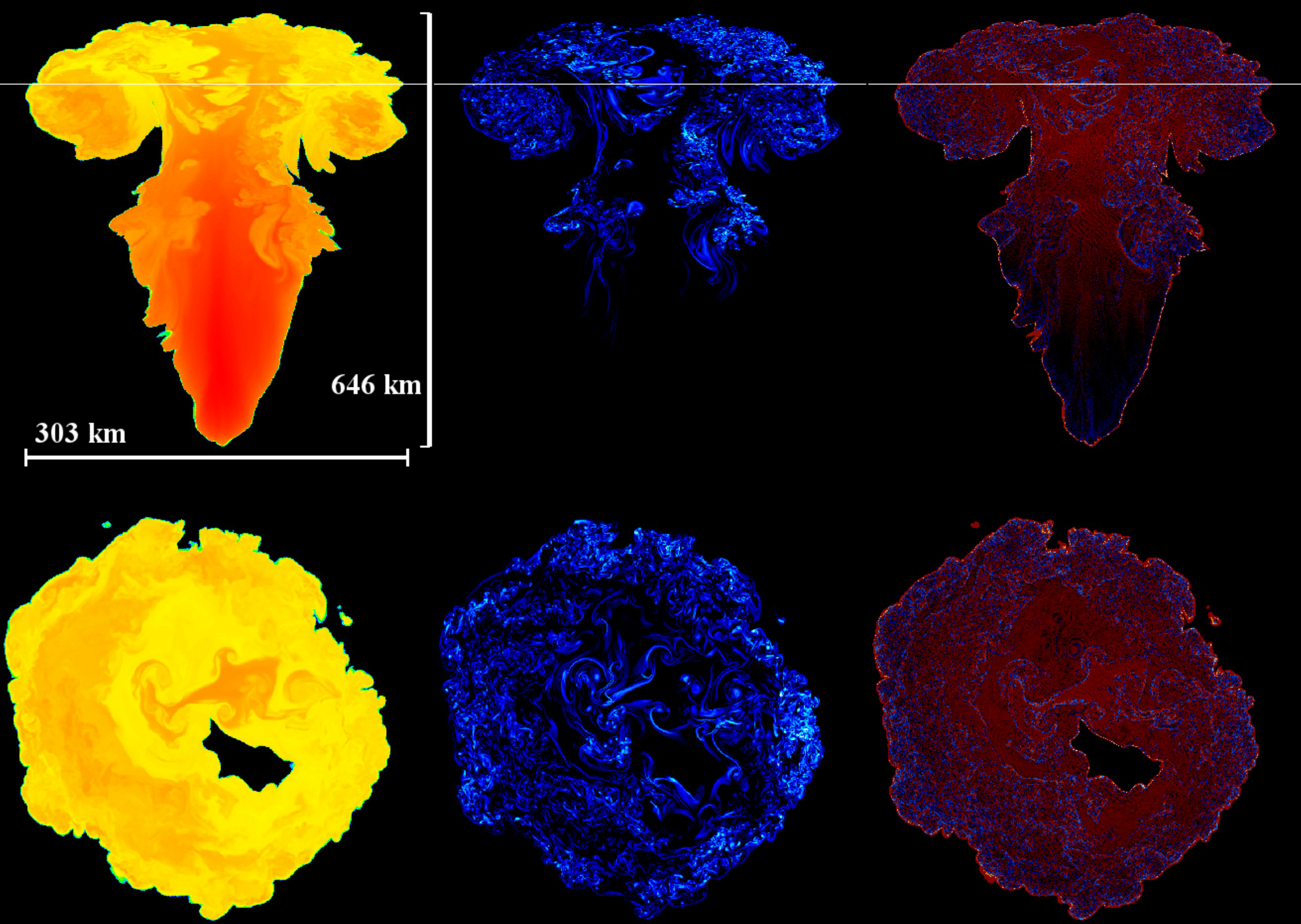}
    \includegraphics[width=0.8\textwidth]{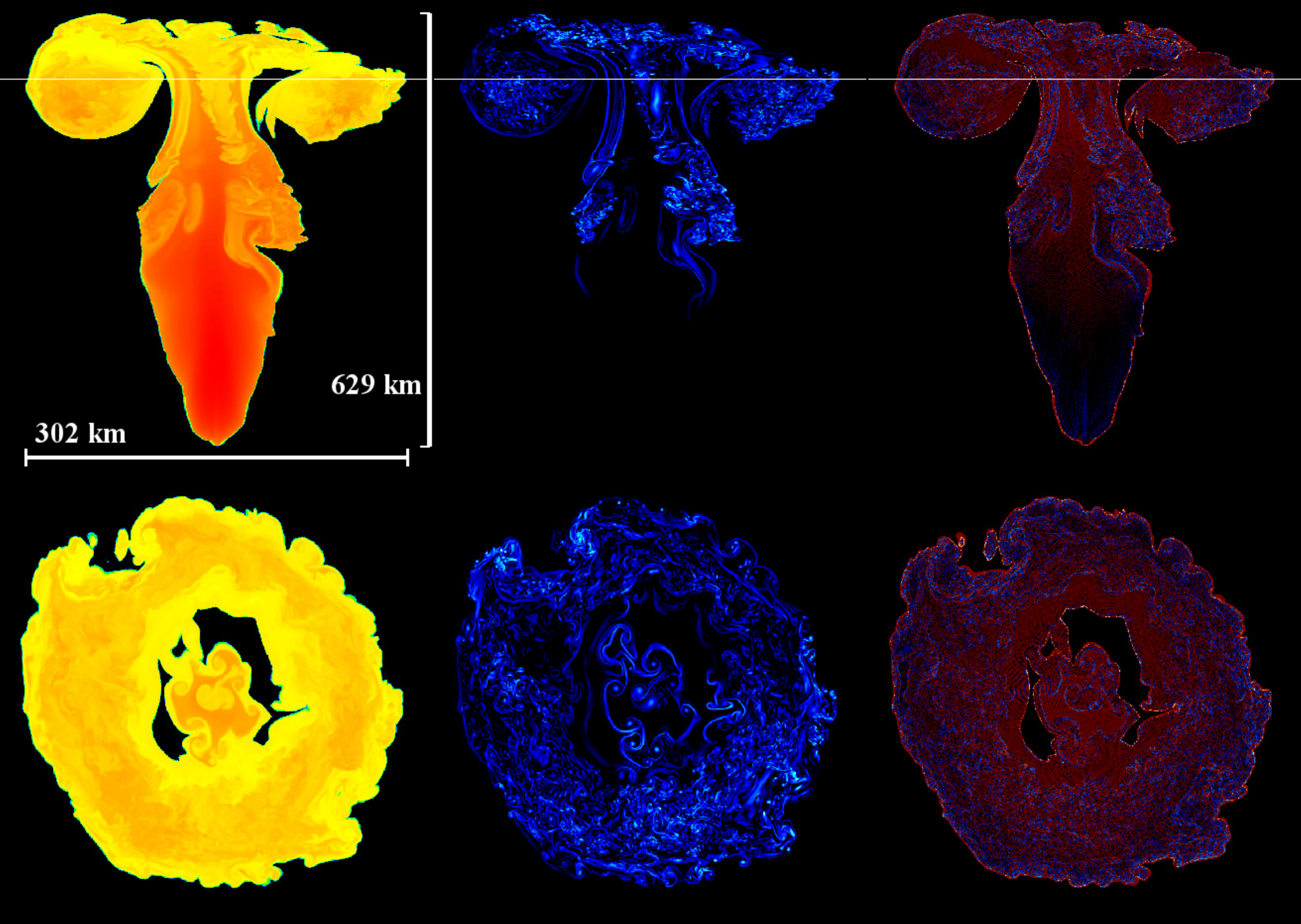}
    \caption{\label{fig:slice5} Comparison of Model AV (top) with
      Model A0 (bottom) at $t=0.5$ s.  The columns are, from left to
      right, temperature (yellow through orange), magnitude of
      vorticity (blue through white) and energy generation rate (blue
      for negative values, red for positive).  The thin, horizontal
      white line in the first row of each panel indicates the location
      of the orthogonal slice shown in the bottom row of each panel.
      The top row of each panel has an aspect ratio given by the
      dimensions shown; the bottom rows have an aspect ratio of unity,
      with the scale given by the width shown.}
  \end{center}
\end{figure}

\clearpage

\begin{figure}
  \begin{center}
    \includegraphics[width=0.8\textwidth]{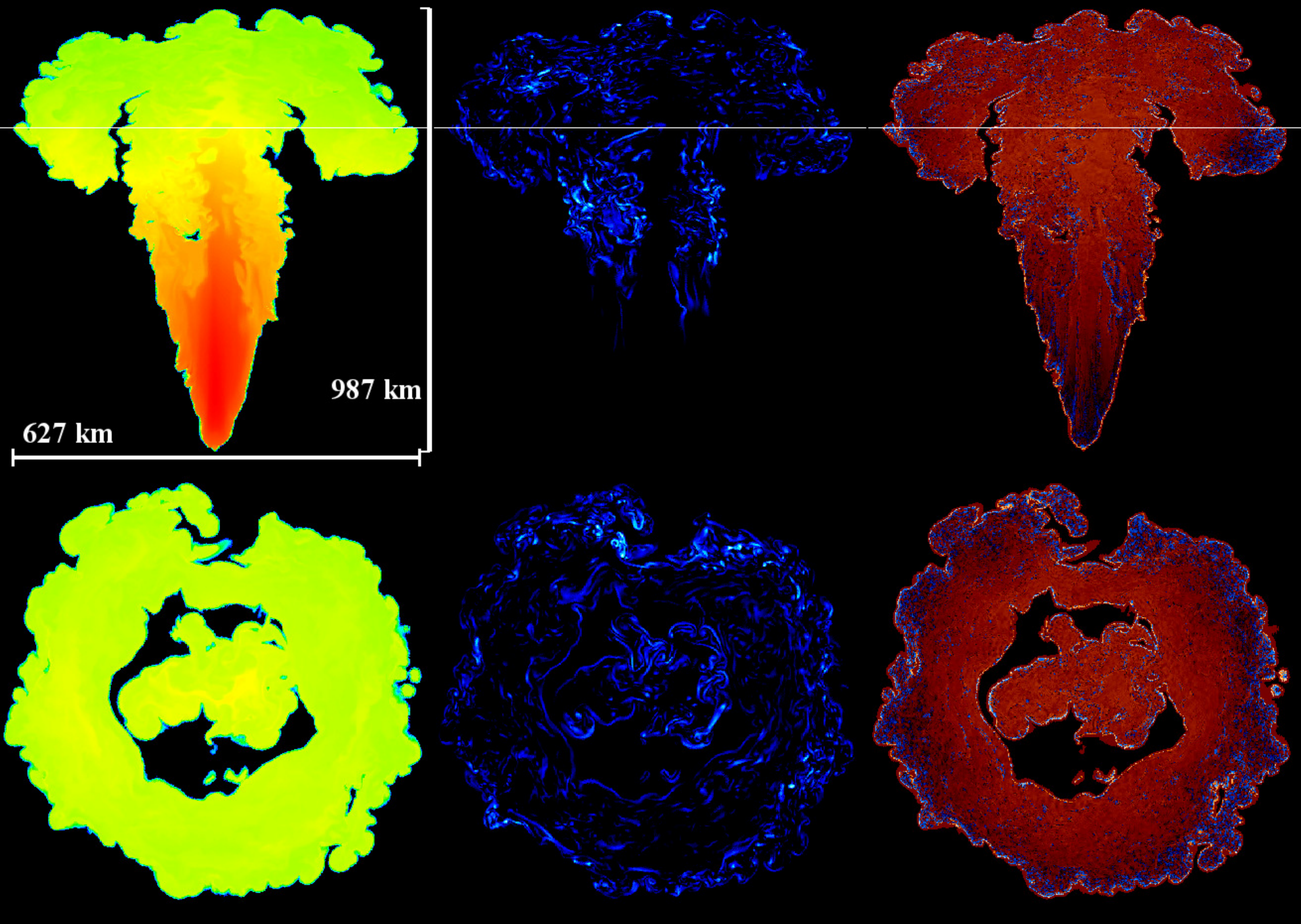}
    \includegraphics[width=0.8\textwidth]{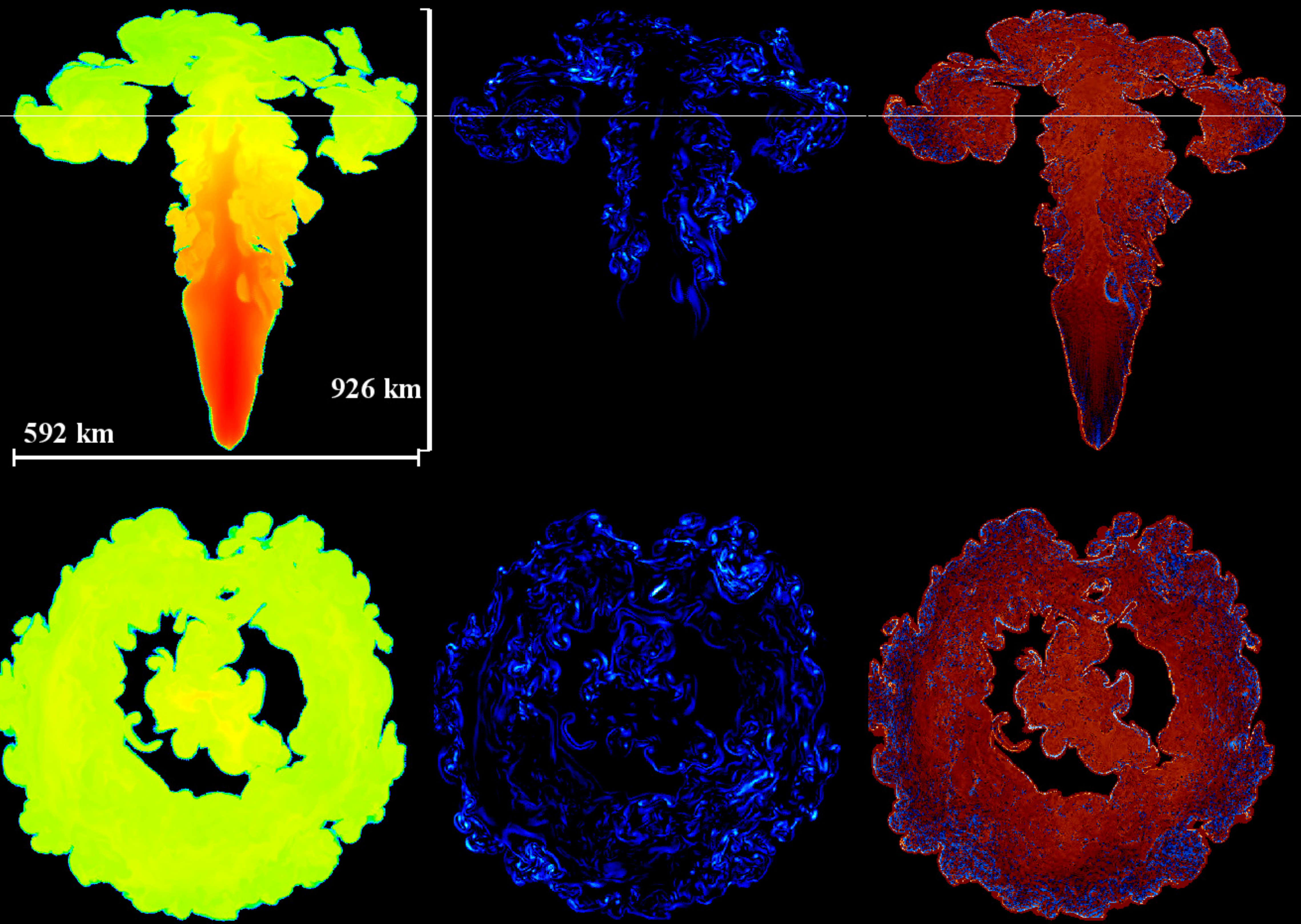}
    \caption{\label{fig:slice6} Same as Figure \ref{fig:slice5},
      comparison of Model AV (top) with Model A0 (bottom), except at
      $t=0.6$ s.  The columns are, from left to right, temperature
      (yellow through orange), magnitude of vorticity (blue through
      white) and energy generation rate (blue for negative values, red
      for positive).  The thin, horizontal white line in the first row
      of each panel indicates the location of the orthogonal slice
      shown in the bottom row of each panel.  The top row of each
      panel has an aspect ratio given by the dimensions shown; the
      bottom rows have an aspect ratio of unity, with the scale given
      by the width shown.}
  \end{center}
\end{figure}

\clearpage

\begin{figure}
  \begin{center}
    \includegraphics[width=0.8\textwidth]{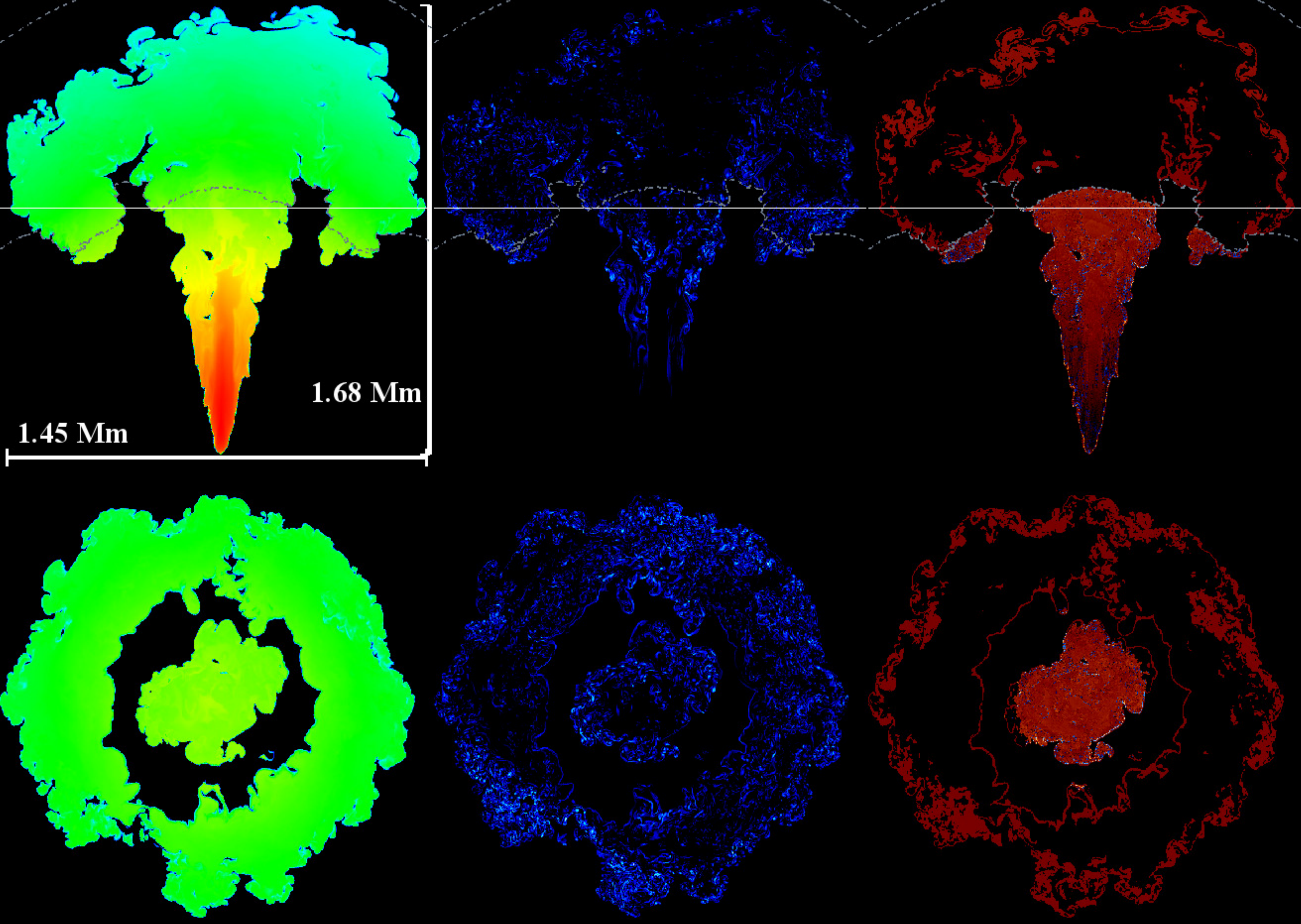}
    \includegraphics[width=0.8\textwidth]{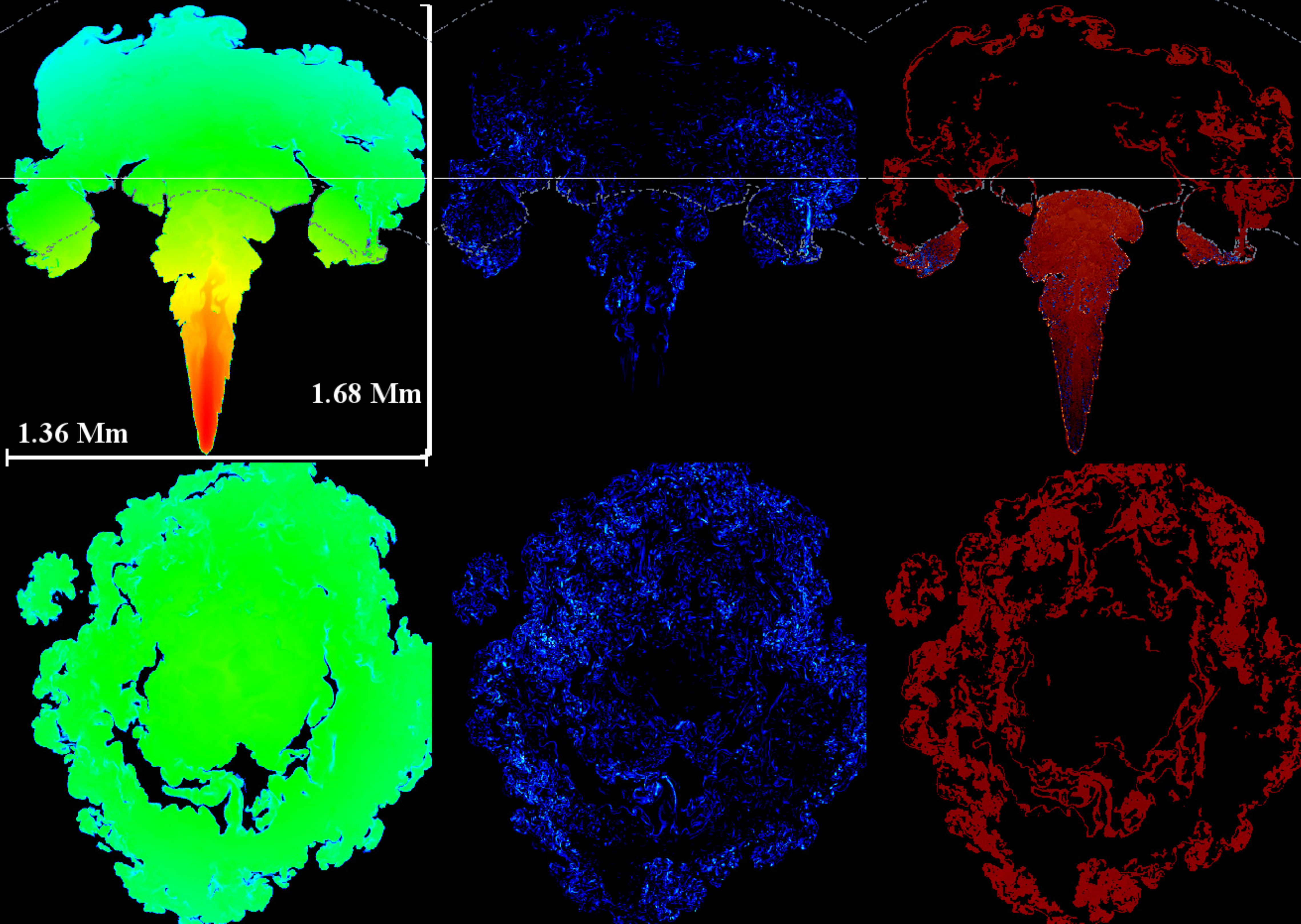}
    \caption{\label{fig:slice75} Same comparison of Model AV (top)
      with Model A0 (bottom) as Figure \ref{fig:slice5} but at
      $t=0.75$ s.  The columns are, from left to right, temperature
      (blue through orange), magnitude of vorticity (blue through
      white) and energy generation rate (blue for negative values, red
      for positive).  The thin, horizontal white line in the first row
      of each panel indicates the location of the orthogonal slice
      shown in the bottom row of each panel.  The top row of each
      panel has an aspect ratio given by the dimensions shown; the
      bottom rows have an aspect ratio of unity, with the scale given
      by the width shown.  The dotted grey lines are contours of
      density with the inner at $\rho=10^8$ g cm$^{-3}$ and the outer
      at $\rho=10^{7}$ g cm$^{-3}$.}
  \end{center}
\end{figure}

\clearpage

\begin{figure}
  \begin{center}
    \includegraphics[width=0.9\textwidth]{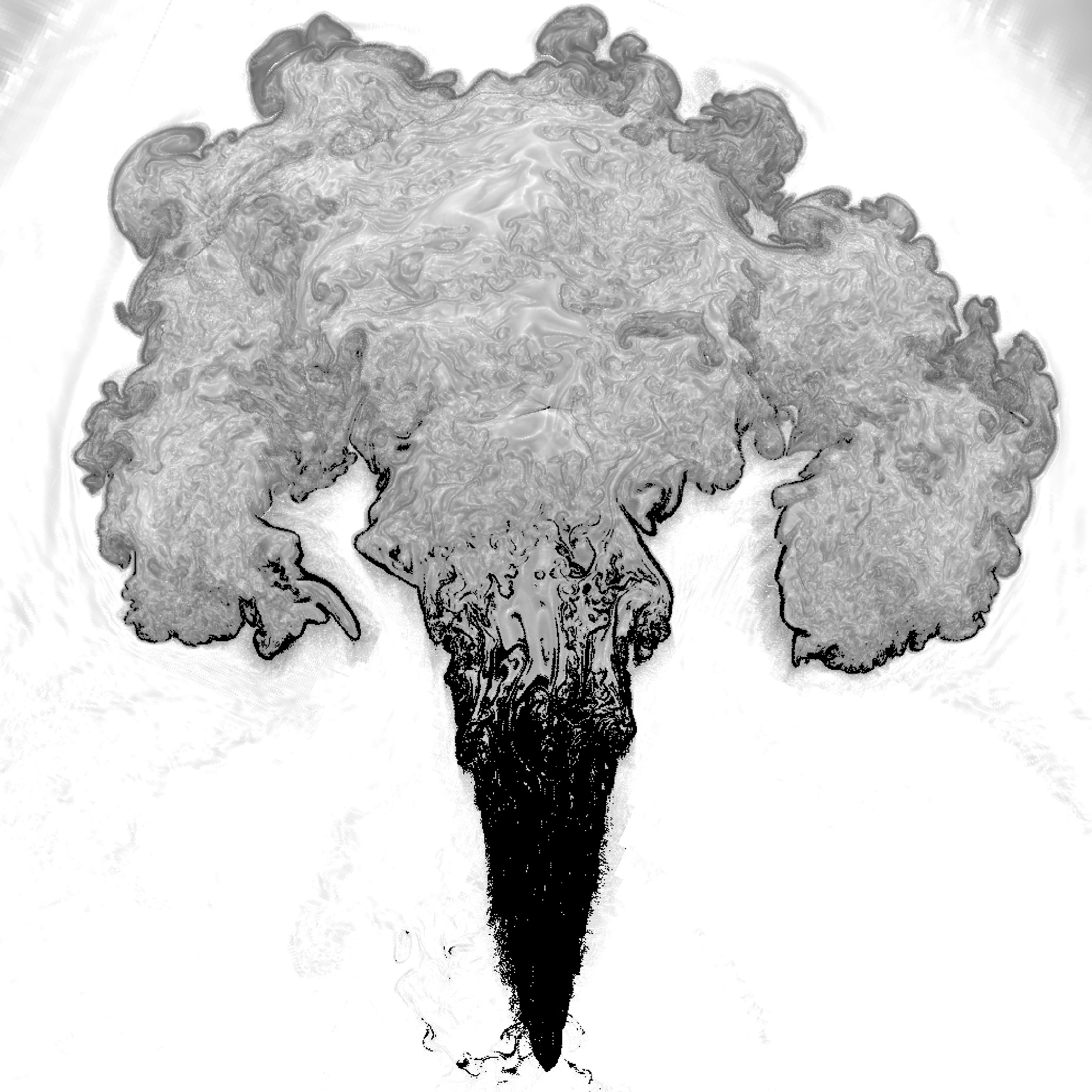}
    \caption{\label{fig:baro} Magnitude of baroclinicity, $\psi$, for
      Model AV at $t=0.8$ s.  The color map goes from less than 1
      (white) to $3.3\times10^6$ s$^{-2}$ (black).  Baroclinicity is
      strongest in the tail and underside of the cap, however the
      relative direction with respect to the flame surface is
      different in these two regions.  In the tail, $\vec\psi$ is
      orthogonal to the flame surface and creates a bulk flow along
      the direction of the plume in addition to buoyancy.  The
      vicinity of the cap sees $\vec\psi$ nearly parallel to the flame
      surface such that fuel is entrained across the flame front. }
  \end{center}
\end{figure}

\clearpage

\begin{figure}
  \begin{center}
    \includegraphics[width=0.8\textwidth]{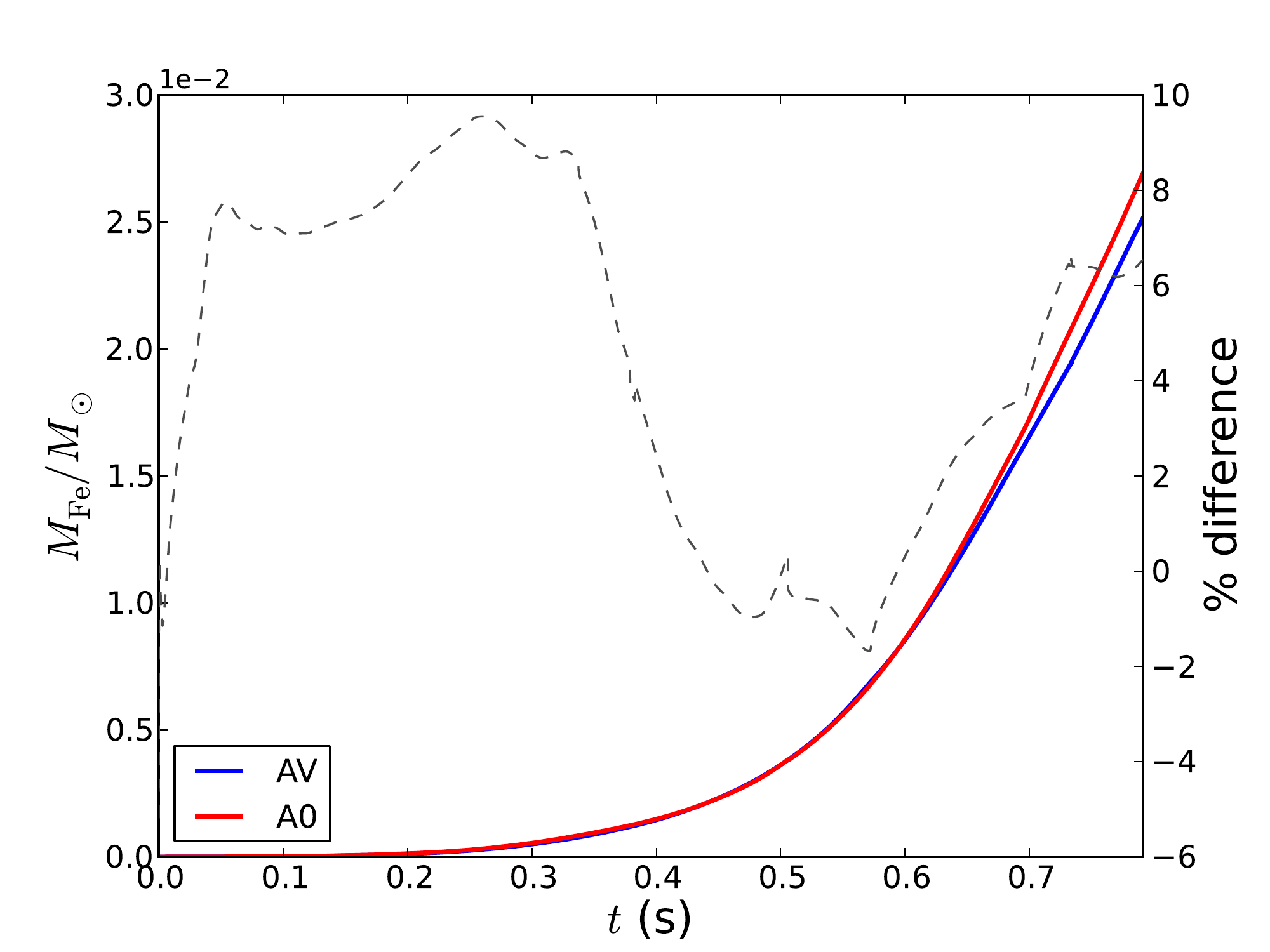}
    \includegraphics[width=0.8\textwidth]{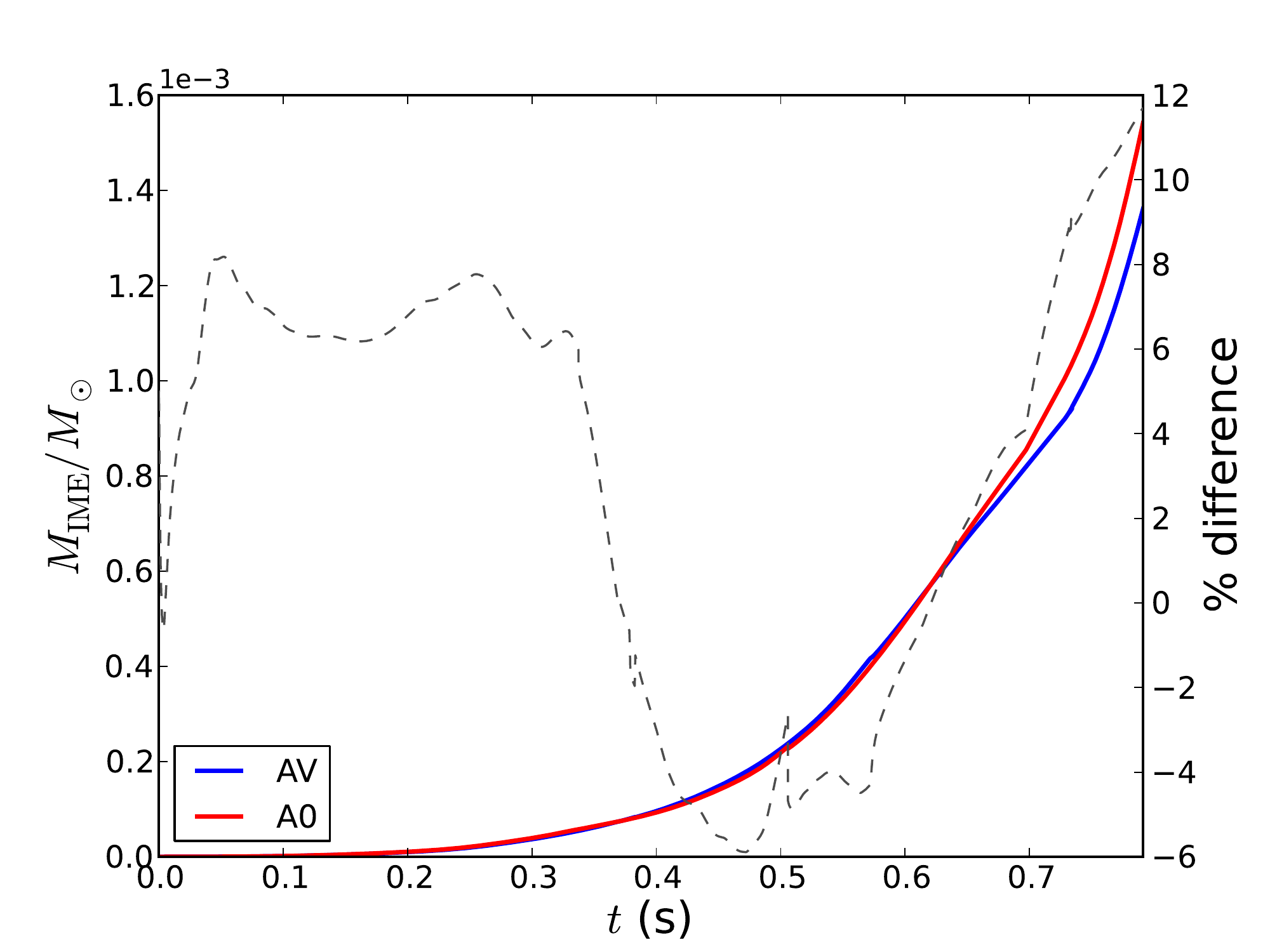}
    \caption{\label{fig:mass-burned} Nucleosynthetic yields (left
      axes) and percent difference (right axes) as a function of time
      for Models AV (blue) and A0 (red).  The top plot shows the
      production of iron-group material, and the bottom panel shows
      IME.  The percent differences are formulated such that a
      positive value indicates more production in Model A0 than Model
      AV.  In general, the model without the background flow field (A0)
      burns more IME and iron-group material.}
  \end{center}
\end{figure}

\clearpage

\begin{figure}
  \begin{center}
    \includegraphics[width=0.49\textwidth]{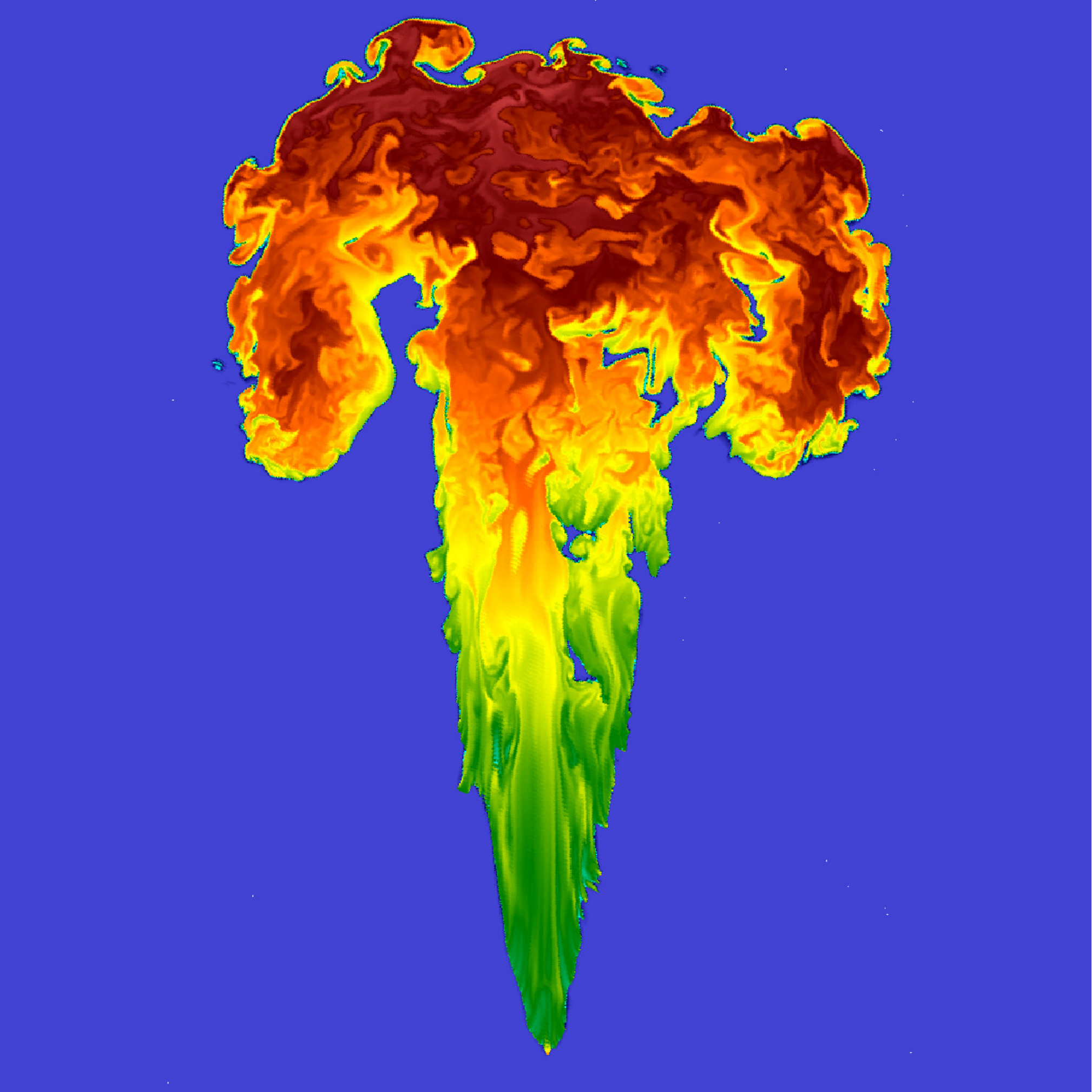}
    \includegraphics[width=0.49\textwidth]{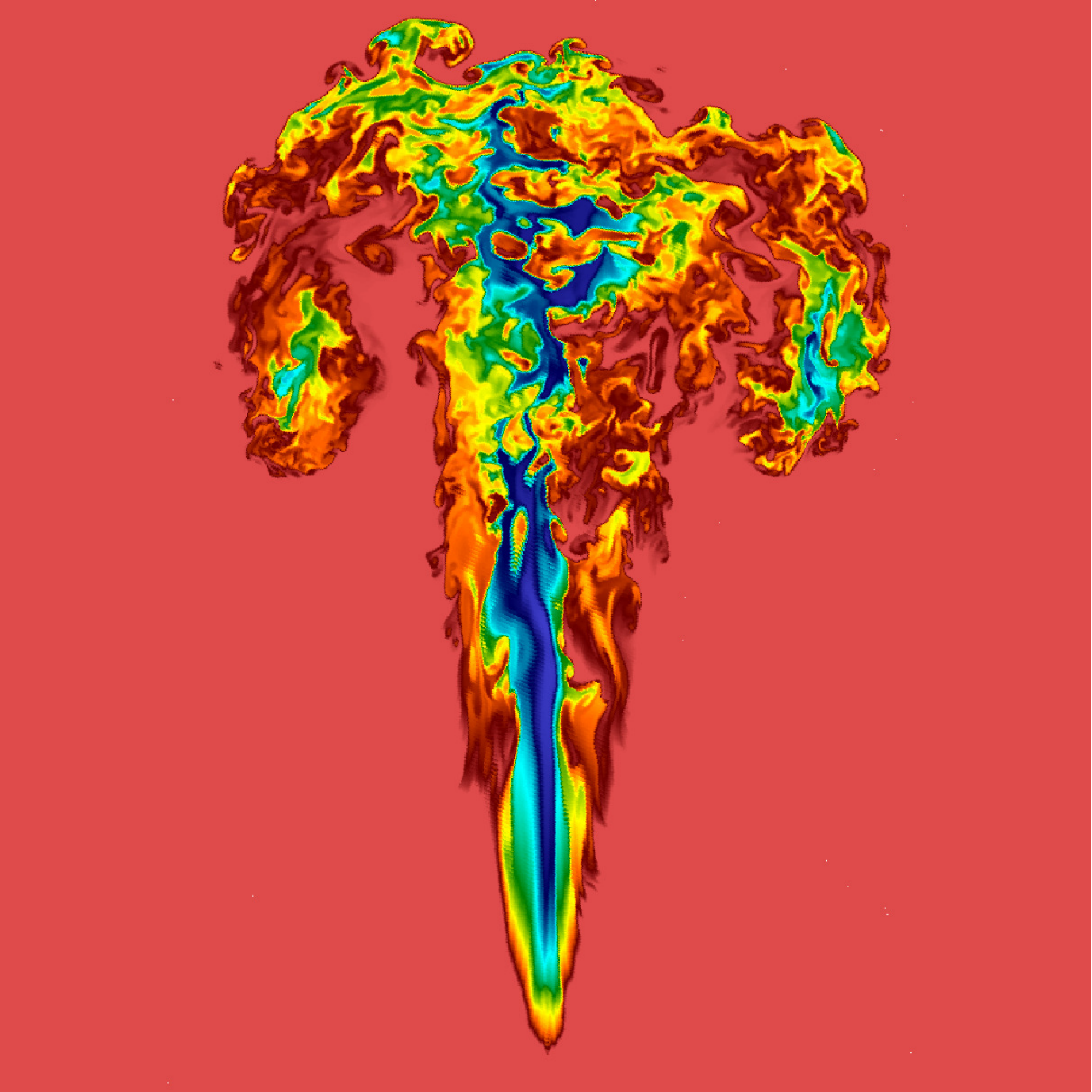}
    \includegraphics[width=0.49\textwidth]{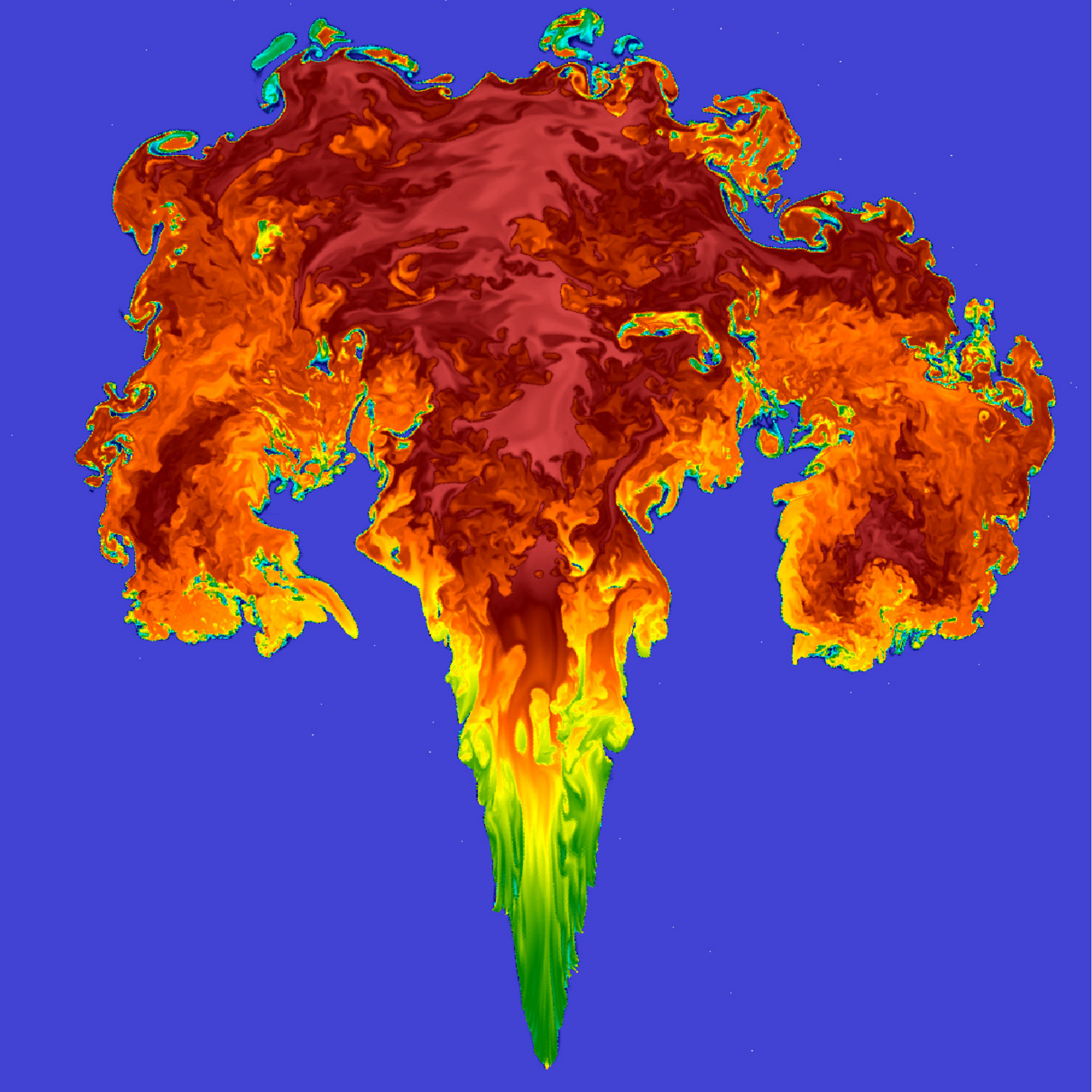}
    \includegraphics[width=0.49\textwidth]{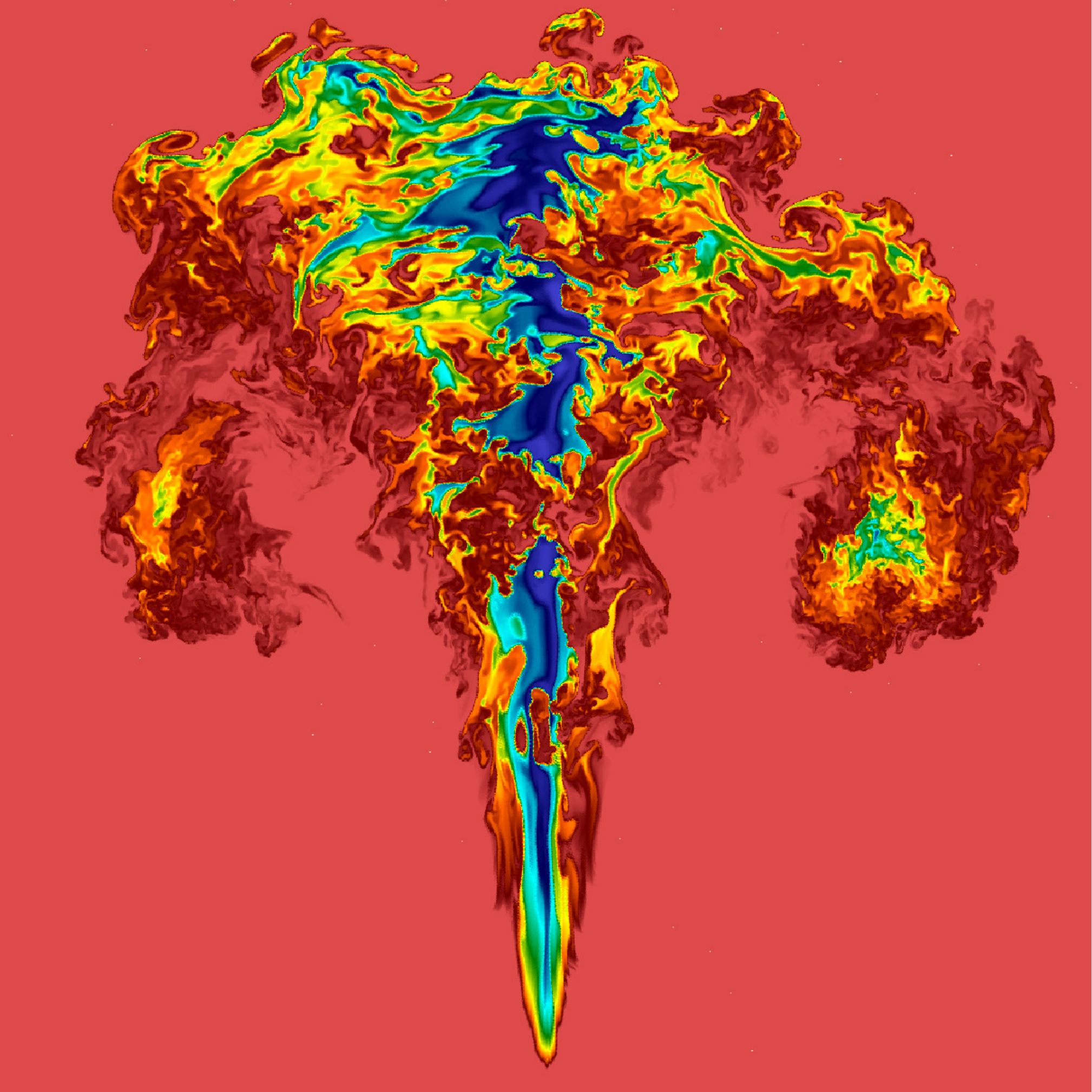}
    \caption{\label{fig:bea_ye} BE/A (left column) and $Y_e$ (right
      column) at $t=0.65$ s (top row) and $t=0.8$ s (bottom row) for
      the standard run AV.  The color maps increase from blue to red;
      for BE/A the range is [7.82,8.68] and for $Y_e$ the range is
      [0.47,0.50].  The frames in the top row are 1200 km on a side
      and those in the bottom row are 2000 km on each side.  At early
      times and in regions closer to the center, the ash temperature
      is high (Figure \ref{fig:t9_he}) and the helium mass fraction is
      nearly equal to the IGE mass fraction.  As the flame progresses
      to lower densities (bottom row), the temperature in the ash
      decreases and helium recombination leads to more tightly bound
      nuclei with larger BE/A.  Electron capture prefers higher
      density, but there is also a significant abundance of
      neutron-rich material in a narrow plume about the plume's
      center, which coincides with the region of tightly bound
      nuclei. The neutron-rich nuclei now in the outer layers were
      made deeper in the star and pushed out by a vigorous central
      flow.}
  \end{center}
\end{figure}

\clearpage

\begin{figure}
  \begin{center}
    \includegraphics[width=0.49\textwidth]{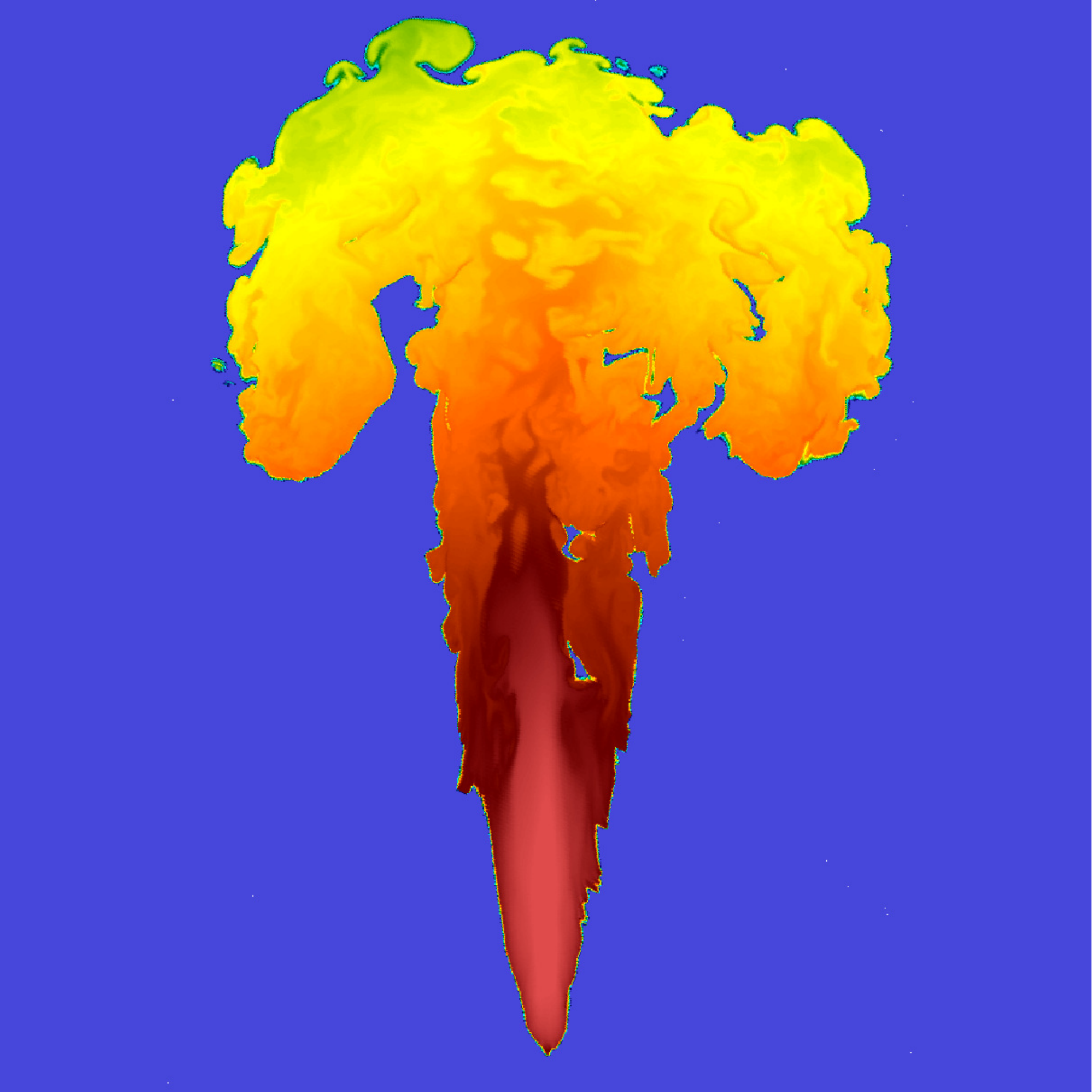}
    \includegraphics[width=0.49\textwidth]{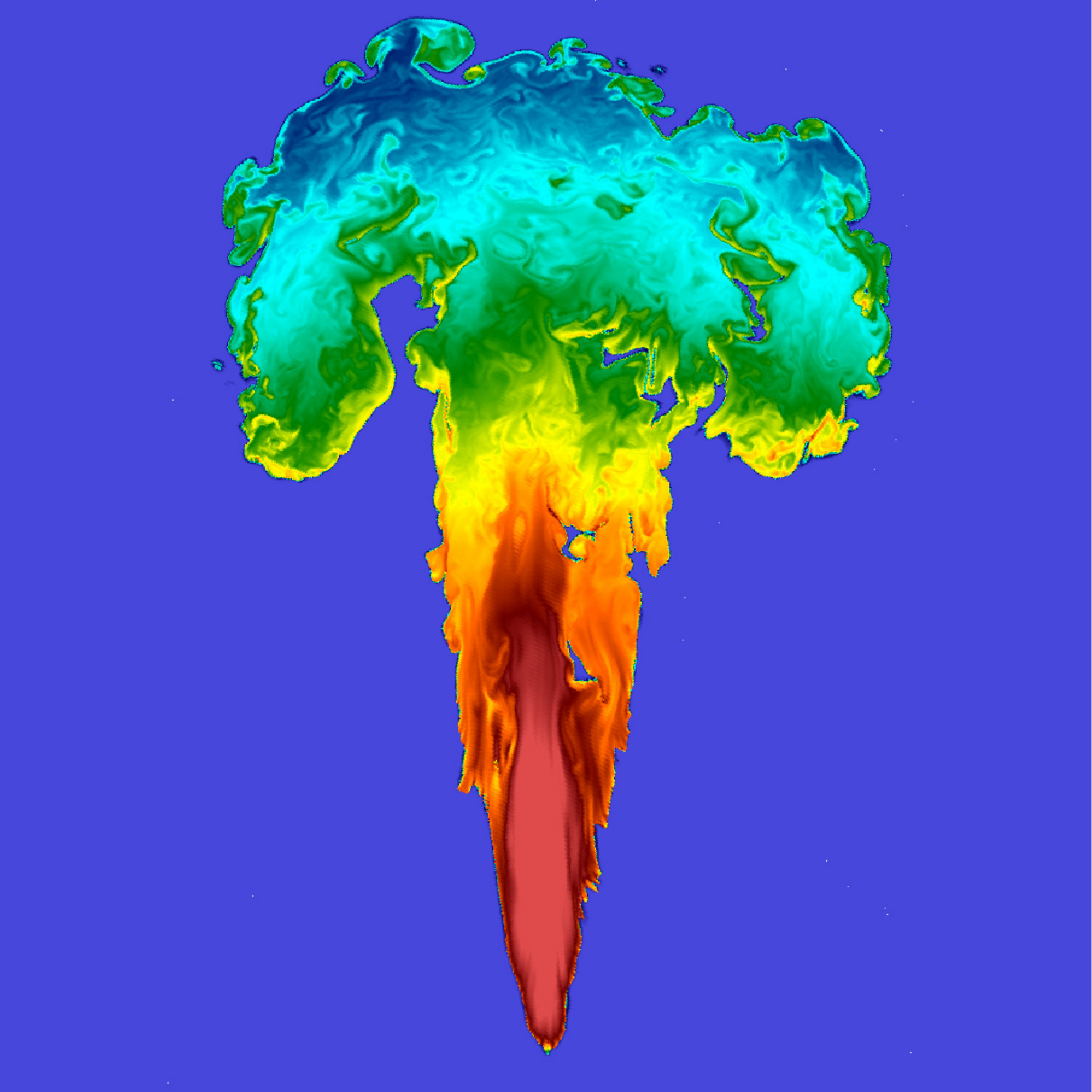}
    \includegraphics[width=0.49\textwidth]{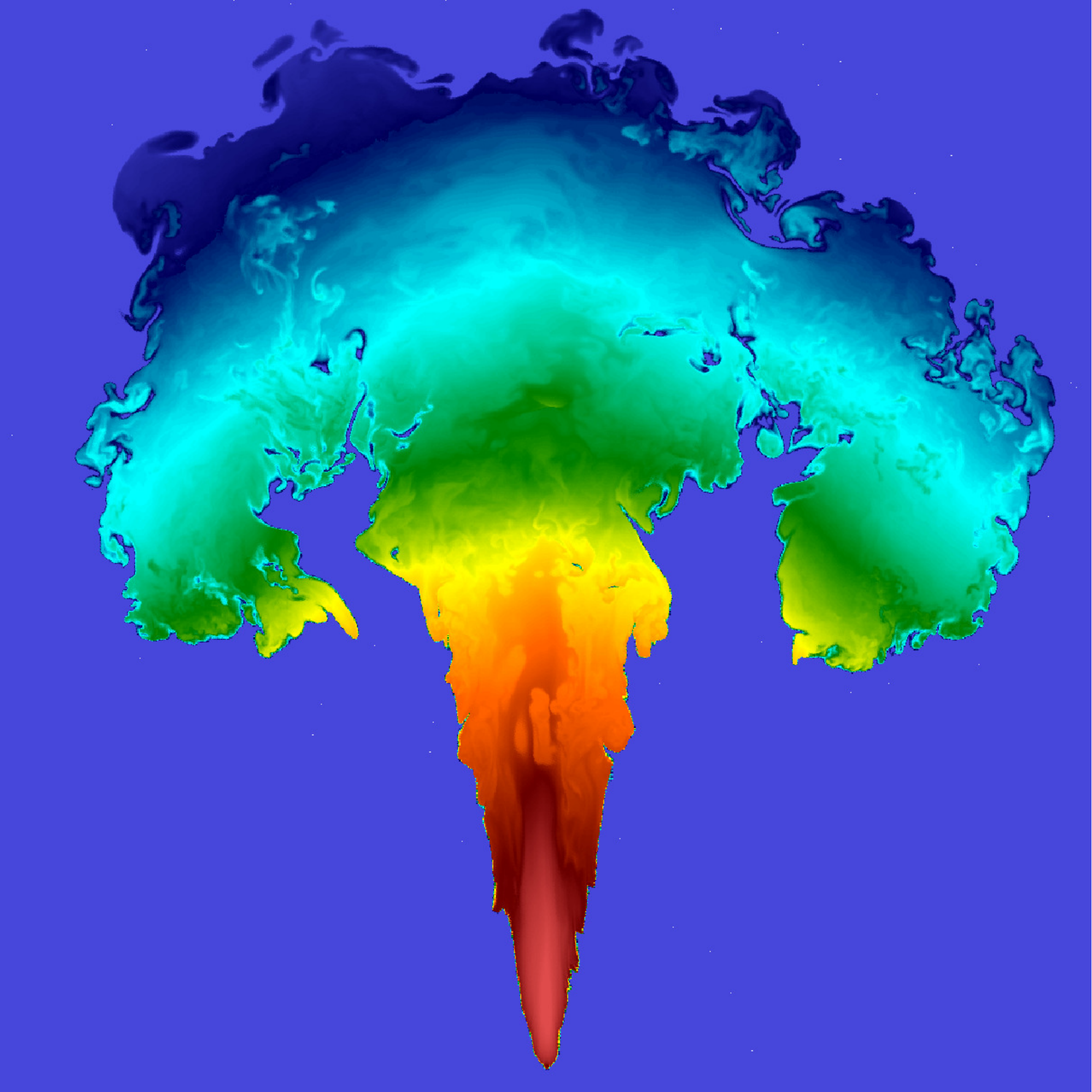}
    \includegraphics[width=0.49\textwidth]{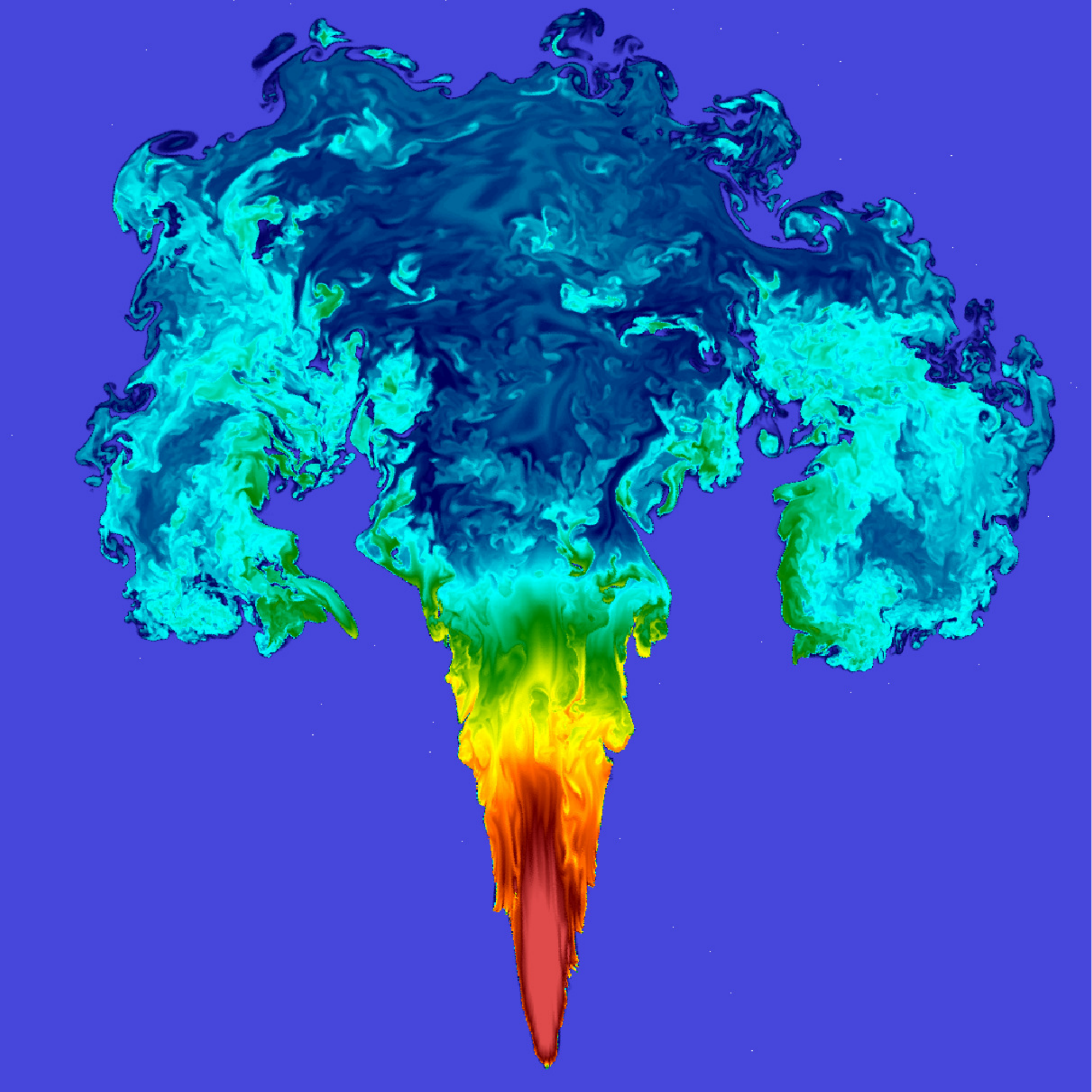}
    \caption{\label{fig:t9_he} $T_9=T / 10^9$ K (left column) and
      $X(^4{\rm He})$ (right column) at $t=0.65$ s (top row) and
      $t=0.8$ s (bottom row) for the standard run AV.
      The color maps increase from blue to red; for $T_9$ the range is
      [2,9] and for $X(^4{\rm He})$ the range is [0,0.25].  The frames
      in the top row are 1200 km on a side and those in the bottom row
      are 2000 km on each side.  Note the high degree of correlation
      between the temperature and helium abundance. In high density
      regions --- either at early times, or in the tail of the flame
      --- large amounts of helium (right column) are produced via
      photodisintegration reactions because of the high temperature
      there.  At lower density and temperature, recombination occurs
      leading to a lower helium mass fraction and more tightly bound
      nuclei (Figure \ref{fig:bea_ye}).  As the flame progresses to
      lower densities (bottom row), the temperature in the ash
      decreases and helium recombines giving extra energy to the ash.}
  \end{center}
\end{figure}

\clearpage

\begin{figure}
  \begin{center}
    \includegraphics[width=2.1in]{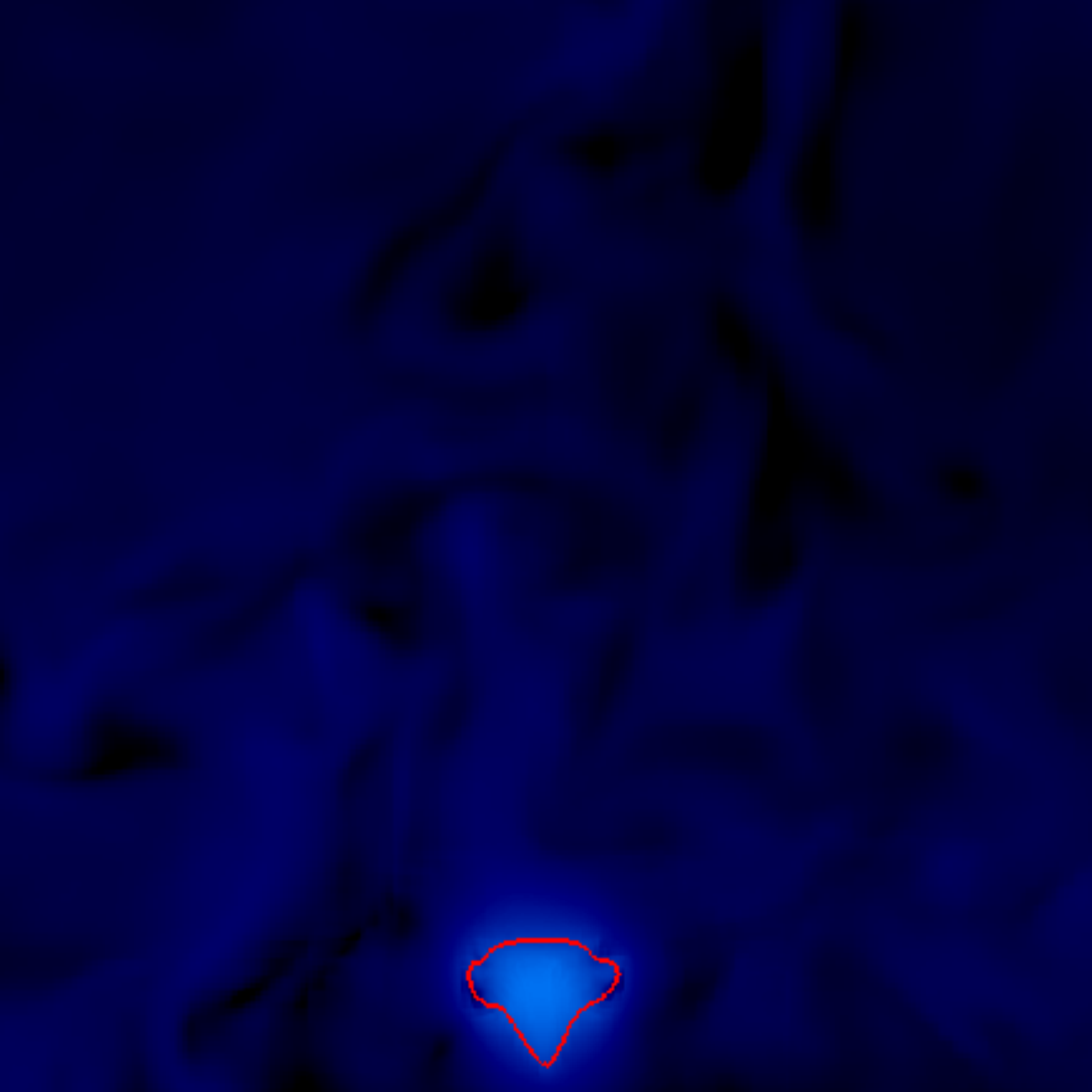}
    \includegraphics[width=2.1in]{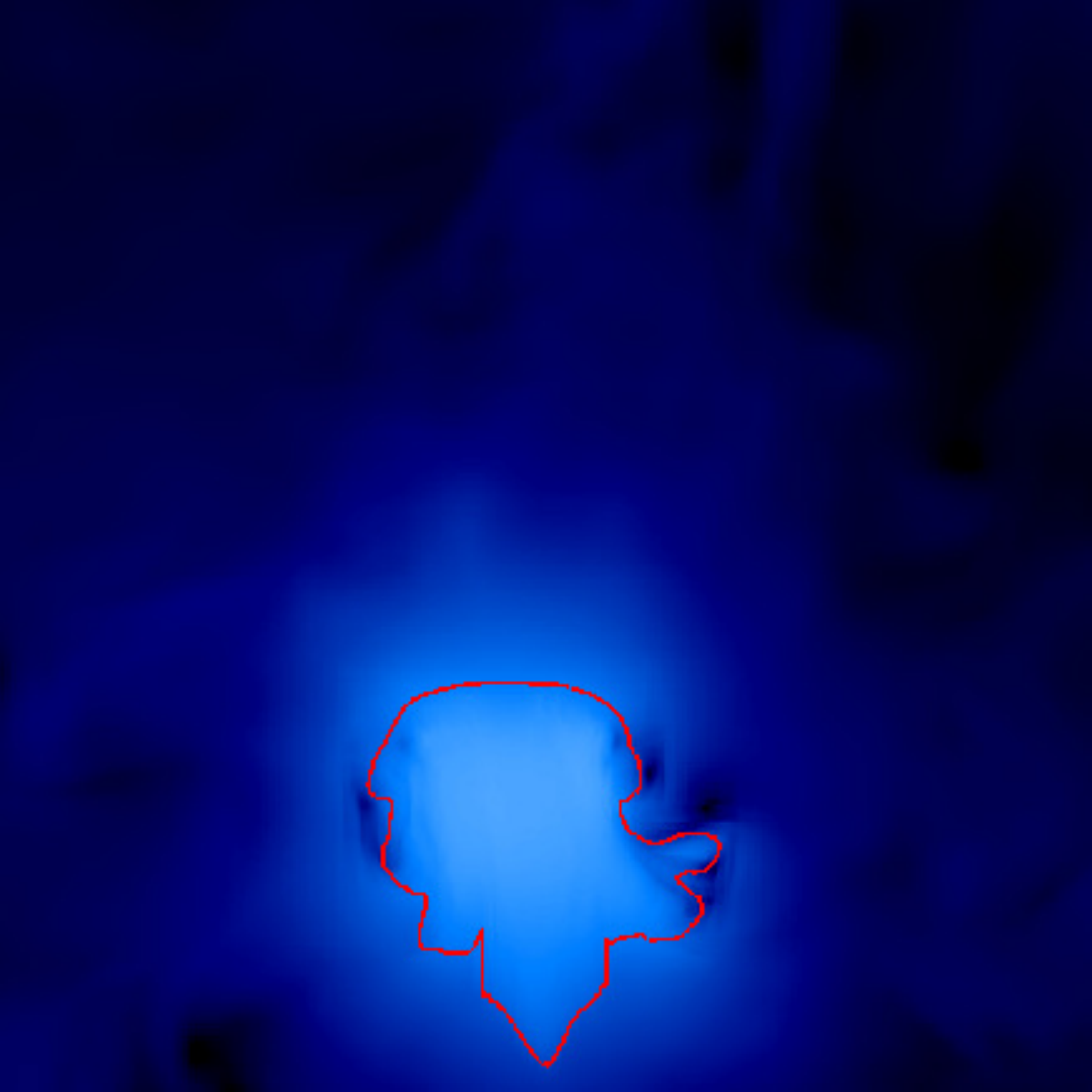}
    \includegraphics[width=2.1in]{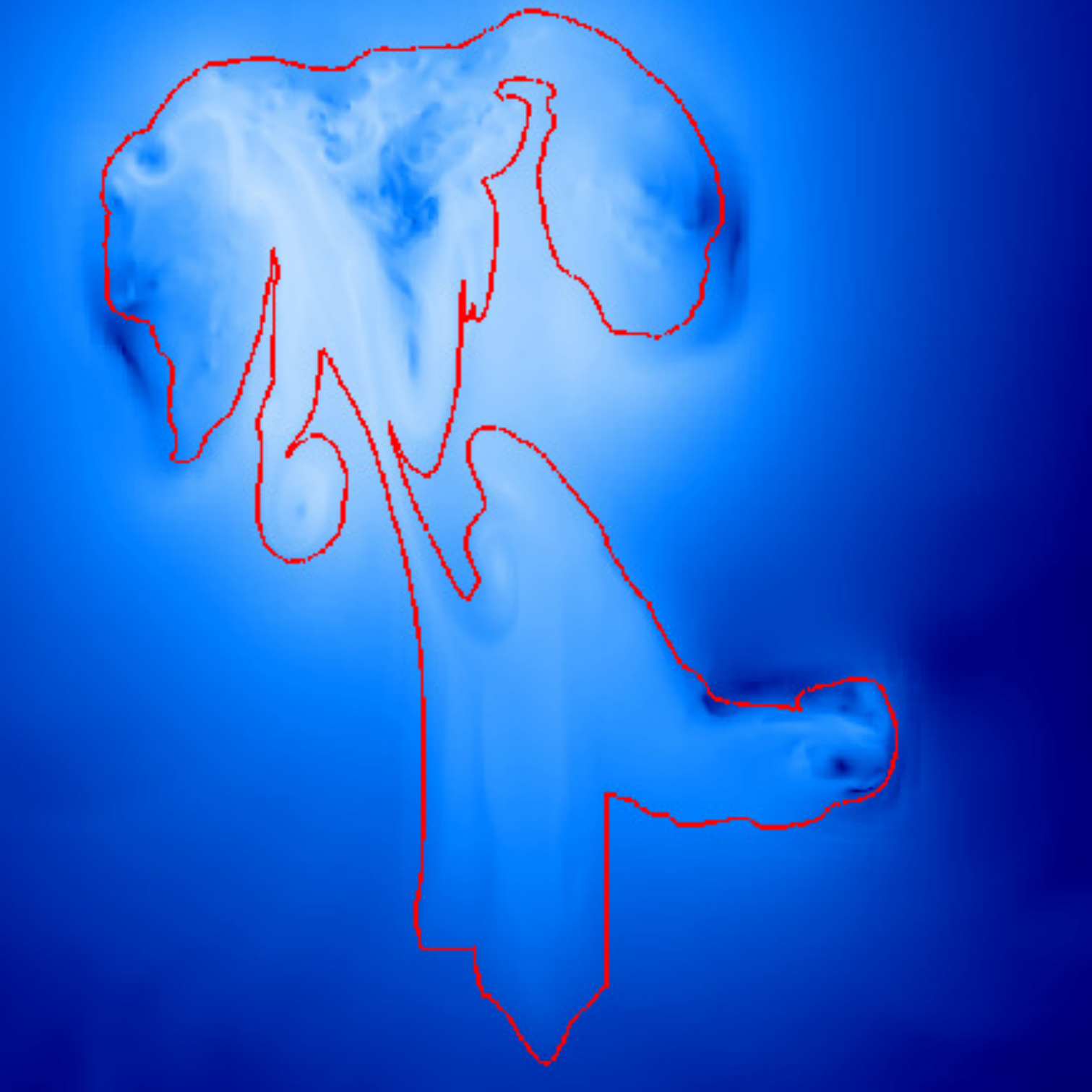}
    \includegraphics[width=2.1in]{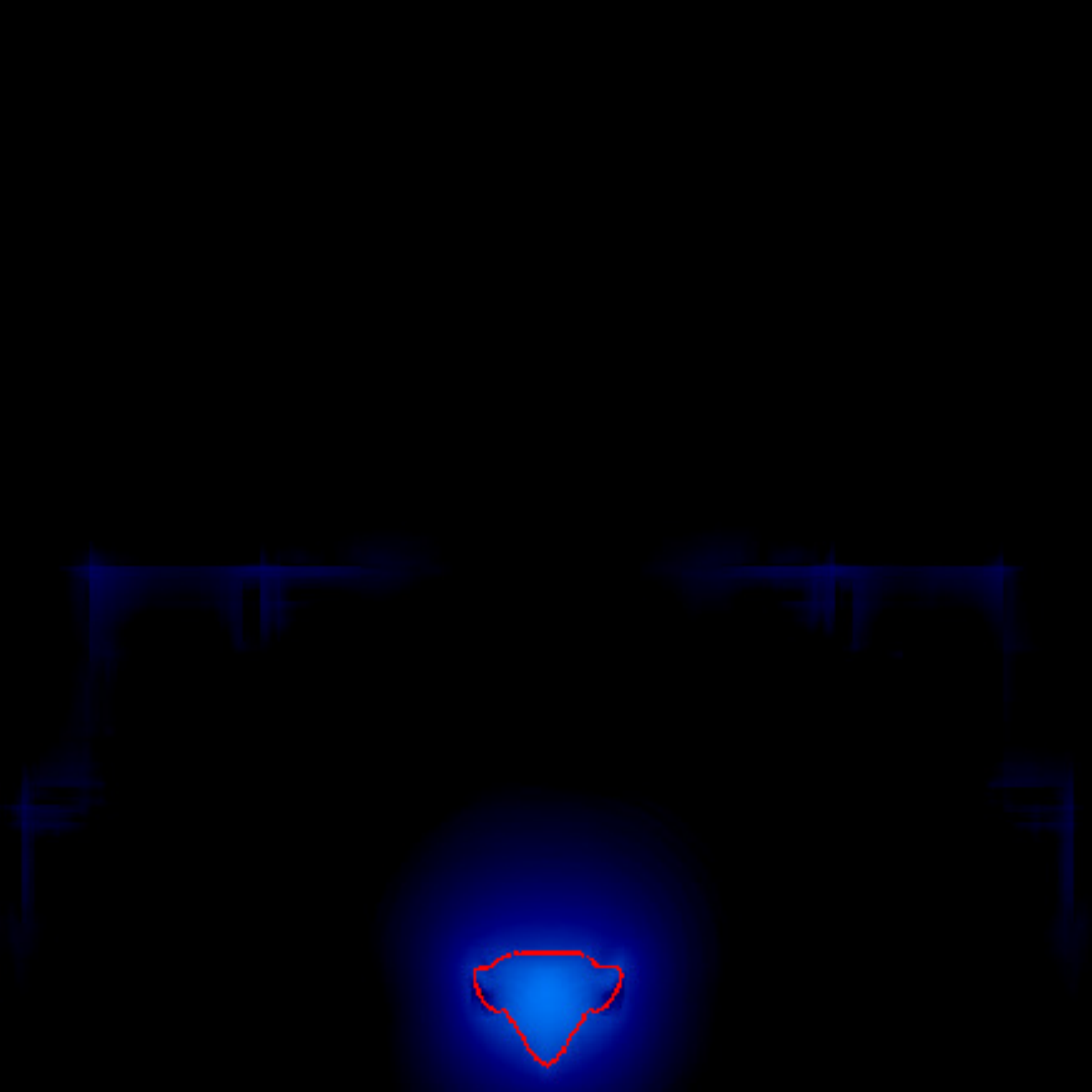}
    \includegraphics[width=2.1in]{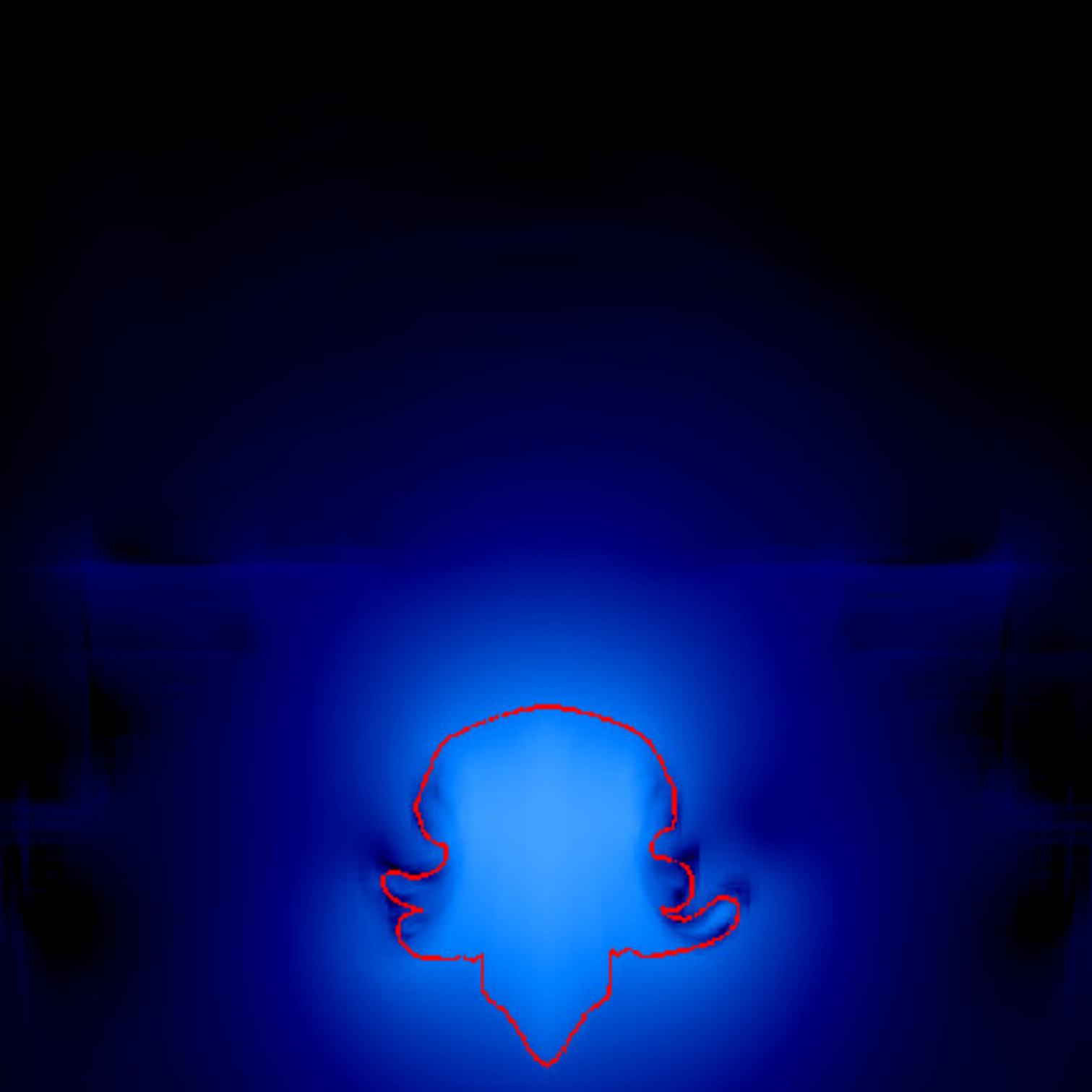}
    \includegraphics[width=2.1in]{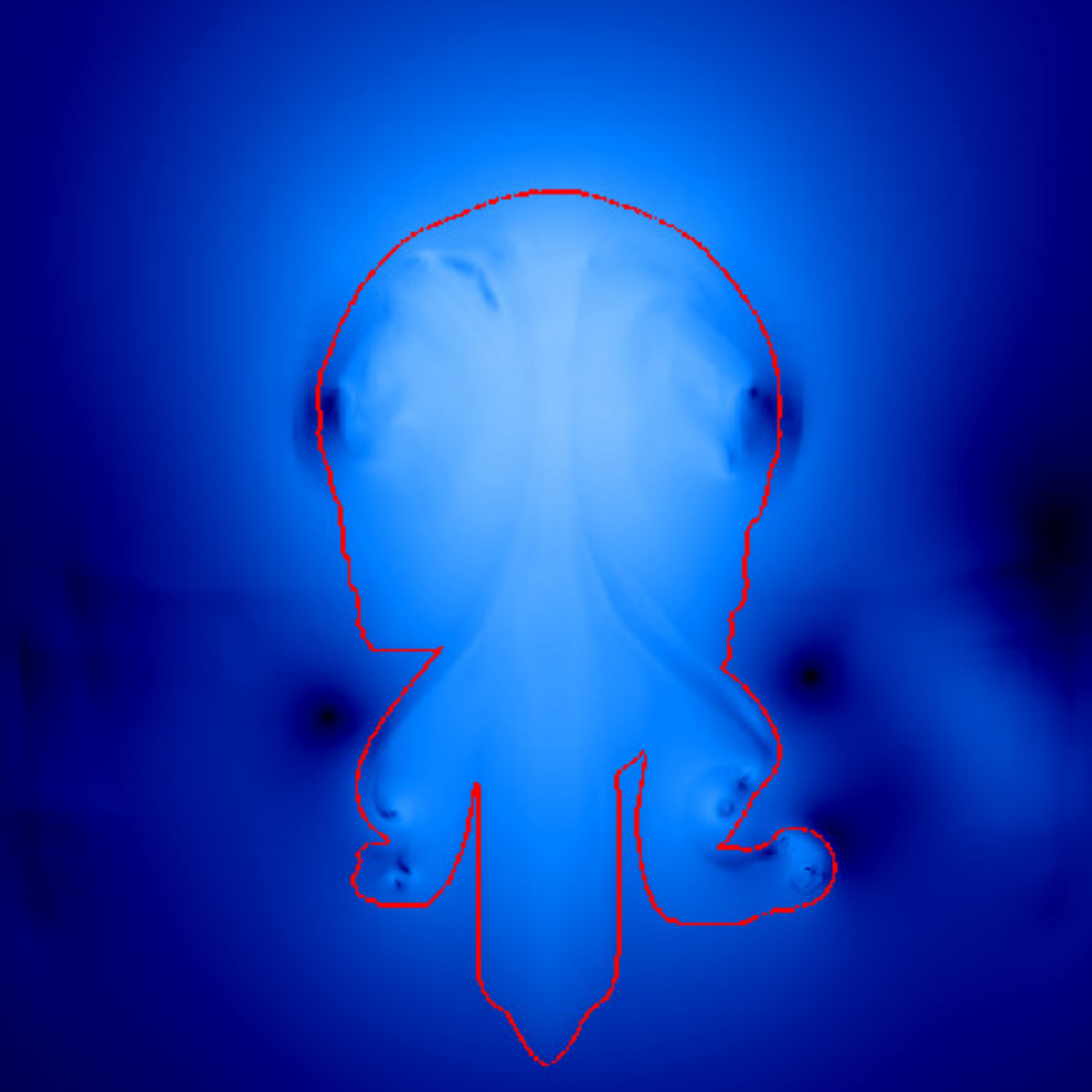}
    \caption{\label{fig:10-slice}Slices in the YZ plane for Models BV
      (top row) and B0 (bottom row) at $t=$0.1, 0.2, and 0.3 s, from
      left to right.  The red line is the $X({\rm C}^{12}) = 0.45$
      isocontour marking the flame surface.  The color plot is the
      fluid speed, which spans, logarithmically, from less than 10 km
      s$^{-1}$ (black) to 10$^4$ km s$^{-1}$ (white).  The center of
      the bottom edge of each panel is at the center of the white
      dwarf, and all panels have a spatial scale of 280 km on a side.
      For the BV models (top row), there is a significant non-radial
      component to the velocity field, as evident by distortion of the
      flame surface to the upper left of the frames.}
  \end{center}
\end{figure}

\clearpage

\begin{figure}
  \begin{center}
    \includegraphics[width=2.1in]{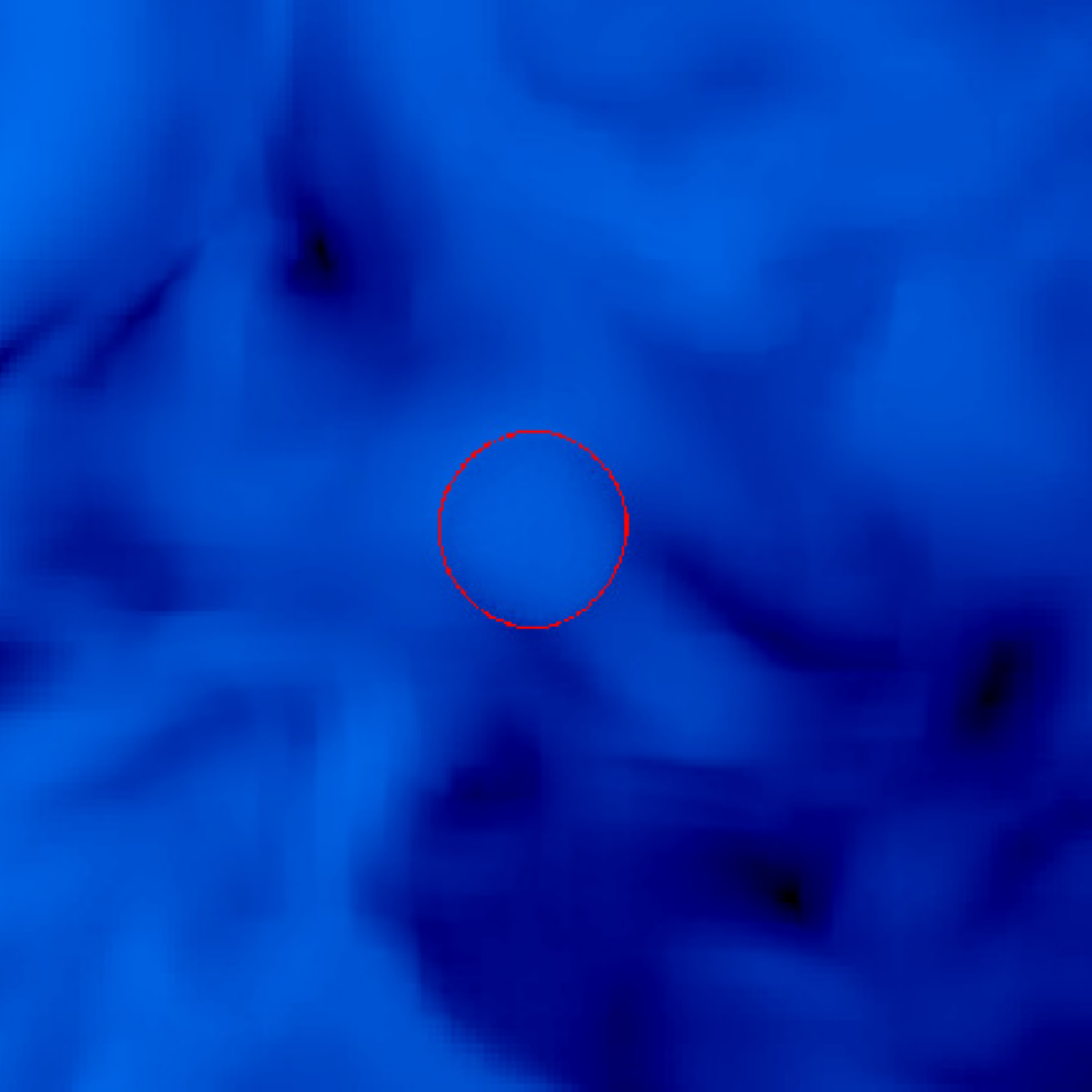}
    \includegraphics[width=2.1in]{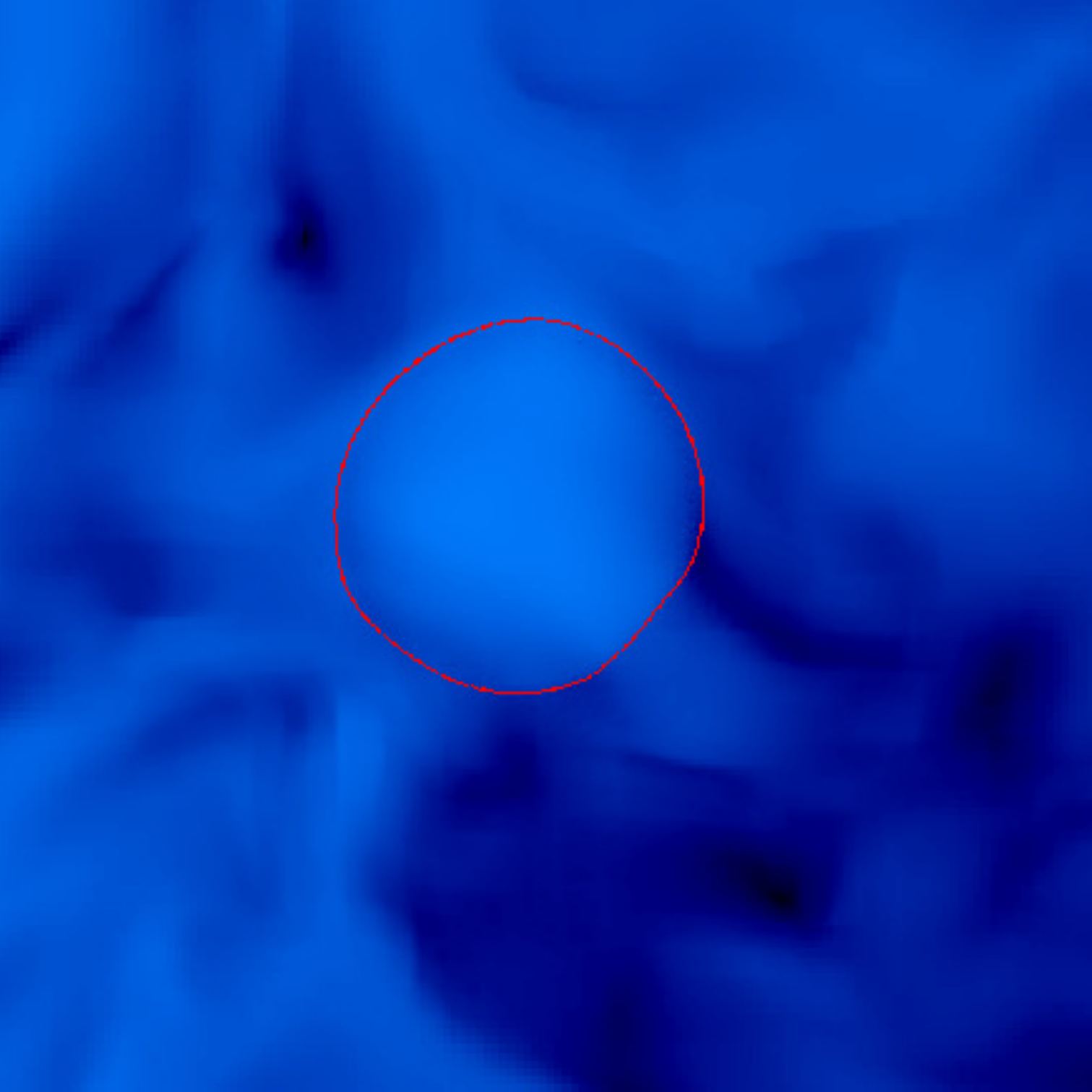}
    \includegraphics[width=2.1in]{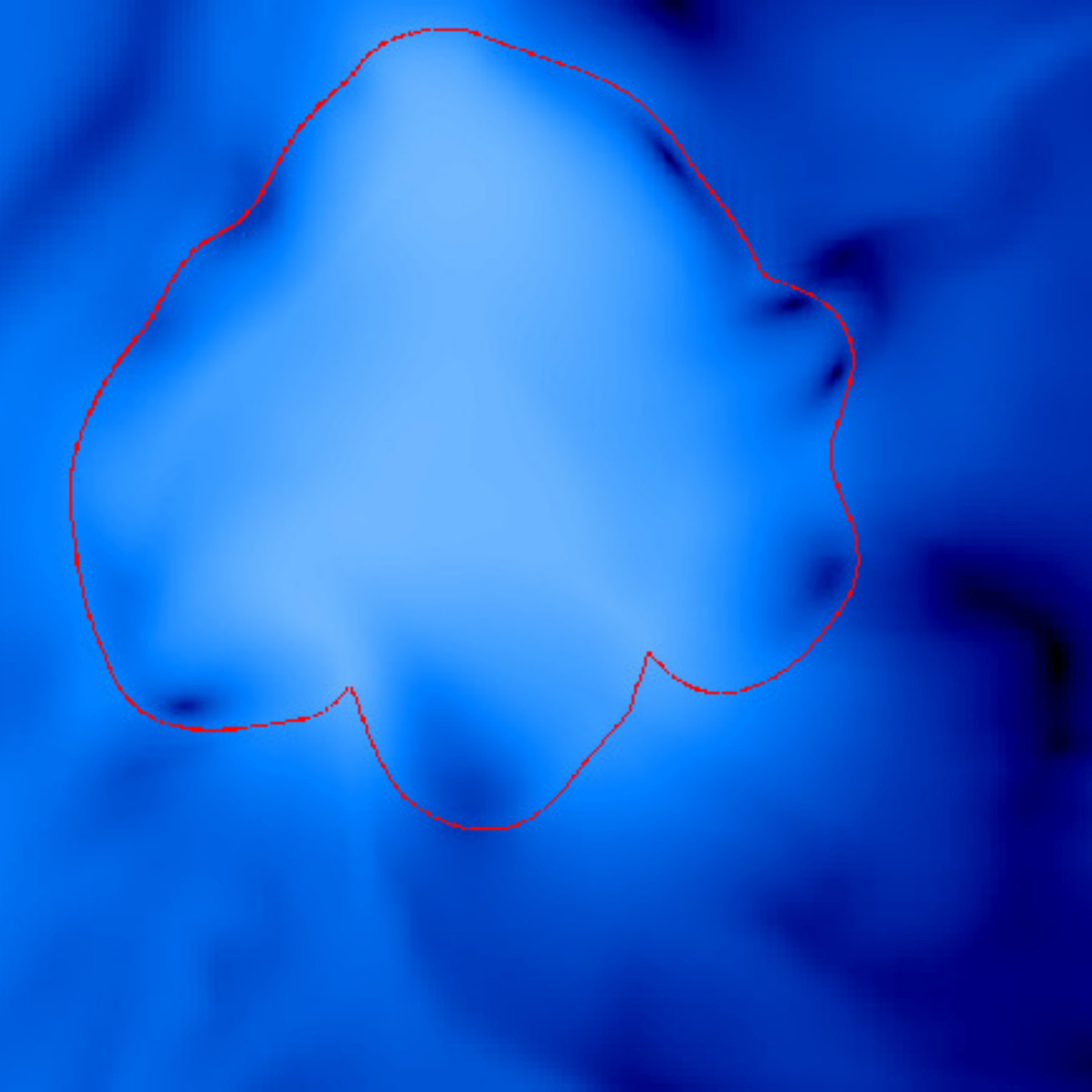}
    \includegraphics[width=2.1in]{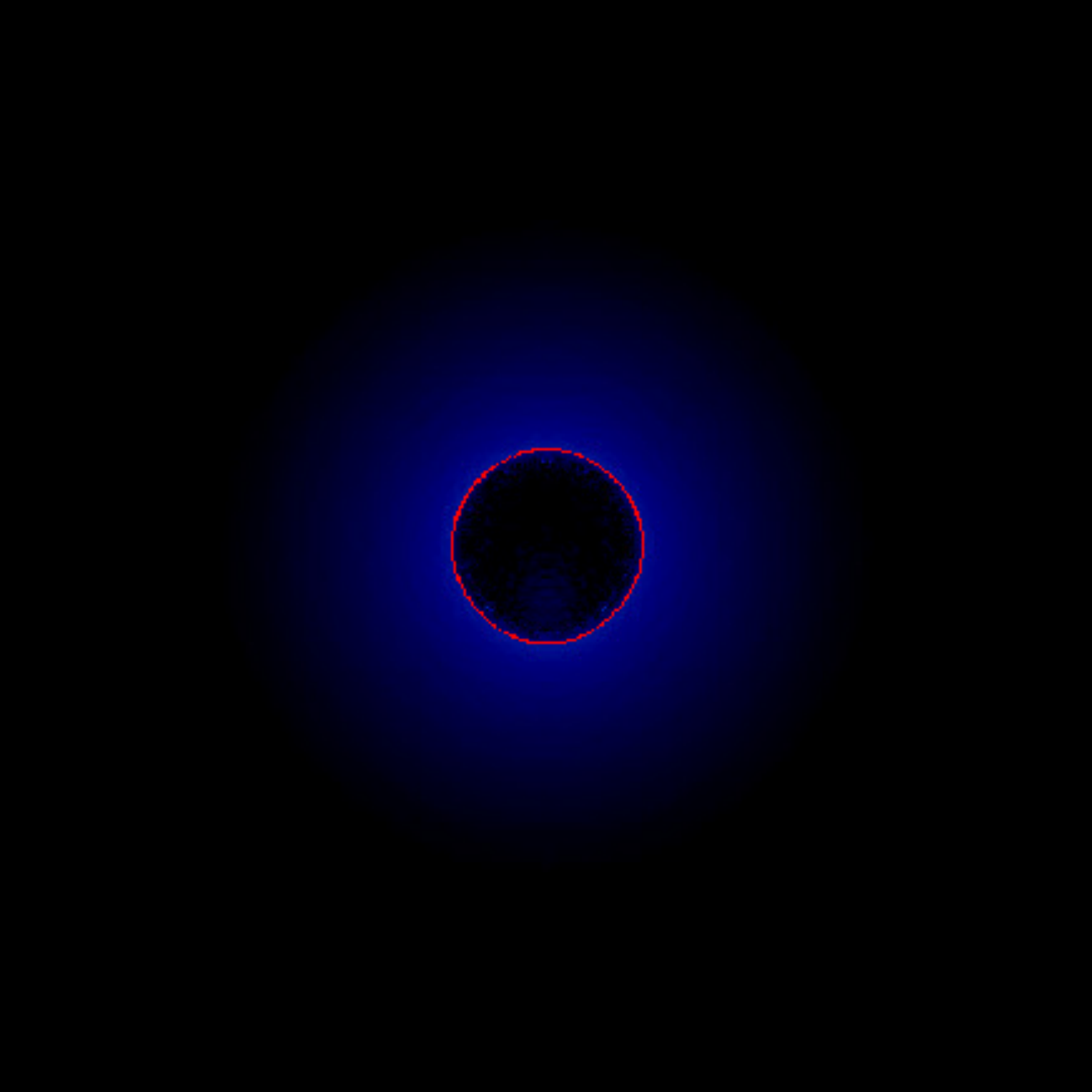}
    \includegraphics[width=2.1in]{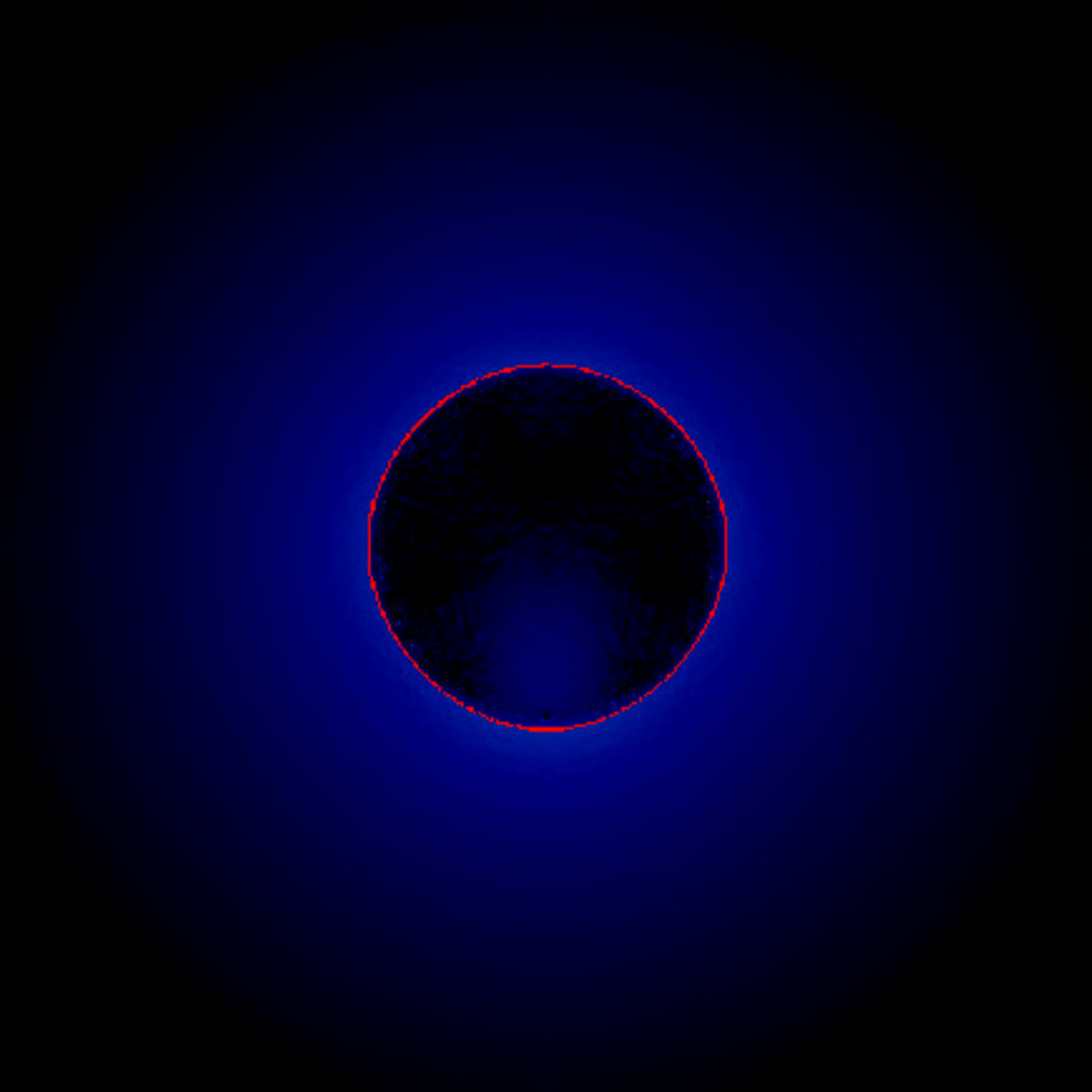}
    \includegraphics[width=2.1in]{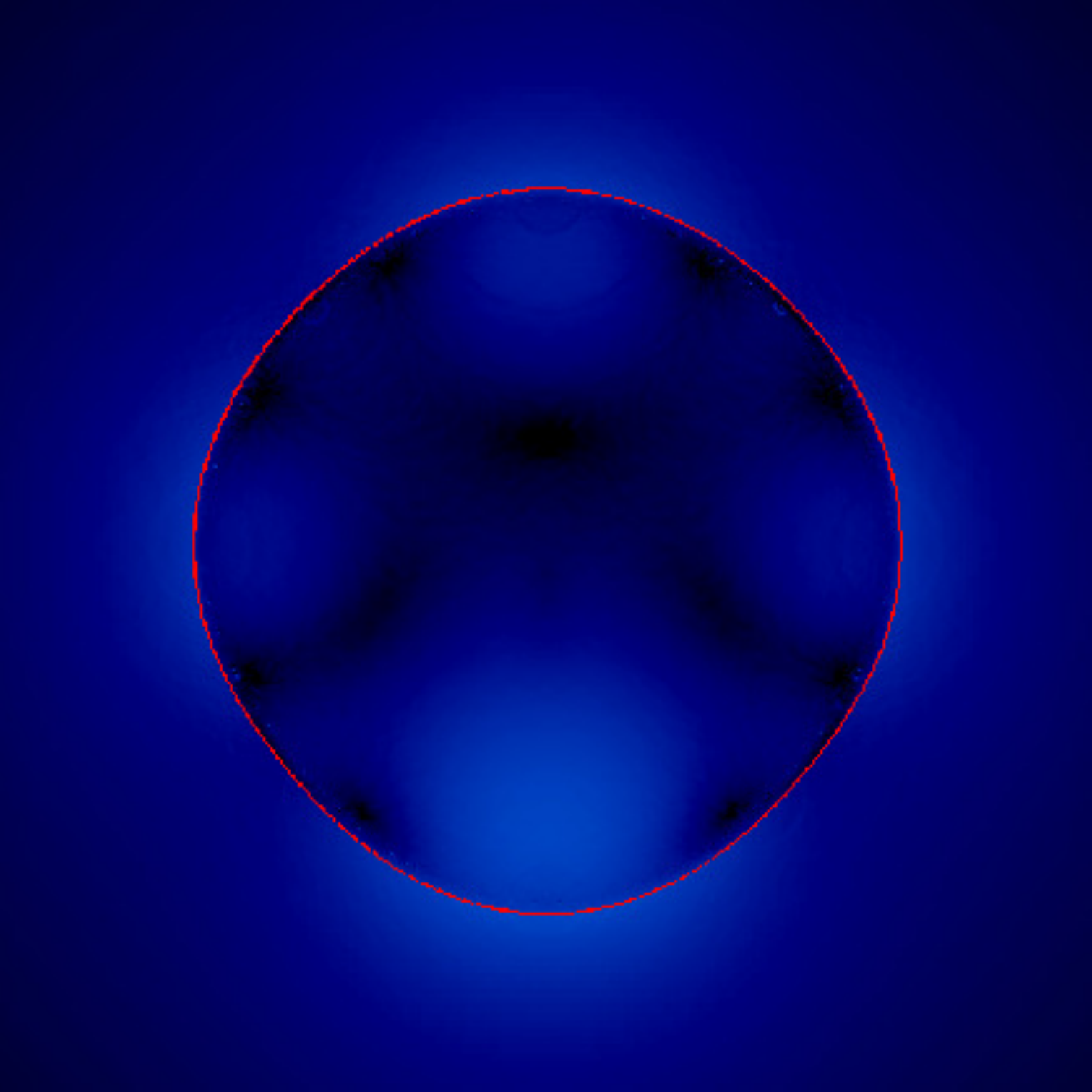}
    \caption{\label{fig:cent-slice}Slices in the YZ plane for Models
      CV (top row) and C0 (bottom row) at $t=$0.1, 0.2, and 0.4 s,
      from left to right.  The red line is the $X({\rm C}^{12}) =
      0.45$ isocontour marking the flame surface.  The color plot is
      the fluid speed, which spans, logarithmically, from less than 1
      km s$^{-1}$ (black) to 10$^3$ km s$^{-1}$ (white).  Note the
      lower maximum speed compared to Figure \ref{fig:10-slice}.  The
      center of each panel is at the center of the white dwarf, and
      all panels have a spatial scale of 150 km on a side.}
  \end{center}
\end{figure}

\clearpage

\begin{figure}
  \begin{center}
    \includegraphics[width=0.45\textwidth]{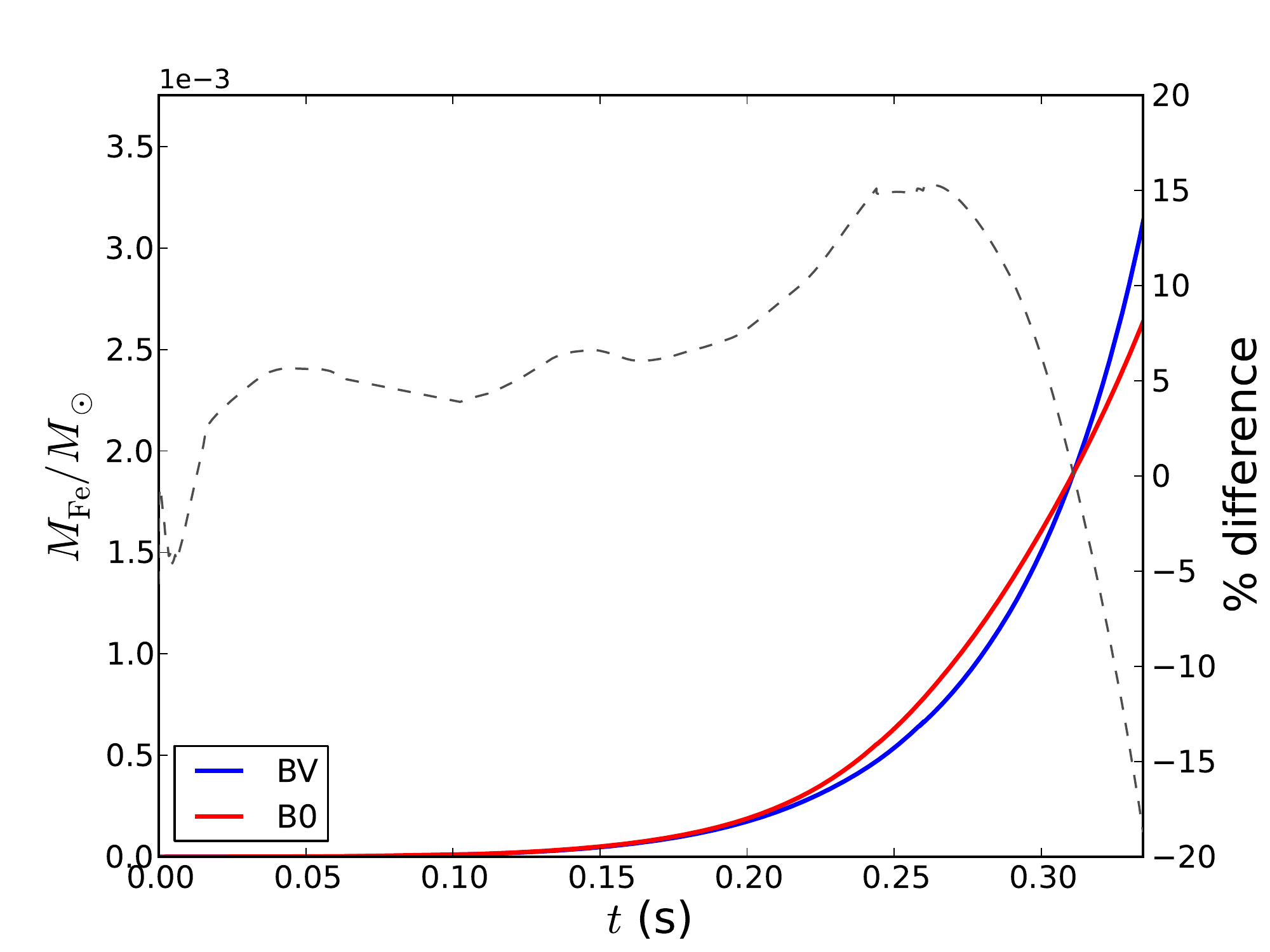}
    \includegraphics[width=0.45\textwidth]{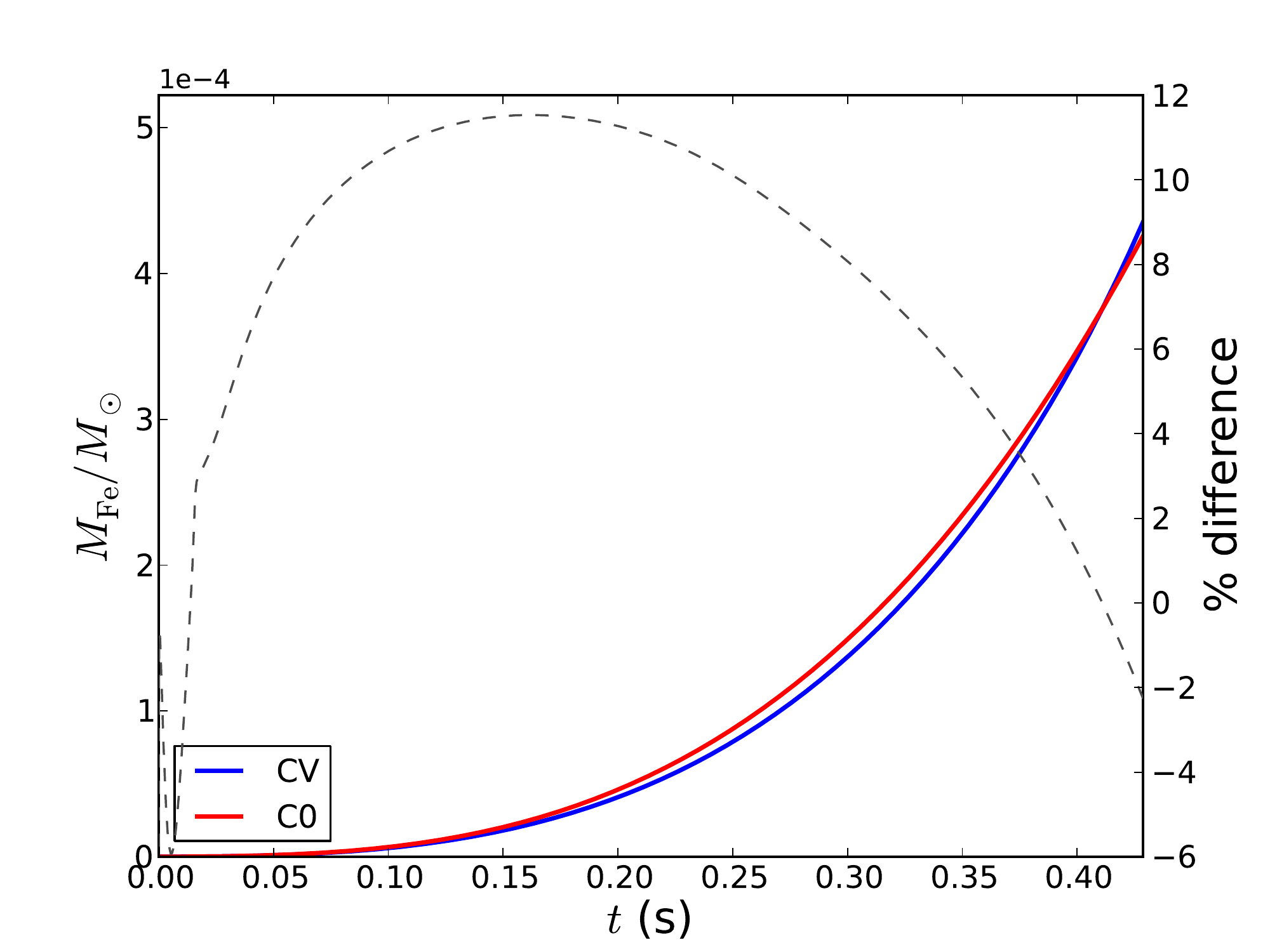}
    \includegraphics[width=0.45\textwidth]{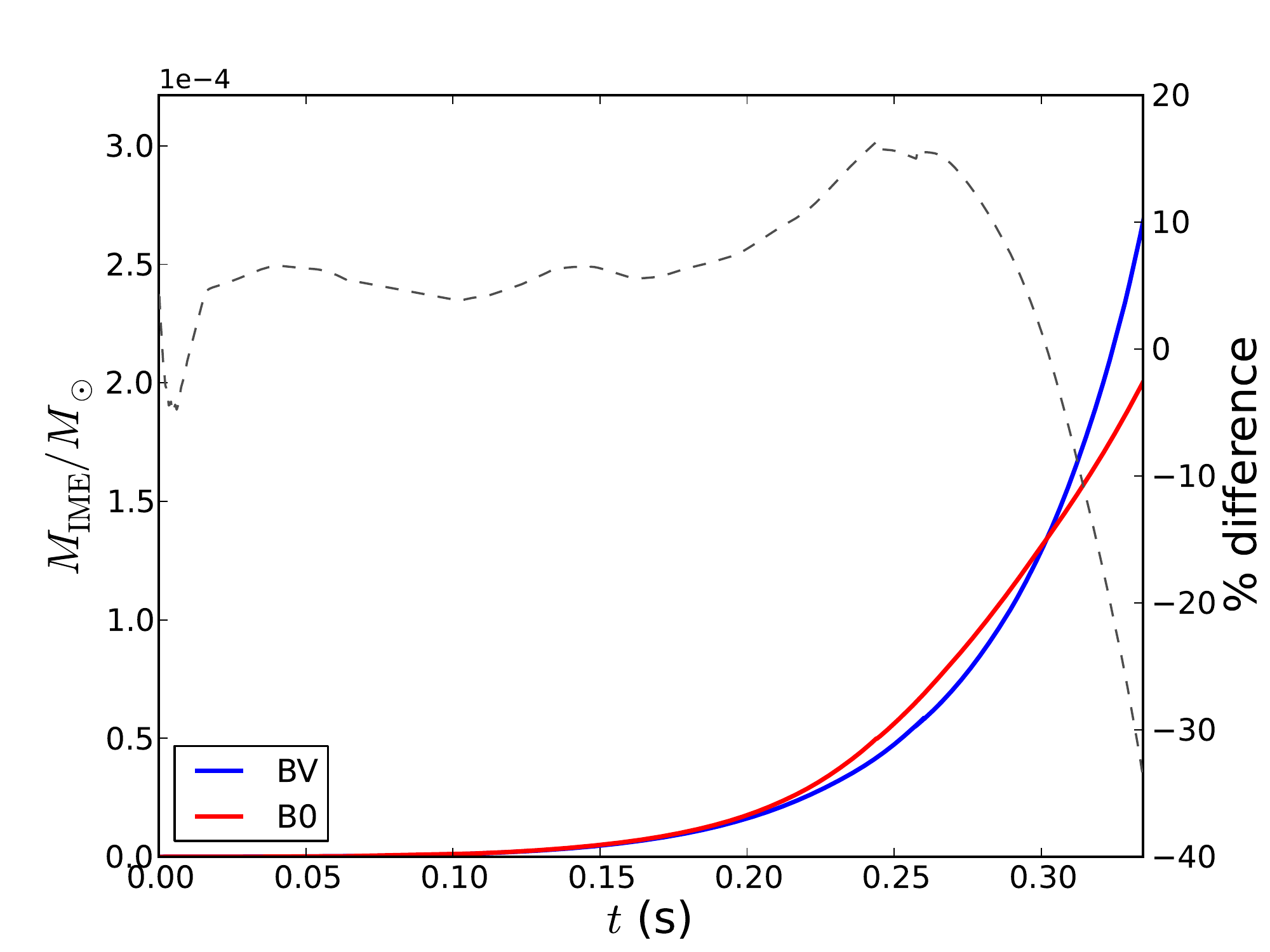}
    \includegraphics[width=0.45\textwidth]{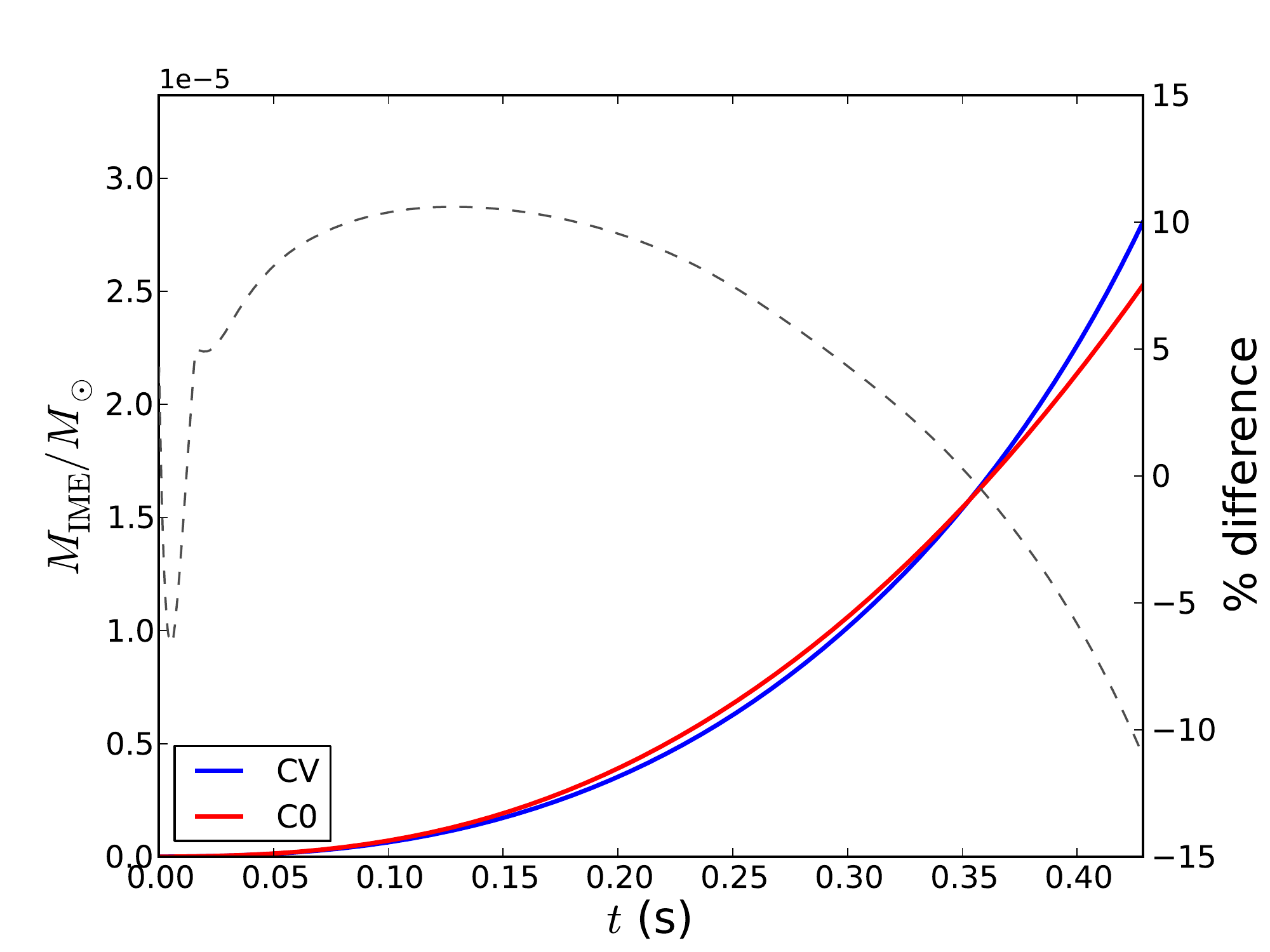}
    \caption{\label{fig:10-cent-massburned}Comparisons of
      nucleosynthetic yields of iron-group elements (top row) and IME
      (bottom row) for both the B series (left column) and C series
      (right column).  The blue(red) lines correspond to models
      with(without) a background convective flow field.  The dashed
      grey line shows the percent difference between models with and
      without the background field; positive values indicate more
      production in the models without the background field.  These
      plots indicate that the more centrally-ignited models with the
      background turbulence burn more material than those without,
      which is in contrast to the further off-center ignition models
      shown in Figure \ref{fig:mass-burned}.}
  \end{center}
\end{figure}

\clearpage

\begin{figure}
  \begin{center}
    \includegraphics[width=0.49\textwidth]{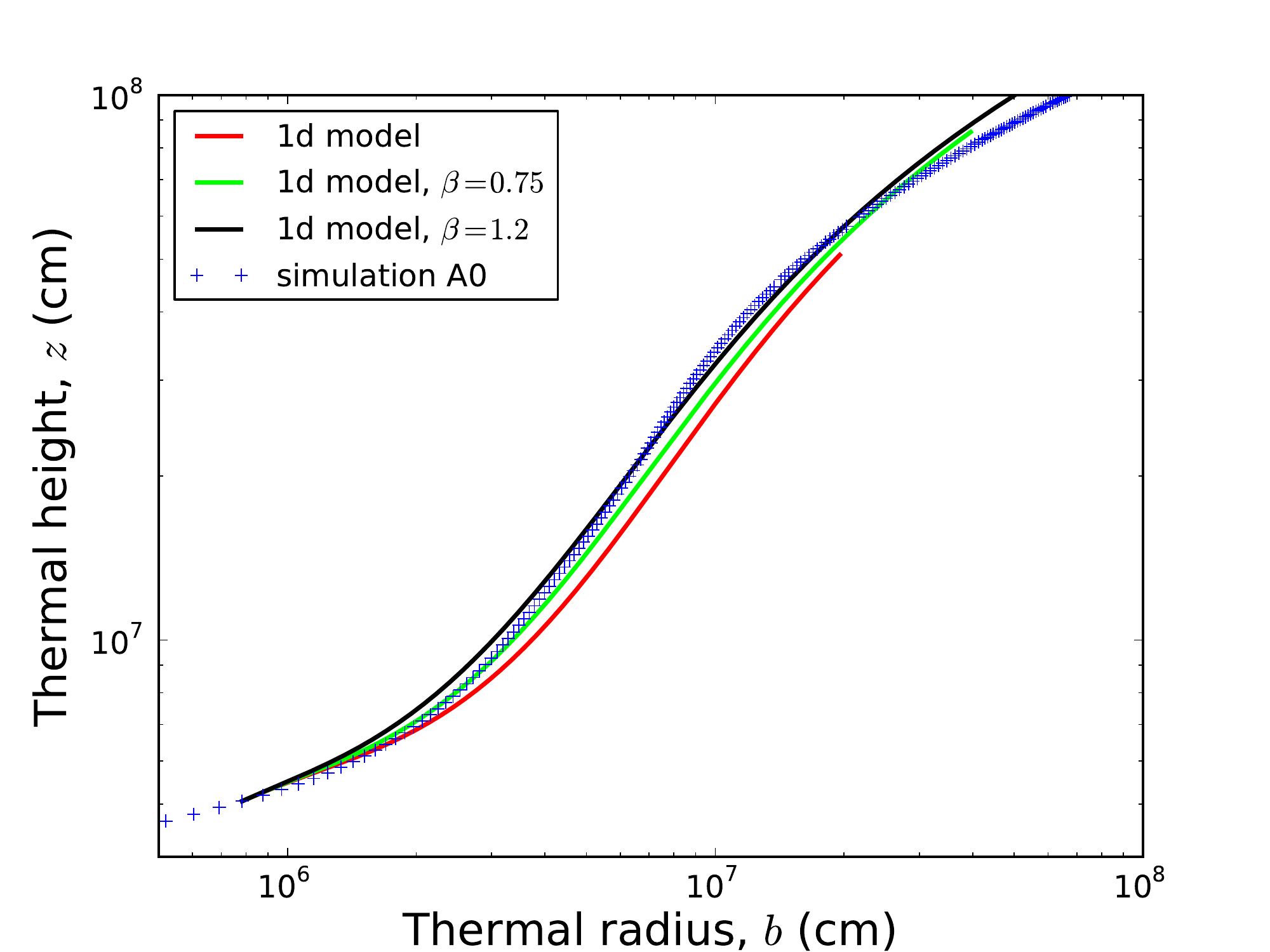}
    \includegraphics[width=0.49\textwidth]{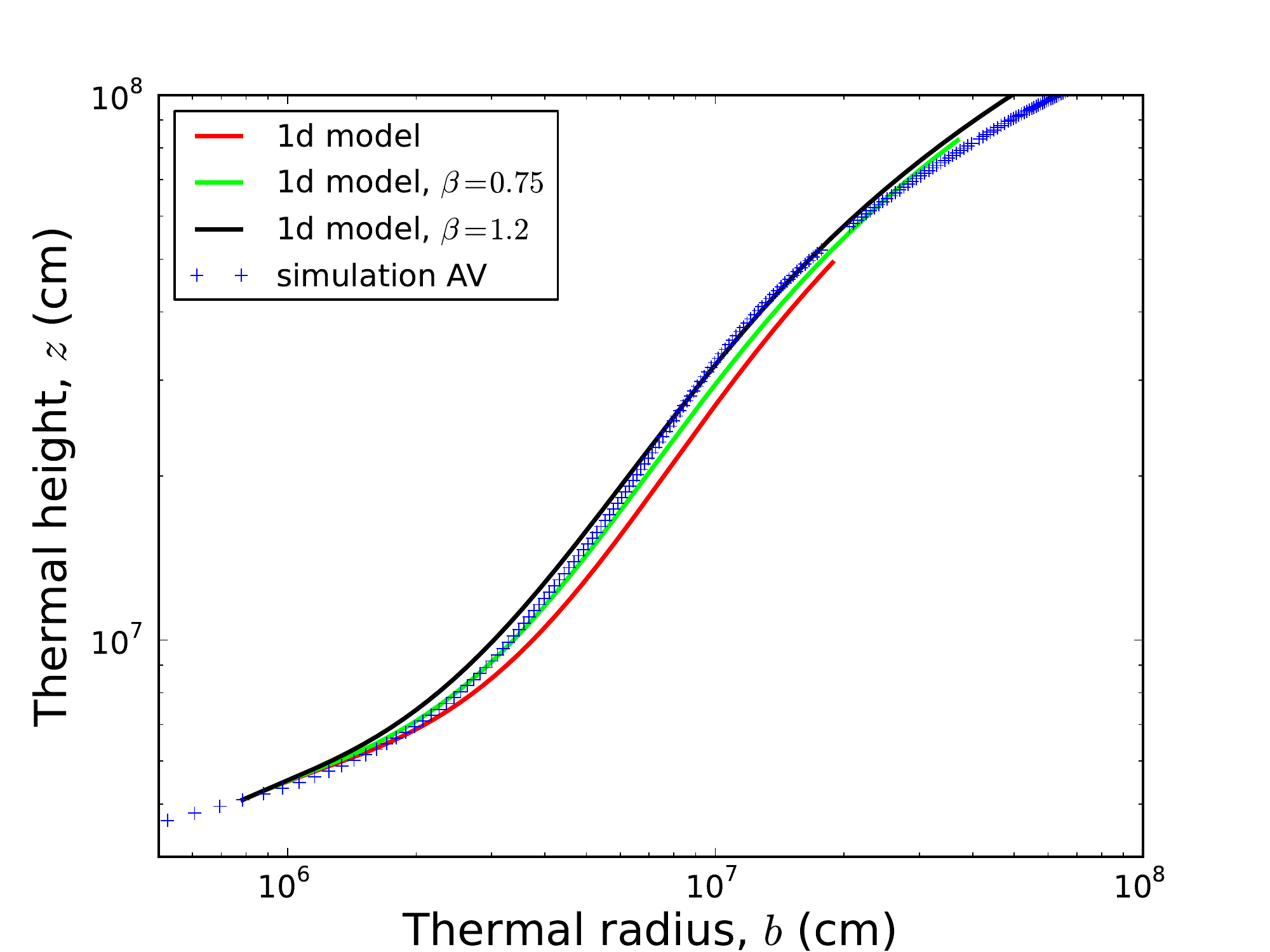}
    \includegraphics[width=0.49\textwidth]{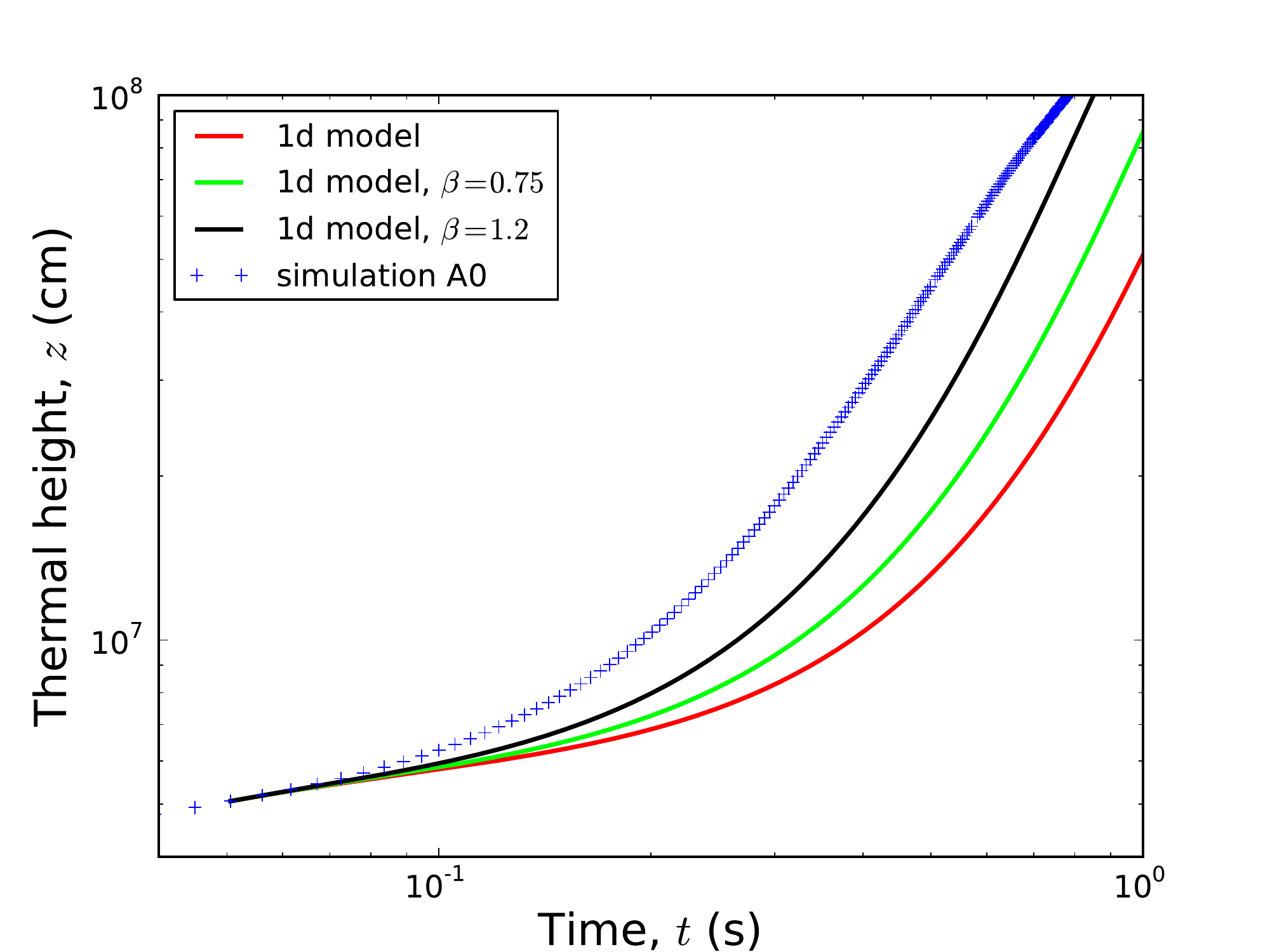}
    \includegraphics[width=0.49\textwidth]{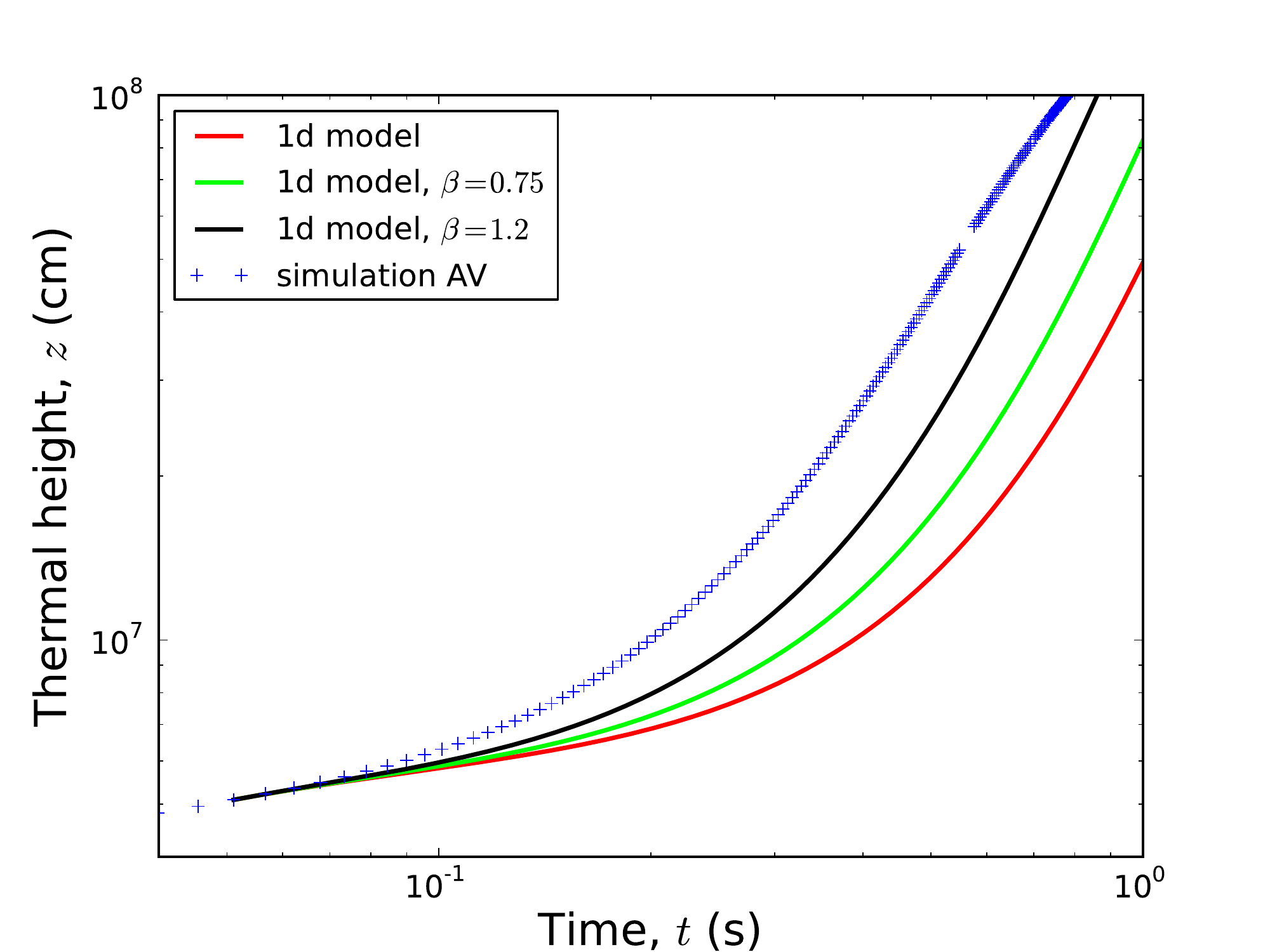}
    \includegraphics[width=0.49\textwidth]{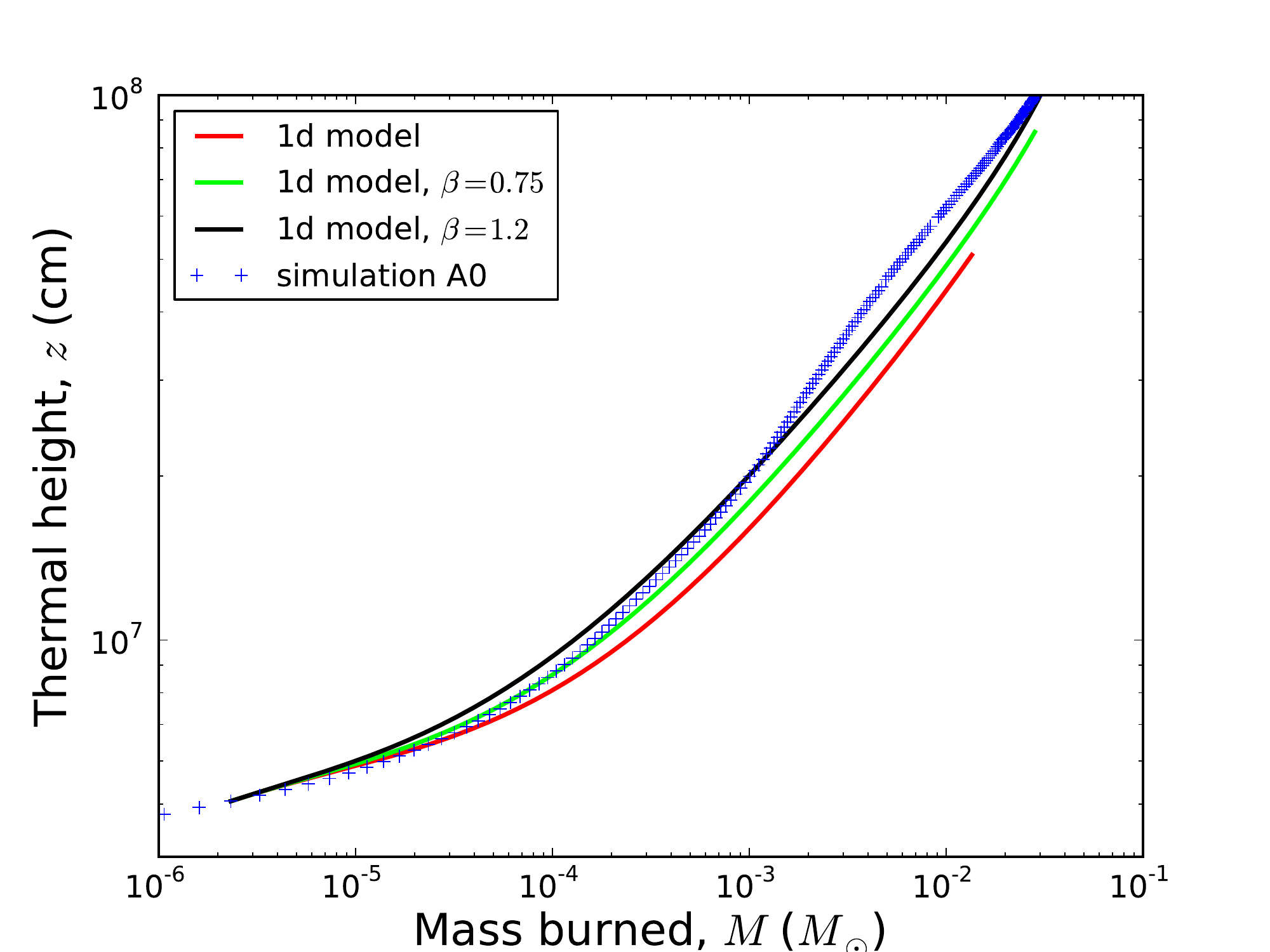}
    \includegraphics[width=0.49\textwidth]{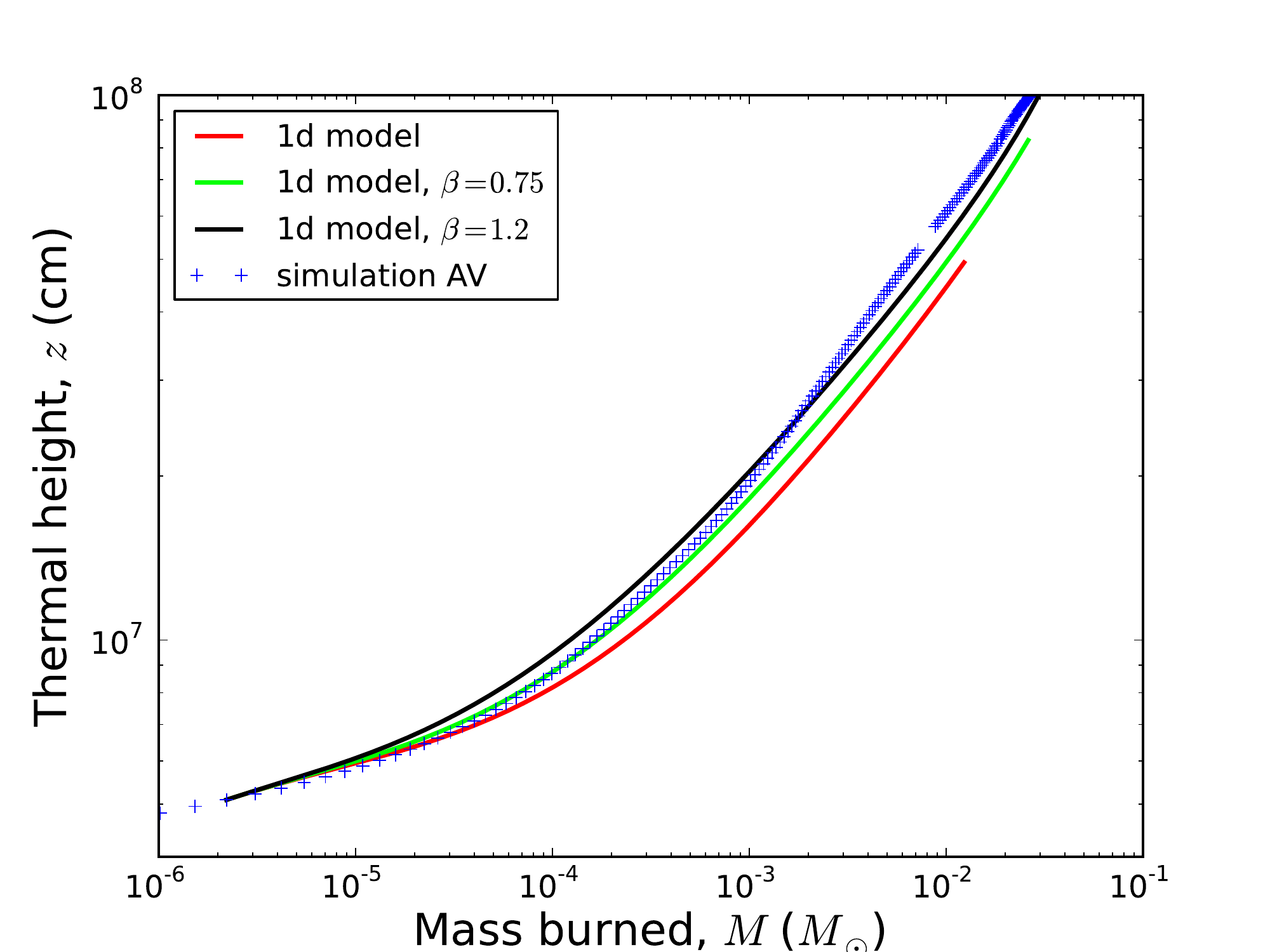}
    \caption{\label{fig:anal-model} Comparisons of simulation data
      (crosses) with the semi-analytic one-dimensional model with
      various parametrizations (lines) for both Model A0 (left
      column) and Model AV (right column).  In all cases, the
      semi-analytic model reproduces the general trend of the
      simulation data, but lacks sufficient buoyancy during the
      evolution.}
  \end{center}
\end{figure}

\begin{table}
  \caption{Initial conditions for semi-analytic model comparison to data in Figure \ref{fig:anal-model}.\label{tab:init-cond}}
  \begin{center}
    \begin{tabular}{cccccccc}
      \hline\hline
      &$t_0$&$z_0$&$u_0$&$b_0$&$\rho_0$&$g_0$&$\rho_{f,0}$\\
      Model&(ms)&(km)&(10$^2$ km s$^{-1}$)&(km)&(10$^9$ g cm$^{-3}$)&(10$^{4}$ km s$^{-2}$)&(10$^9$ g cm$^{-3}$)\\
      \hline
      A0 & 50.59 & 50.61 & 2.37 & 7.83 & 2.30 & 3.57 & 2.50\\
      AV & 51.10 & 50.89 & 2.42 & 7.86 & 2.16 & 3.76 & 2.65\\
      \hline
    \end{tabular}
  \end{center}
\end{table}

\clearpage

\begin{figure}
  \begin{center}
    \includegraphics[width=0.48\textwidth]{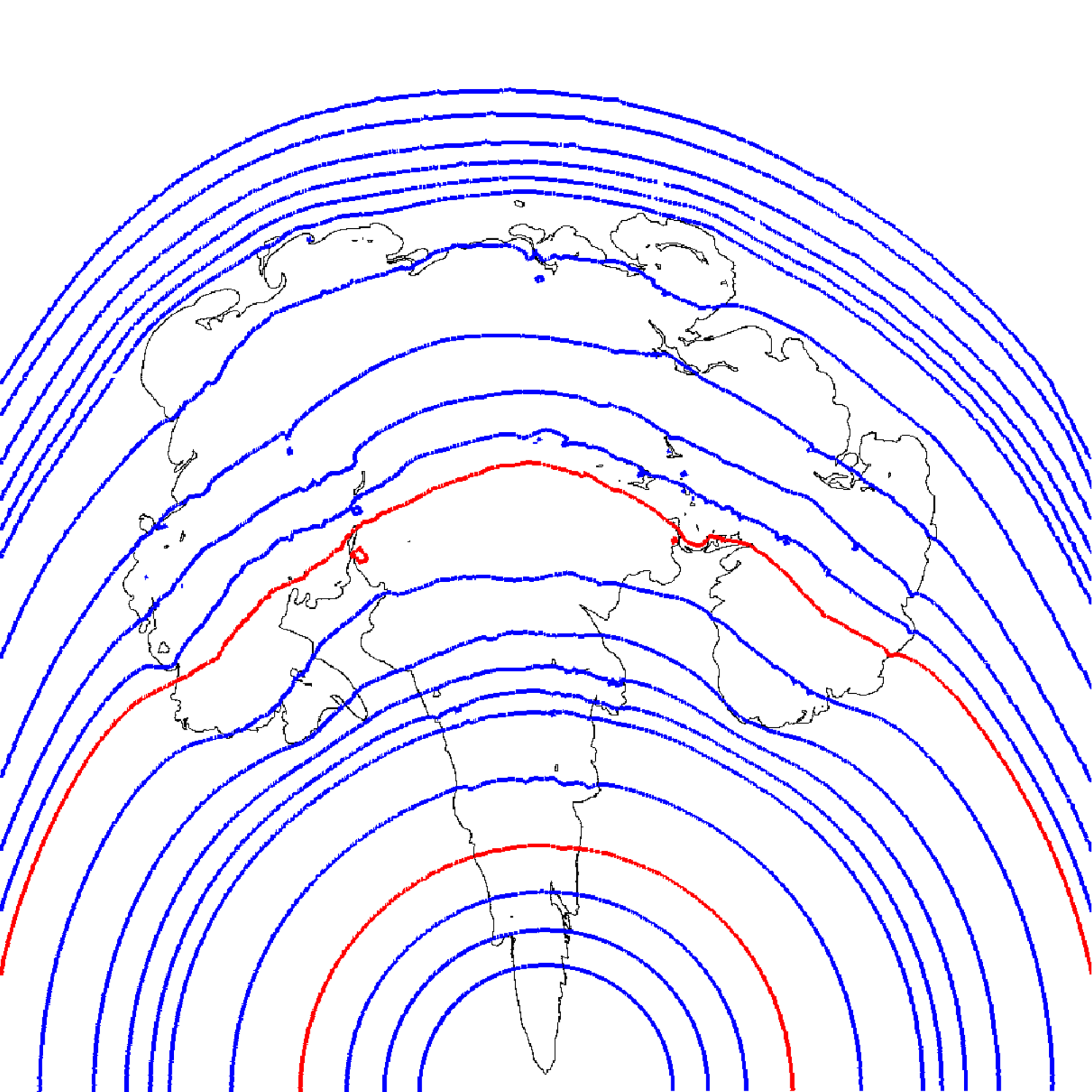}
    \includegraphics[width=0.48\textwidth]{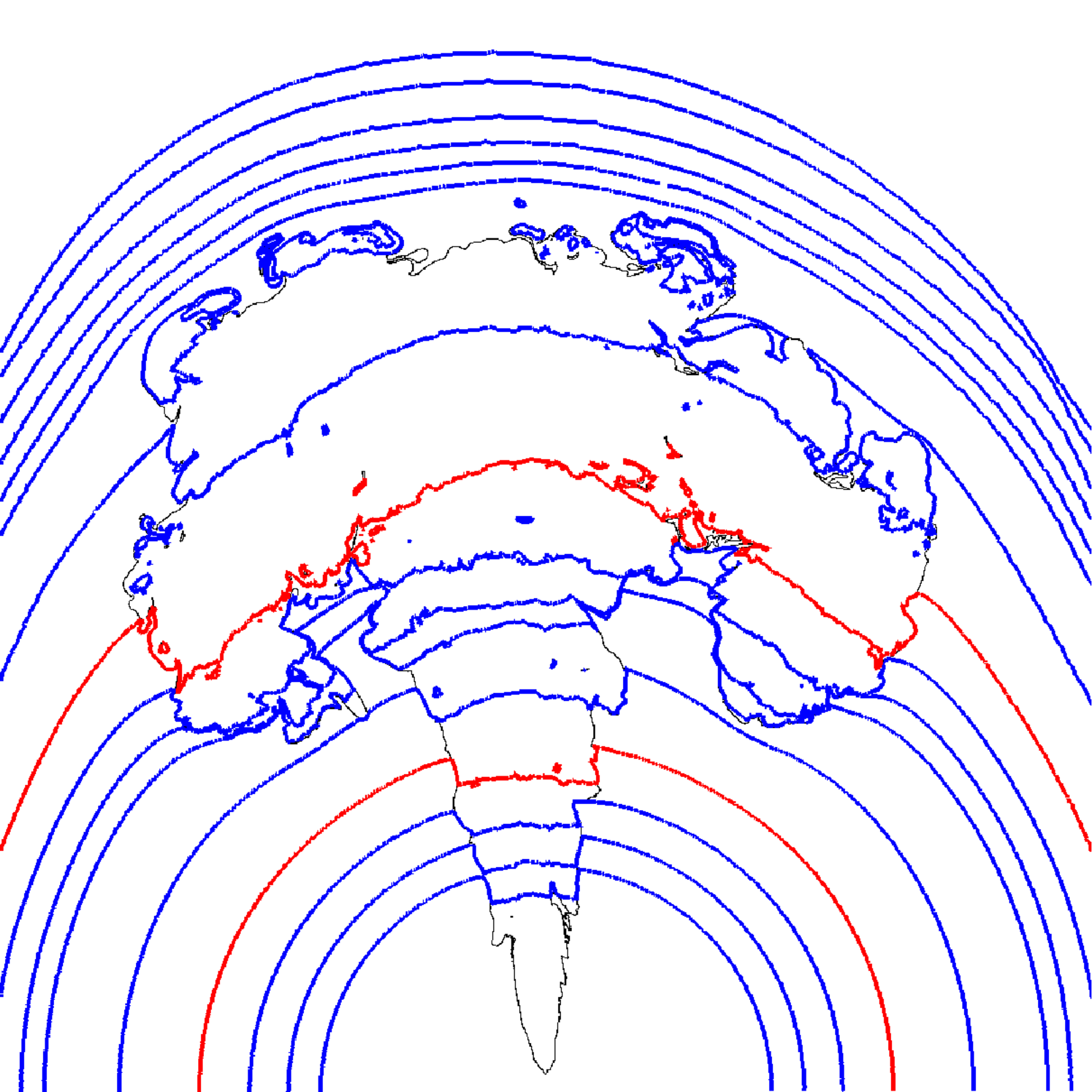}
    \caption{\label{fig:contours}Pressure (left) and density (right)
      contours for Model AV at $t=0.8$s.  The thin black contour marks
      the flame surface.  The red contours are highlighted to aid the
      eye.  Each slice is 2400 km on a side.  Deep in the star, where
      the pressure scale height is large, pressure contours lie ontop
      of gravitational equipotentials, which are lines of constant
      radius.  Equivalently, the lateral sound crossing time of the
      plume is short compared to the bubble rise so that pressure
      balance is maintained across the flame.  At larger radii, the
      pressure scale height and sound speed decrease, while the
      bubble's lateral extent and float speed increase.  These changes
      lead to a situation where sound waves can no longer maintain
      pressure balance across the flame.  Lateral pressure gradients
      form, which accelerate the lateral expansion of the flame.}
  \end{center}
\end{figure}

\clearpage

\begin{figure}
  \begin{center}
    \includegraphics[width=\textwidth]{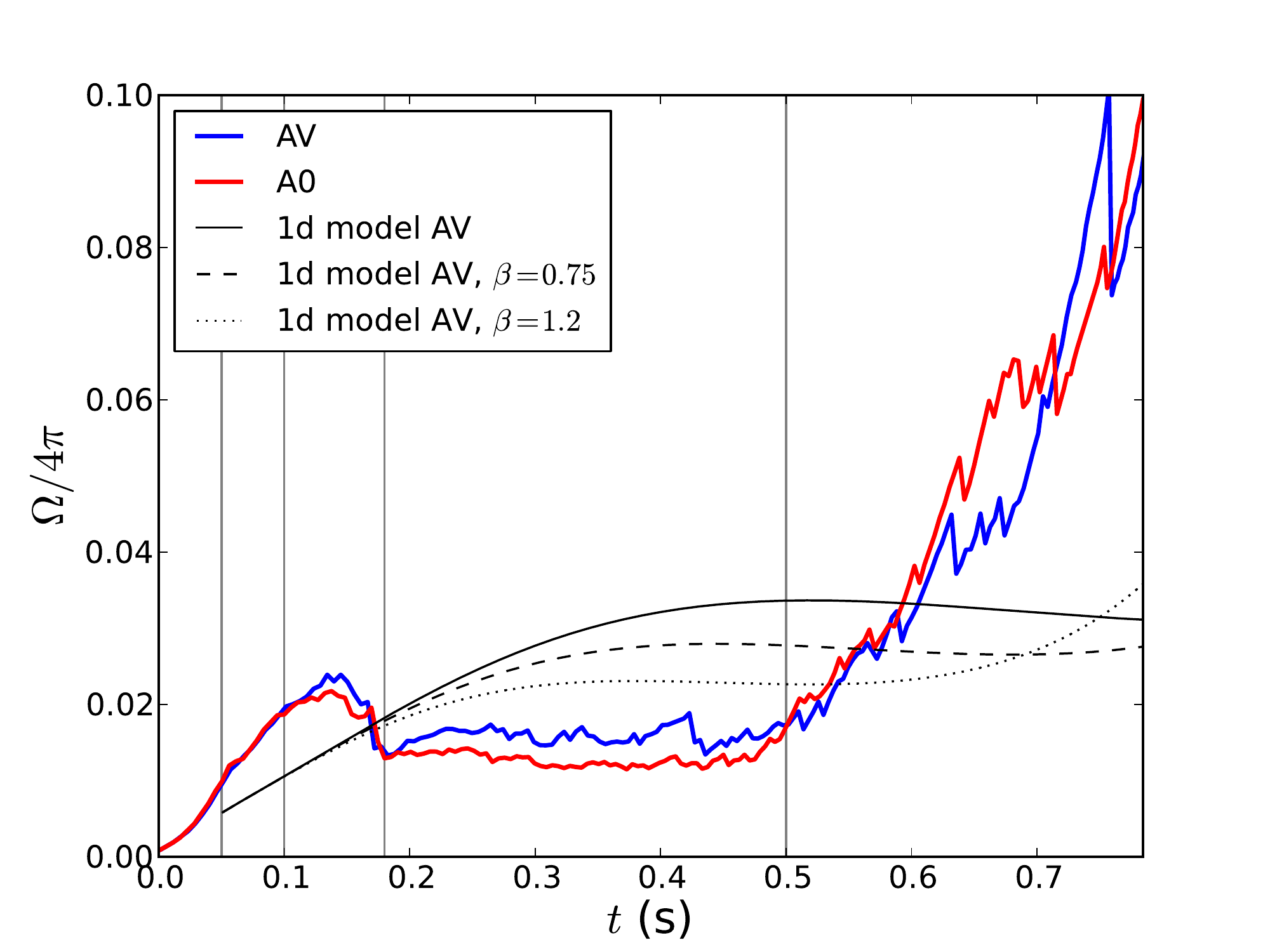}
    \caption{\label{fig:geo} Solid angle of the bubbles as seen from
      the center of the star for Model AV (blue), Model A0 (red), and
      the one-dimensional semi-analytic model based on Model AV.  The
      solid angle --- after laminar burning through the initial
      perturbations --- was $\Omega/4\pi = 4.4\times10^{-3}$.  The thin
      grey lines mark the different stages in the bubble's evolution
      as discussed in the text.  Briefly, these phases can be
      delineated as: laminar burn, floatation, vorticity formation,
      transition to turbulence, and lateral expansion, from left to
      right.}
  \end{center}
\end{figure}

\clearpage

\begin{figure}
  \begin{center}
    \includegraphics[width=\textwidth]{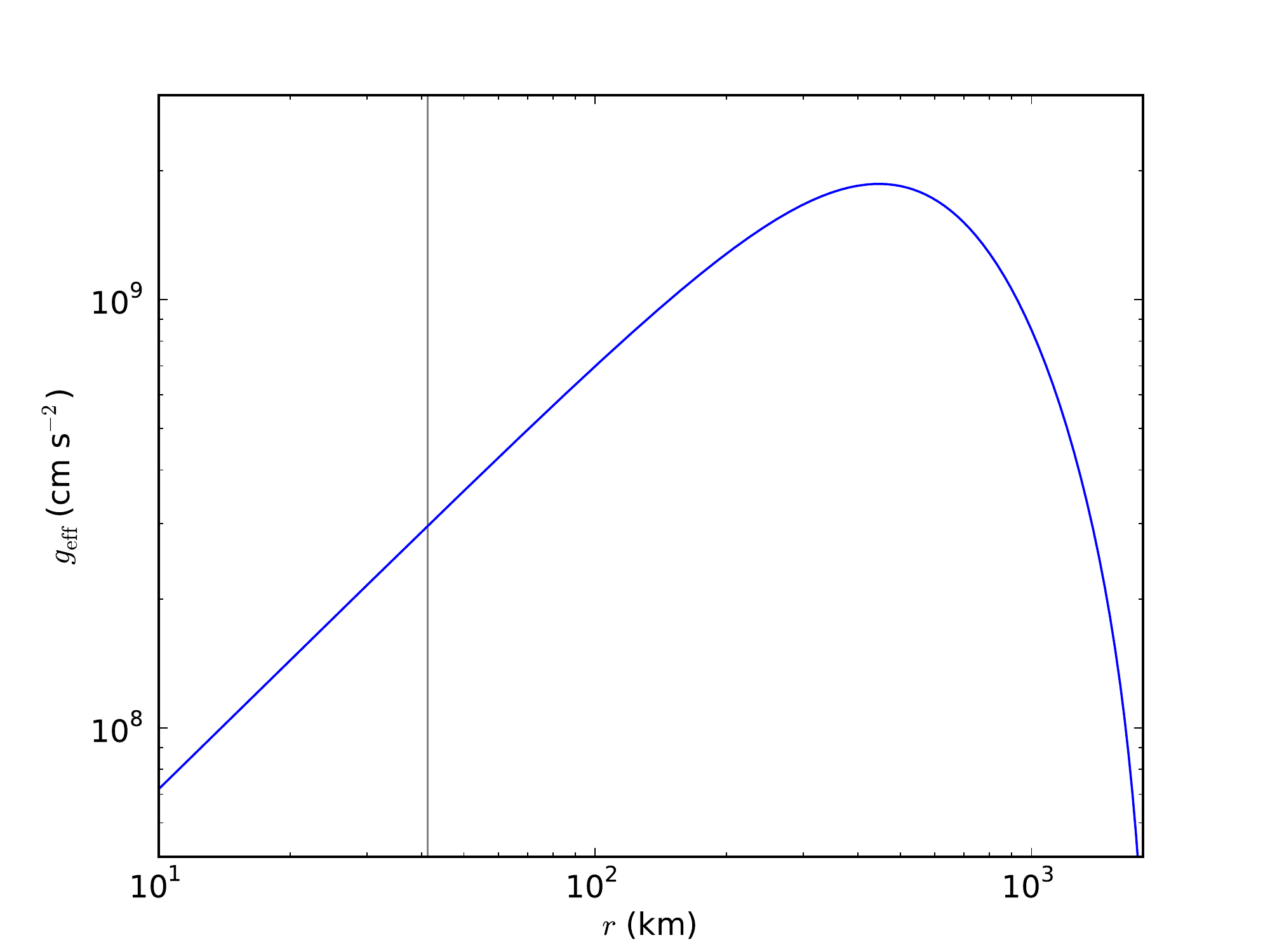}
    \caption{\label{fig:geff} Effective gravity as a function of
      radius for our density profile and Atwood number determined from
      the flame data in \cite{timmeswoosley1992}.  The grey vertical
      line marks the location of ignition for our model A runs.  For
      radii less than about 300 km, $g_{\rm eff}\propto r^{0.92}$; for
      larger radii, the decrease in density overcomes the increase in
      Atwood number, and $g_{\rm eff}$ plumets. }
  \end{center}
\end{figure}

\clearpage


\clearpage
\appendix

\section{Isotropic Burn Flame Test Problem}
\label{sec:isoburn}
We wanted to make sure that the flame was burning properly in all
directions, such that the thermal diffusion was behaving isotropically
for a given thermodynamic state.  To that end, we set up a constant
density ($\rho = 2\times10^9$ g cm$^{-3}$), constant temperature ($T =
5\times10^8$ K) atmosphere without gravity and added an ignition point
of the same size as that used in the simulations presented in this
paper.  This problem was performed on 576$^3$ zones with a resolution
of 100 m zone$^{-1}$.

\begin{figure}
  \begin{center}
    \includegraphics[width=0.45\textwidth]{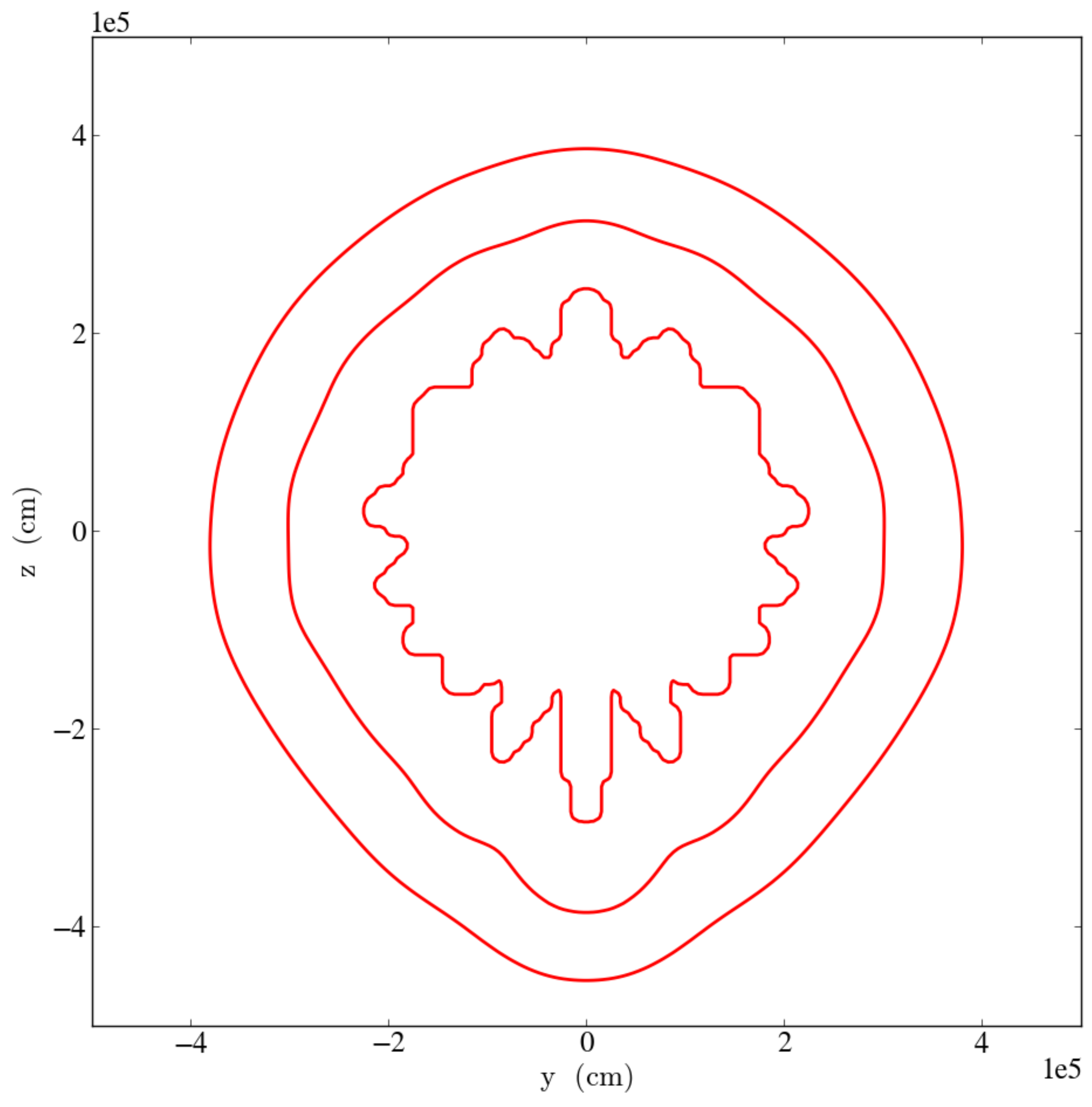}
    \includegraphics[width=0.45\textwidth]{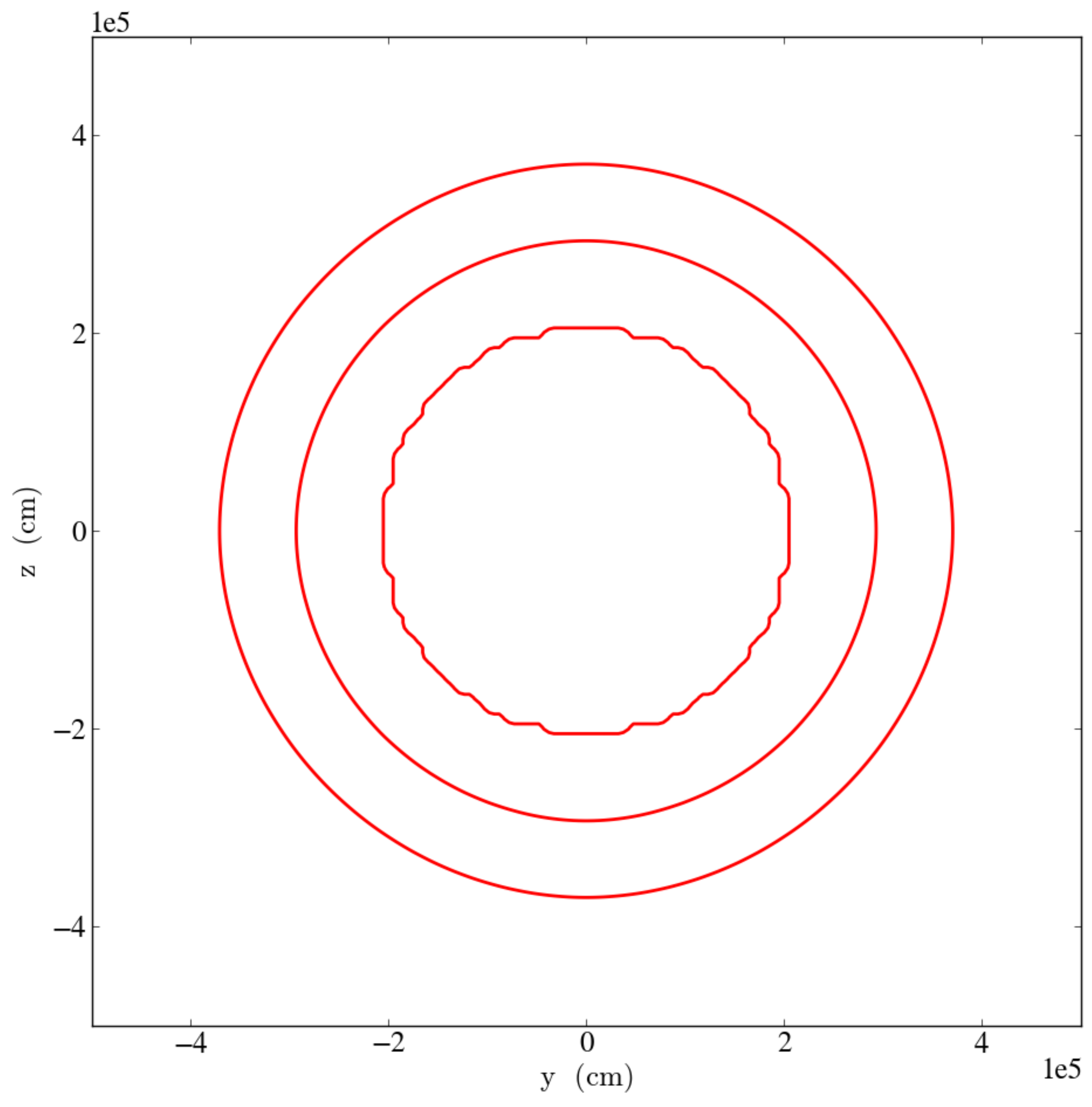}
    \caption{\label{fig:isoburn}Isotropic burn test problem for two
      different initial conditions --- perturbed sphere (left) and
      sphere (right).  Each of the contours marks the
      $X\left(^{12}\textrm{C}\right) = 0.45$ isocontours at $t=0,
      0.01,$ and $0.02$ s, with the initial conditions being the
      central contour.  The spacing between contours is consistent
      with the laminar speed used to move the flame, $v_l = 76$ km
      s$^{-1}$.}
  \end{center}
\end{figure}

We set up two different initial conditions for the shape of the
ignition spot as shown in the slices of Figure \ref{fig:isoburn}.  The
perturbed sphere in the left plot has the same shape and size as that
used in the simulations presented in this paper; the ignition spot in
the right plot is a perfect sphere, given our resolution.  Each of the
contours marks where $X\left(^{12}\textrm{C}\right)=0.45$, and they
are separated by $0.01$ s of evolution with the initial conditions
shown as the central contour.  Indeed, there is no preferred
direction to the flame propagation, and the initial perturbations are
burned through very quickly, as is the case in the main simulations of
this paper (e.g. Figure \ref{fig:early-burn}).  The constant flame
speed used in this test problem was set low so that the flame moved at
the minimum speed, the laminar speed, $v_l = 76$ km s$^{-1}$ at this
density.  The distance between contours in these plots should then be
760 m, which is consistent with the results of the test problem.

\section{Galilean Invariance of the Flame}
The stem of the flame experiences a strong crosswind --- the bulk
velocity in the vicinity of the flame's stem  is nearly parallel to the flame
surface, whereas the flame propagates normal to its surface.  We
wanted to check that this crosswind was not affecting the propagation
of the flame in an unexpected manner; we wanted to make sure the flame
propagation was Galilean-invariant.  This problem was performed on a
$128\times1024$ grid.

\begin{figure}
  \begin{center}
    \includegraphics[width=0.45\textwidth]{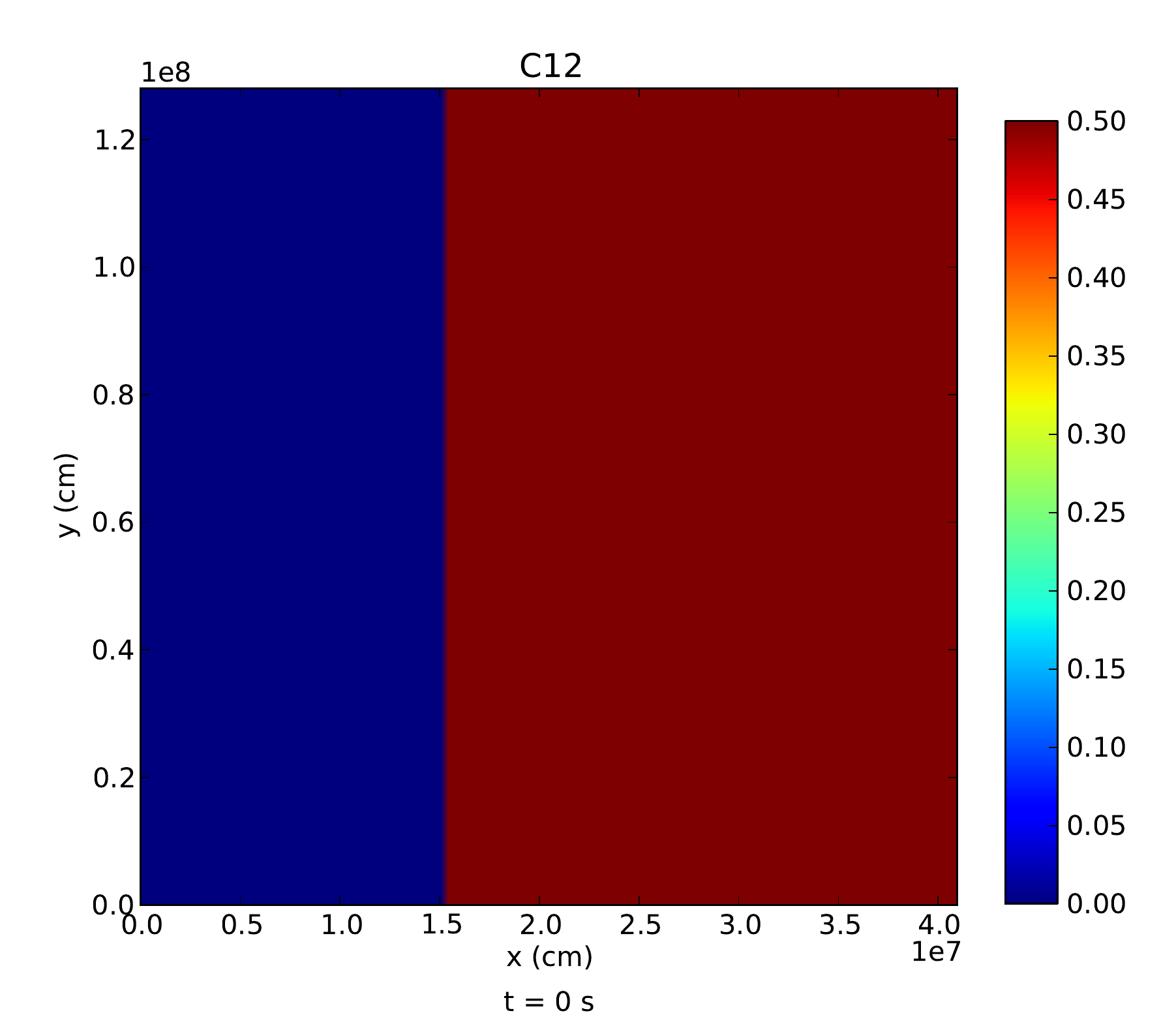}
    \includegraphics[width=0.45\textwidth]{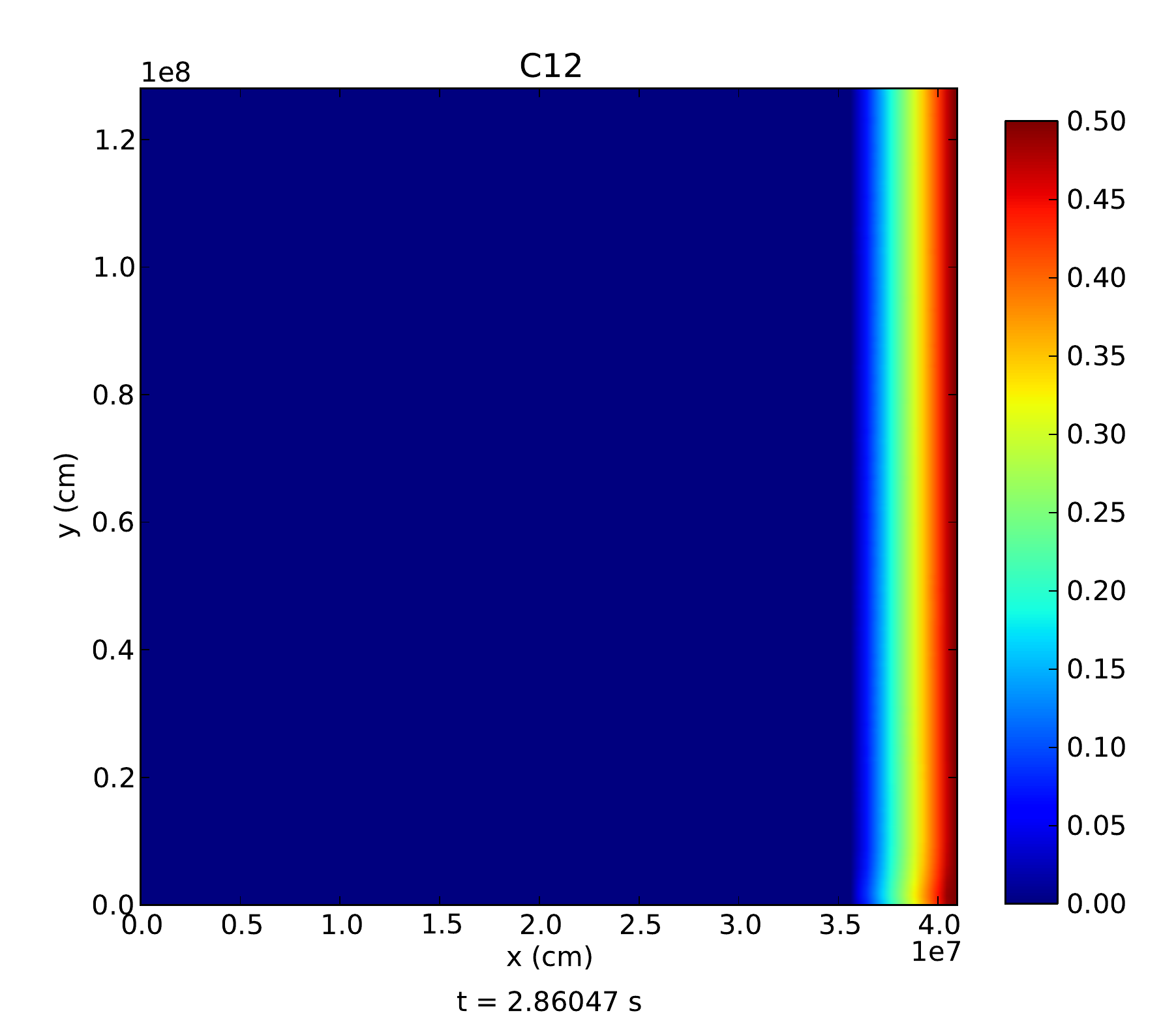}
    \includegraphics[width=0.45\textwidth]{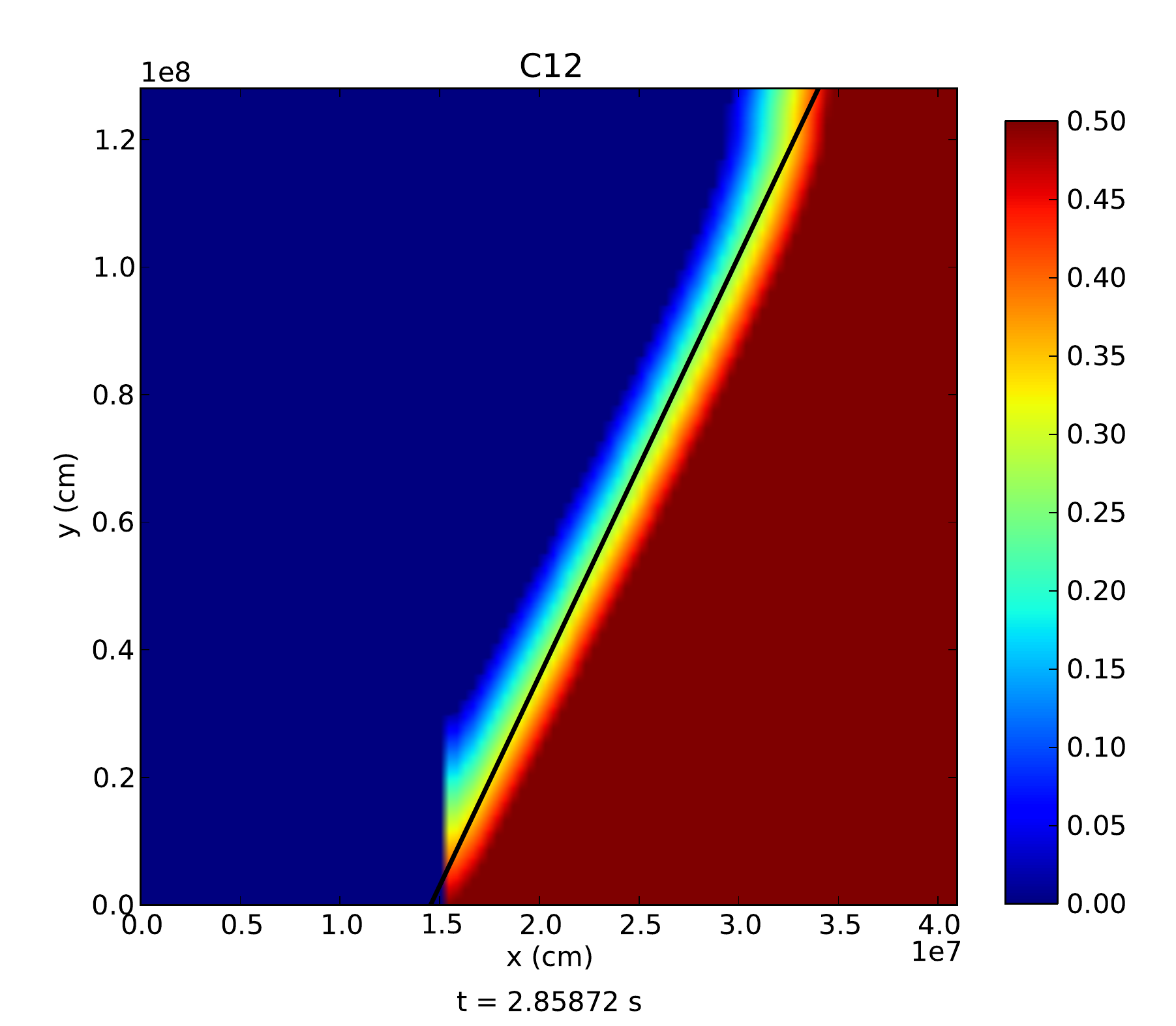}
    \includegraphics[width=0.45\textwidth]{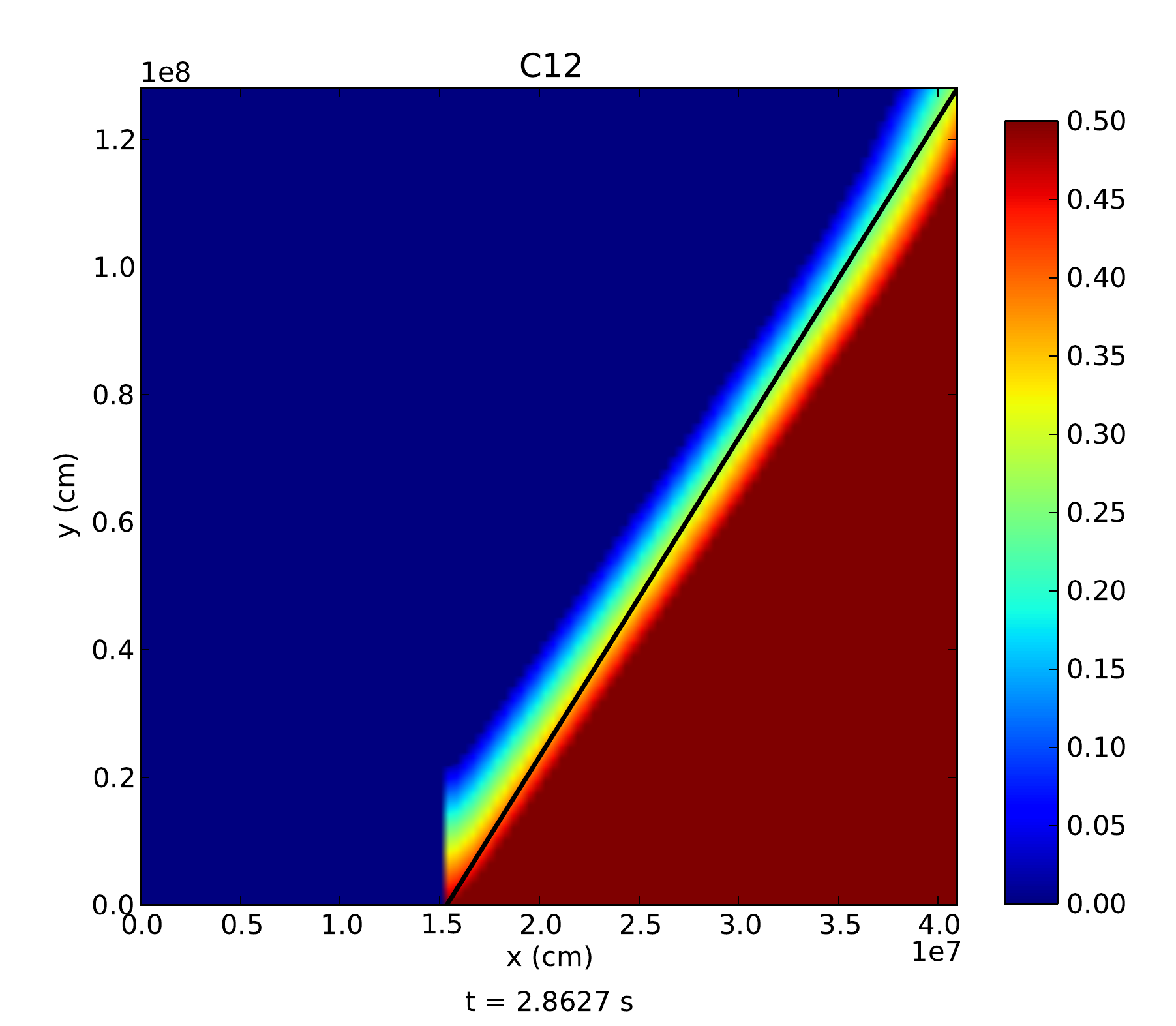}
    \caption{\label{fig:gal-test}Galilean invariance flame test
      problem.  Color maps show the $^{12}$C mass fraction, with the
      top left panel showing the initial conditions.  The boundary
      conditions are such that the initial composition is inflowing
      through the bottom at speed $v_{\textrm{in}}$ and the flame is
      burning to the right at speed $v_f$.  The top right panel is for
      $v_{\textrm{in}}=0$ and $v_f = v_l = 76$ km s$^{-1}$,
      i.e. laminar flame with no cross wind..  The bottom two panels
      have $v_\textrm{in}=500$ km s$^{-1}$, with the left panel using
      $v_f = v_l$ and the right $v_f = 100$ km s$^{-1}$.  The black
      lines in the bottom panels show the expected steady-state slope
      of the flame-ash interface.  Neglecting curvature effects from
      the inflow boundary condition, the expected behavior is very
      nearly reproduced. }
  \end{center}
\end{figure}

We used the same initial background conditions as in Section
\ref{sec:isoburn}: constant density ($\rho = 2\times10^9$ g
cm$^{-3}$), constant temperature ($T = 5\times10^8$ K) atmosphere in
the absence of gravity.  We isobarically turned the left 30\% of the
domain into hot ash; the initial conditions are shown as the upper
left panel in Figure \ref{fig:gal-test}.  The lower $x$ boundary
condition was set to a symmetry wall, and the upper $x$ and $y$
boundaries were set to outflow.  At the lower boundary, we inject the
same initial composition profile with a velocity,
$\vec{v_{\textrm{in}}} = \left(0,v_{\textrm{in}}\right)$, with the
inflow speed $v_{\textrm{in}}$ a free parameter.  In steady-state, a
Galilean-invariant flame surface should constitute a line with slope
$v_{\textrm{in}}/v_f$, with the limiting case of no inflow being a
vertical line.

Figure \ref{fig:gal-test} shows in the top right the $^{12}$C mass
fraction of a pure laminar ($v_l = 76$ km s$^{-1}$) flame with no
cross wind.  The bottom two panels experience a $v_{\textrm{in}} =$500
km s$^{-1}$ cross wind with the left panel using the laminar speed and
the right panel using $v_f = 100$ km s$^{-1}$.  All panels are at
about the same time, when the flame first hits the right wall in the
laminar case.  Note that in the bottom left panel, the flame also
first touches the right wall at this time, but it occurs off the top
of this panel; the images were cropped to keep the aspect ratio
manageable. Excepting an initial transient and adjustment at the
inflow boundary, the flames trace a straight line.  The black lines in
the bottom two panels show the expected behaviour with slopes of
$5/0.76$ for the left and 5 for the right.

\clearpage

\clearpage

\end{document}